\documentclass[11pt]{article}
\usepackage[utf8]{inputenc}
\usepackage{natbib}
\usepackage{array}
\usepackage{amsfonts}
\usepackage{bm}
\usepackage{caption}
\usepackage[mathscr]{euscript}
\usepackage{enumerate}
\usepackage{amsmath}
\usepackage{graphics}
\usepackage{graphicx}
\graphicspath{{./figures/}}
\usepackage{subfig}
\usepackage{algorithm}
\usepackage[noend]{algpseudocode}
\usepackage{authblk}
\usepackage{booktabs}
\usepackage[labelsep=period]{caption}
\usepackage[super]{nth}
\usepackage{multirow}
\usepackage{hyperref}
\usepackage{amsthm}
\usepackage{linegoal}
\usepackage{chngcntr}
\usepackage[margin=1in]{geometry}
\usepackage[noprefix]{nomencl}
\usepackage{array}
\usepackage{xpatch}
\usepackage{multicol}

\newcommand{\argmaxE}{\mathop{\mathrm{argmax}}}

\newtheorem{assumption}{Assumption}
\newtheorem{theorem}{Theorem}
\newtheorem{lemma}{Lemma}
\newtheorem{definition}{Definition}

\newcolumntype{L}[1]{>{\raggedright\arraybackslash}p{#1}}
\newcolumntype{C}[1]{>{\centering\arraybackslash}p{#1}}
\newcolumntype{R}[1]{>{\raggedleft\arraybackslash}p{#1}}

\makeatletter
\def\BState{\State\hskip-\ALG@thistlm}
\makeatother

\newcommand{\argmin}{\mathop{\rm arg~min}\limits}
\newcommand\norm[1]{\left\lVert#1\right\rVert}

\newcommand\fnsep{\textsuperscript{,}}

\makenomenclature

\title{\bf An axiomatic nonparametric production function estimator: Modeling production in Japan's cardboard industry}

\author[1]{Daisuke Yagi}
\author[2]{Yining Chen}
\author[1,3]{Andrew L. Johnson}
\author[3]{Hiroshi Morita}
\affil[1]{Department of Industrial and Systems Engineering\\Texas A\&M University}
\affil[2]{Department of Statistics\\London School of Economics and Political Science}  
\affil[3]{Department of Information and Physical Sciences\\Osaka University}

\date{\today}

\begin{document}
\def\spacingset#1{\renewcommand{\baselinestretch}%
	{#1}\small\normalsize} \spacingset{1}

\maketitle

\begin{abstract}
We develop a new approach to estimate a production function based on the economic axioms of the Regular Ultra Passum law and convex non-homothetic input isoquants. Central to the development of our estimator is stating the axioms as shape constraints and using shape constrained nonparametric regression methods. 
We implement this approach using data from the Japanese corrugated cardboard industry from 1997--2007. Using this new approach, we find most productive scale size is a function of the capital-to-labor ratio and the largest firms operate close to the largest most productive scale size associated with a high capital-to-labor ratio. We measure the productivity growth across the panel periods based on the residuals from our axiomatic model. 
We also decompose productivity into scale, input mix, and unexplained effects to clarify the sources the productivity differences and provide managers guidance to make firms more productive. 

\end{abstract}


\newpage
\spacingset{1.5}
\section{Introduction}
\label{sec:intro}
 
How does scale and input mix affect a firm's productivity? This question is vital to any models that aim to study the effects of automation. Since productivity is a scalar measure defined as the ratio of output to input, a fundamental challenge to answering this question lies in modeling how firms aggregate inputs.

The standard approach is to use growth accounting methods which calculates the parameters of implied parametric production function, see for example \cite{barro2004economic}. Similarly if production function estimation is to be performed, the Cobb--Douglas production function is the most common specification.\footnote{Many extensions of the Cobb--Douglas function have been developed, the most widely known is the Trans-log production function which is a second order Taylor series expansion at a point of the Cobb--Douglas \citep{christensen1973transcendental}. However, the Trans-log function inherits certain drawbacks from the Cobb-Douglas production function, including the parametric limitation.} However, the Cobb--Douglas function has several restrictive characteristics. Specifically, it implies that the input isoquants are homothetic, the elasticity of substitution between inputs is one, and the function can have either increasing or decreasing returns-to-scale, but not both. While the Trans-log relaxes both the later two of these restrictions, it often does not satisfy even basic economic axioms such as convex input isoquants or it may not have positive marginal product estimates. Perhaps for these reasons, the Cobb--Douglas production function, whether implied or estimated, remains the work horse for empirical research on productivity \citep{syverson2011determines}. 

The goal of this paper is to develop a new approach that is less dependent on functional form assumptions to estimate a production function while maintaining basic economic axioms. We use nonparametric local averaging methods, but augment these methods with shape constraints that reflect economic axioms. Nonparametric local averaging methods without shape constraints would avoid the potential for functional form misspecification and flexibly capture the nuances of the data, but would be difficult to interpret economically and would not satisfy some commonly accepted economics theory, e.g. positive rates of marginal substitution or non-negative marginal products. Thus, we can use a minimal set of economic axioms which are unlikely to be violated 
while providing additional structure. The axioms we impose are the Regular Ultra Passum (RUP) law as the scaling property\footnote{As explained below, we will actually use an S-shape restriction which requires a single inflection point, but otherwise generalizes the RUP law. Under this condition the most productive scale size is equivalent to the minimum efficient scale of production. See \cite{aksaray2017density}.} and that input isoquants are convex but could be potentially non-homothetic. This new modeling approach estimates the most productive scale size conditional on input mix.

The RUP law \citep{frisch1964theory} states that along any expansion path, the production function should first have increasing returns-to-scale followed by decreasing returns-to-scale. Intuitively, when a firms is small it tends to face increasing returns-to-scale because it can increase productivity easily through specialization and learning \citep{bogetoft1996dea}. In contrast, as the scale size becomes larger, a firm tends to have decreasing returns-to-scale due to scarcity of ideal production inputs and challenges related to increasing span of control. Firms in competitive markets should operate close to the most productive scale size in the long-run to minimize the cost per unit and assure positive profits. The RUP law with a single inflection point will assure we have a well-defined marginal products and most productive scale sizes.

Convex input isoquants, which are a standard assumption in production theory, are motivated by the argument that there are optimal proportions in which inputs should be used for production and that deviations from the optimal proportion by decreasing the level of one input, such as capital, will require more than a proportional increase in another input, such as labor \citep{petersen1990data}. Relaxing the homotheticity of input isoquants allows the optimal proportions to depend on the output level. For example, the optimal proportion of inputs for low output levels could be more labor intensive than at higher output levels. Further, non-homothetic isoquants allows for the most productive scale size measured with different input mix to exist at different output levels. Non-homothetic isoquants allows us to more easily capture the empirical fact that productivity levels are a function of capital-to-input ratio. 

The axiomatic approach is critical for interpreting the estimates of a production function to gain managerial insights. The production function is often used to estimate firm expansion behavior including how many resources need to be added to expand output or how automation (i.e. changing the capital-to-labor ratio) can be used to achieve larger scales of production. Without data and production function estimates, managers are left to make these decisions based on a firms historical behavior or rules-of-thumb or other approximations. The analysis of firm as a whole allows for the accounting of synergies between inputs in the production process. 

We implement our approach using data from Japan's corrugated cardboard industry. As classified in the Japanese Census of Manufactures, the cardboard industry\footnote{In the \emph{Japan Standard Industrial Classification} (JSIC), the corrugated cardboard industry is indexed as Industry 1453.} includes both cardboard manufacturers and cardboard box manufacturers. The latter sector is not particularly capital intensive nor does it require technical know-how to enter, thus firms tend to focus more on customer service and lead-times. Overall, the industry has a few large firms and many smaller firms, which is typical for a mature manufacturing industry. The largest firms in the industry are vertically integrated and include cardboard production, box making, and paper making.\footnote{In the Census of Manufacturing, establishments are classified by industry based on the primary product produced in the establishment. Paper making establishments are typically specialized and do not appear in our data set. However, vertically integrated firms that own paper producing establishments typically have larger cardboard and box making establishments.}

In the cardboard industry, like most industries, firms enter the market as small firms and must expand over time taking advantage of capital and labor specialization or other characteristics of the technology to be more productive \citep{haltiwanger2013creates,foster2016slow}. Recently, medium and large sized firms in the industry have been acquiring smaller firms and reducing the combined input levels without significant reductions in the combined output levels, leading to higher productivity levels. In particular, since the medium size firms are operating below the most productive scale size, they have the potential for significant increase in productivity by increasing their scale of production, thus mergers are attractive to medium sized firms. Unlike previous models, our models motives mergers by making the productivity benefits of increasing scale size explicit. 


Several nonparametric shape constrained estimators have been proposed that combine the advantage of avoiding functional misspecification with improving the interpretability of estimation results relative to unconstrained nonparametric methods, see for example \cite{kuosmanen2015stochastic} or \cite{yagi2018shape}. However, existing methods only allow the imposition of simple shape constraints such as  monotonicity and concavity \citep{seijo2011nonparametric,lim2012consistency}. These structures exclude economic phenomena such as increasing returns to scale due to specialization, fixed costs, or learning. Thus, more general functional structures, like the model proposed in this paper, are desirable. 

There have been two previous attempts to develop estimators that impose the RUP law as shape constraints. \cite{olesen2014maintaining} develop an algorithm to estimate a Data Envelopment Analysis (DEA)-type estimator satisfying the RUP law and impose homotheticity on the input isoquants. Noise is not modeled in DEA estimators and all deviations from the estimated function are one-sided and negative. \cite{hwangbo2015power} introduce noise and estimate a scaling function using nonparametric shape constrained methods. However, they also assume homothetic input isoquants and do not provide statistical properties for their estimators. In conclusion, these estimation methods place structure on production function, but the homothetic assumption is not flexible enough to capture a variety of realistic and potential production structures. These drawbacks are to be addressed in our approach.

For the data analysis, we will use our production model to provide a description of the supply-side of the Japanese cardboard industry as we report most productive scale size, productivity evolution and decomposition. We find most productive scale size is dependent on the capital-to-labor input factor ratio and the largest firms operate close to the largest most productive scale size associated with a high capital-to-labor ratio. 

We also decompose the productivity into the scale and input mix productivity to clarify the source of productivity differences. This decomposition provides critical managerial insights for scale and input mix of each firm. Specifically, we find that large capital intensive firms get benefits from both scale and input mix while small capital intensive firms need either expansion of scale size or adjustment of input mix to improve productivity. These scale and mix effects account for significant portion of the productivity estimated by a conventional methods resulting in a much smaller component of unexplained productivity variation.

The remainder of this paper is as follows. Section \ref{sec:model} introduces the proposed production function model and its assumptions. Section \ref{sec:iter}  explains the ideas behind the two-step estimation procedure and the algorithm for our estimator. All the details can be found in Appendix \ref{app:algo}. Statistical properties of the estimator is investigated in Section \ref{sec:property}. Section \ref{sec:simulation} discusses the Monte Carlo simulation results under several different experimental settings. Section \ref{sec:application} applies our estimator to estimate a production function for the Japanese cardboard industry. We conclude in Section \ref{sec:conclusion} with future research directions. Proofs of all the theorems are deferred to  Appendix \ref{app:proof}.

\section{Model framework}
\label{sec:model}

To facilitate our discussion, in this section, we consider the following production function model in the noiseless setting.
\begin{equation}
    \label{eq:prod1}
    y=g_0(\bm{x}),
\end{equation}
where $\bm{x}=(x_1,x_2,\ldots,x_d)'$ is $d$-dimensional input vector, $y$ is an output scalar, and $g_0: \mathbb{R}_+^d \rightarrow \mathbb{R}_+$ is a production function.

\nomenclature[$y$]{$y$}{Output scalar}
\nomenclature[$x$]{$\bm{x}$}{Input vector}
\nomenclature[$g$]{$g_0(\bm{x})$}{True production function}
\nomenclature[$d$]{$d$}{Number of inputs}
\nomenclature[$\phi$]{$\phi(y,\bm{x})$}{Transformed production function}


\begin{definition}
    An input isoquant $\bar{V}(y) = \{\bm{x}: g_0(\bm{x})=y\}$ be the sets of input vectors capable of producing each output $y$. 
	\label{ass:isoq1}
\end{definition}

\nomenclature[$v$]{$\bar{V}(y)$}{Input isoquant at output level $y$}

We write 
\begin{equation}
    \label{eq:prod2}
    \phi(y,\bm{x})=y-g_0(\bm{x}).
\end{equation}
and make the following assumptions on $g_0$ and $\phi$:
\begin{assumption}
\label{ass:1}
\leavevmode
    \begin{enumerate}[(i)]
        \item $g_0(\cdot)$ is a strictly monotonically increasing and Lipschitz function.
        \item $\phi(\cdot,\cdot)$ is a twice-differentiable function.
    \end{enumerate}
\end{assumption}
Under Assumption \ref{ass:1}, by the implicit function theorem, there exists an implicit function $\mathscr{H}_{0,k}$ such that
\begin{equation}
    \label{eq:isoq}
    x_k = \mathscr{H}_{0,k}(x_1,\ldots,x_{k-1},x_{k+1},\ldots,x_d;y) \equiv \mathscr{H}_{0,k}(\bm{x}_{-k};y)~~~\mbox{ for all } k=1,\ldots,d,
\end{equation}
where $\bm{x}_{-k}=(x_1,\ldots,x_{k-1},x_{k+1},\ldots,x_d)'$ is an input vector without the $k$-th input. 
\nomenclature[$k$]{$k$}{Index of input dimensions: $k=1,\ldots,d$}
\nomenclature[$h$]{$\mathscr{H}_{0,k}(\bm{x}_{-k};y)$}{Implicit isoquant function}
\nomenclature[$x$]{$\bm{x}_{-k}$}{Input vector without the $k$-th input}

We are interested in estimating a production function $g_0$ having both convex input isoquants for all output levels and that satisfies an augmented version of the RUP law. The input convexity implies the following conditions on $\mathscr{H}_{0,k}$:
\begin{definition}
\label{def:inputconvex}
    An input isoquant is input-convex if for every $k=1,\ldots,d$, any pair of arbitrary input vectors $\bm{x}_a,\bm{x}_b\in \mathbb{R}_+^{d-1}$, $y \in \mathbb{R}_+$ (where $\mathscr{H}_{0,k}(\bm{x}_a;y)$ and $\mathscr{H}_{0,k}(\bm{x}_b;y)$ are well-defined) and $\lambda \in [0,1]$,
	\leavevmode
	\label{ass:isoq}
	\begin{enumerate}[(i)]
		\item (Convex input isoquant) \\ 
		$\lambda \mathscr{H}_{0,k}(\bm{x}_a;y)+(1-\lambda)\mathscr{H}_{0,k}(\bm{x}_b;y)\geq \mathscr{H}_{0,k}(\lambda \bm{x}_a+(1-\lambda)\bm{x}_b;y)$
		\item (Monotone decreasing input isoquant) \\
		If $\bm{x}_a\leq \bm{x}_b\mbox{, then } \mathscr{H}_{0,k}(\bm{x}_a;y)\geq \mathscr{H}_{0,k}(\bm{x}_b;y)$.
	\end{enumerate}    
\end{definition}
\noindent Intuitively, input convexity implies the existence of an optimal ratio of inputs. Deviations from the optimal input ratios by decreasing the use of a particular input will result in more than a proportional increase in other inputs. Further, larger deviations from the optimal ratio will require larger increases in input consumption to maintain the same output level. Finally, it can be shown that to verify Definition~2, it suffices to check that it holds for any particular $k\in \{1,\ldots,d\}$.

Next, we define the elasticity of scale\footnote{This variable was referred to as the passum coefficient in the seminal work of \cite{frisch1964theory}, but is now commonly referred to as the elasticity of scale.}, $\varepsilon(\bm{x})$, relative to a production function $g_0(\bm{x})$ as
\begin{equation}
    \label{eq:elasticity}
    \varepsilon(\bm{x})=\sum_{k=1}^{d}\frac{\partial g_0(\bm{x})}{\partial x_k}\frac{x_k}{g_0(\bm{x})}.
\end{equation}
\nomenclature[$\epsilon$]{$\varepsilon(\bm{x})$}{Elasticity of scale}
The Regular Ultra Passum (RUP) law was originally proposed by \cite{frisch1964theory}. A version of its extension is given as follows:
\begin{definition}
\label{def:rup}
    (\cite{forsund2004all}) A production function $g_0(\bm{x})$ obeys the Regular Ultra Passum law if $\frac{\partial\varepsilon(\bm{x})}{\partial x_k}<0$ for every $k=1,\ldots,d$, and for some input $\bm{x_a}$ we have $\varepsilon(\bm{x}_a)>1$, and for some input $\bm{x}_b$ we have $\varepsilon(\bm{x}_b)<1$, where $\bm{x}_b>\bm{x}_a$.\footnote{$\bm{x}_a$ and $\bm{x}_b$ are vectors such that the inequality implies that every component of $\bm{x}_b$ is greater than or equal to every component of $\bm{x}_a$.}\fnsep\footnote{This definition of the RUP law modifies \cite{frisch1964theory}'s original definition. This definition does not require the passus coefficient to drop below 0, thus implying congestion or that the production function is not monotonically increasing. This characterization allows for a monotonically increasing production function. Also note that although a concave production function nests within this definition,the definition does not require that the function is  ``nicely concave" as defined in \cite{ginsberg1974multiplant}.}
\end{definition}

Intuitively, for any ray from the origin, a production function $g_0$ has increasing returns to scale followed by decreasing returns to scale. However, note that in both \cite{forsund2004all} and Frisch's original definition, neither rules out the possibility of multiple inflection points; see Appendix \ref{app:ce_rup} for a more detailed explanation. Furthermore, because the RUP law is defined in terms of the elasticity of scale, the law does not allow the function, $g_0$, to grow at an exponential rate. To overcome these issues, we introduce the following definition of an S-shape function.

\begin{definition}
\label{def:s-shape}
    A production function $g_0: \mathbb{R}_+^d \rightarrow \mathbb{R}_+$ is S-shaped if for any $\bm{v}\in\mathbb{R}_+^d$ defining a ray from the origin in input space $\alpha\bm{v}$ with $\alpha>0$, $\nabla_{\bm{v}}^2 g_0(\alpha\bm{v}) > 0$ for $\alpha\bm{v}<\bm{x}^*$, and $\nabla_{\bm{v}}^2 g_0(\alpha\bm{v})< 0$ for $\alpha\bm{v}> \bm{x}^*$ along a ray from the origin, where $\nabla_{\bm{v}}^2 g_0$ is the directional second derivative of $g_0$ along $\bm{v}$. This implies that for any ray from the origin of direction $\bm{v}$, there exists a single inflection point $\bm{x}^*$ that $\nabla_{\bm{v}}^2 g_0(\bm{x^*}) = 0$.\footnote{Note this definition is consistent with \cite{ginsberg1974multiplant} definition of a convex-concave function. See also \cite{baumol1983contestable}.} 
    
\end{definition}

\begin{figure}[ht]
	\centering
	\subfloat[one-input]{\includegraphics[width=0.5\textwidth]{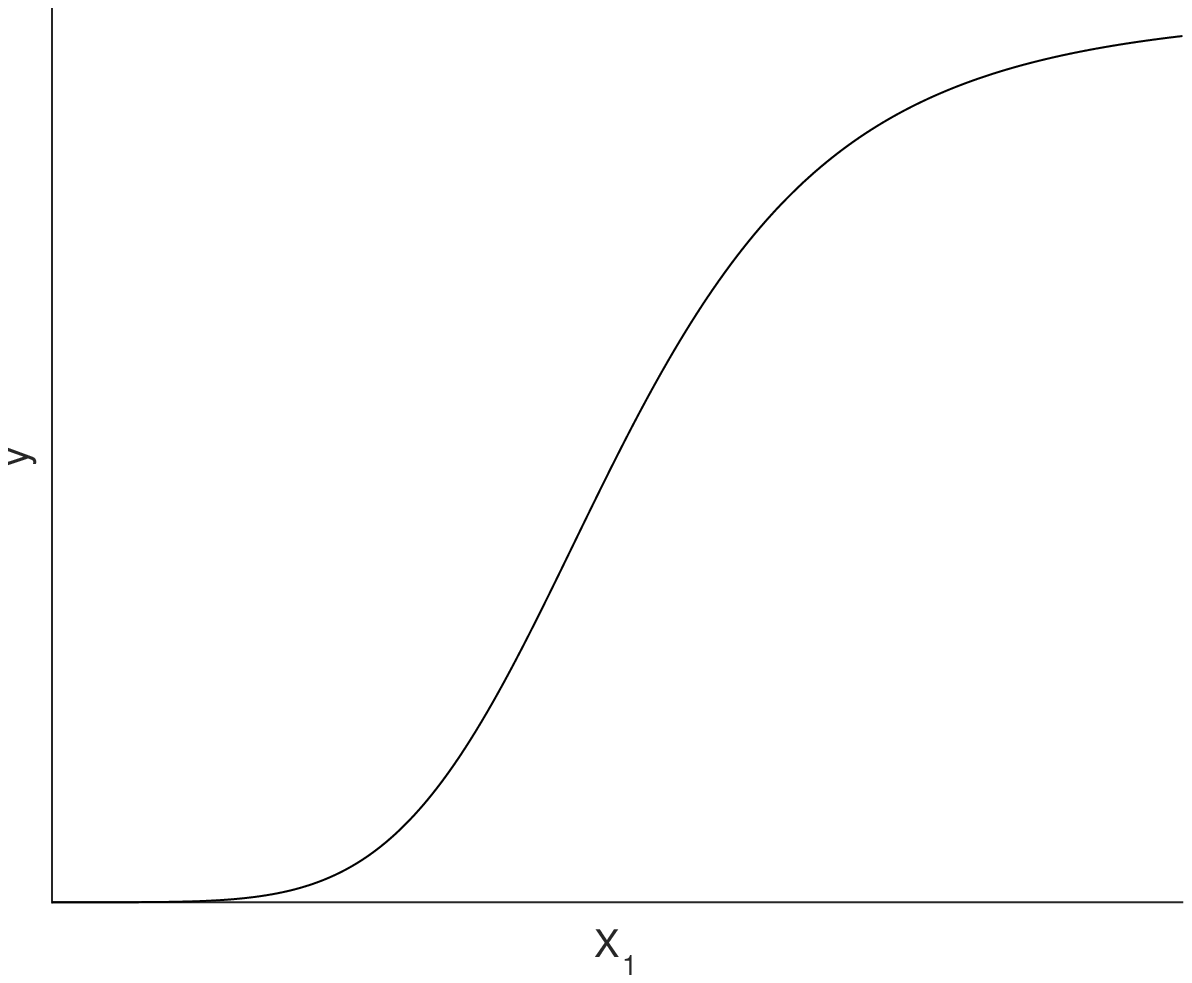}\label{fig:RUP2d}}
	\hfill
	\subfloat[two-input]{\includegraphics[width=0.5\textwidth]{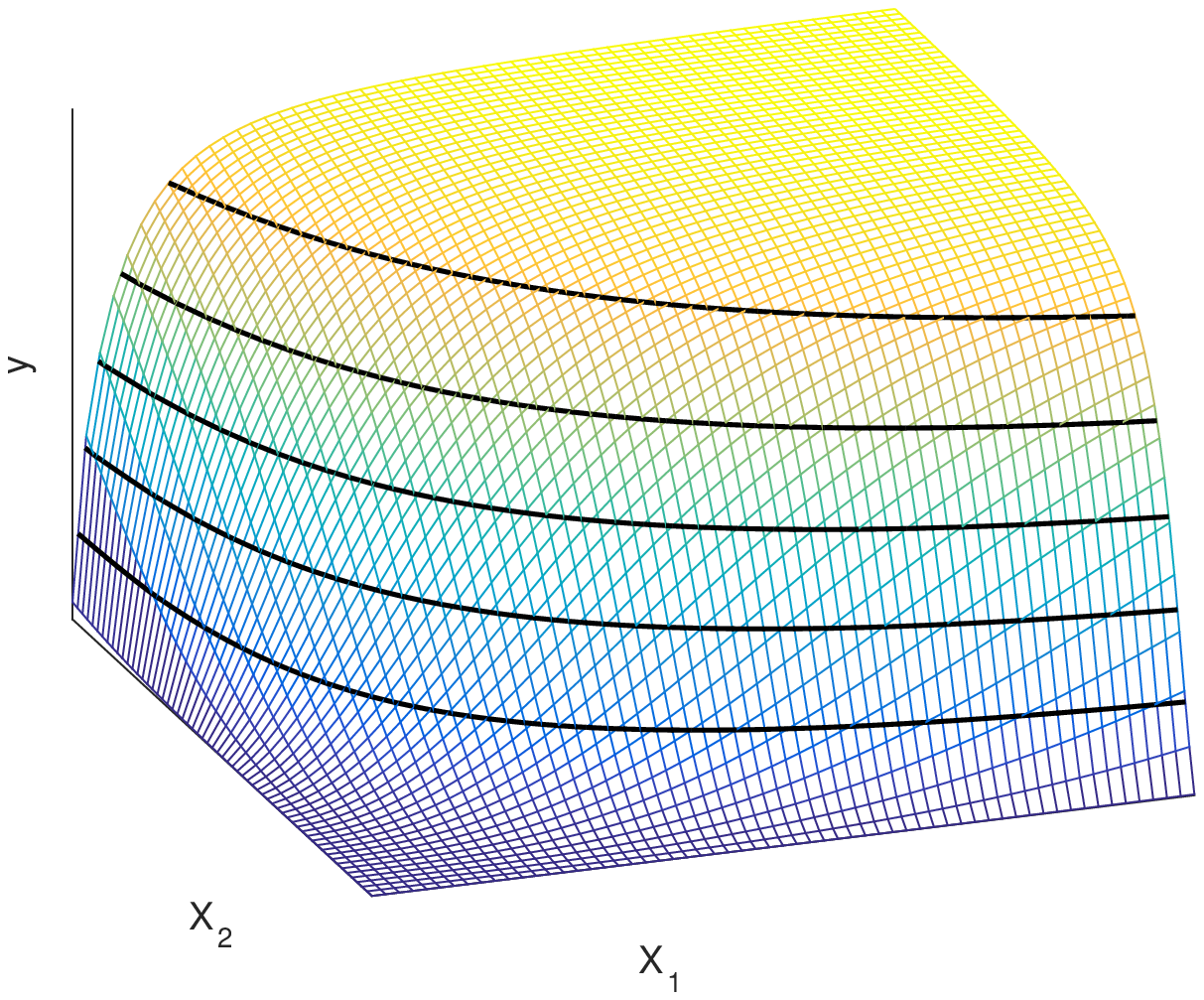}\label{fig:RUP3d}}
	\caption{Production functions satisfying both the RUP law and S-shape definition}
	\label{fig:RUP}
\end{figure}

Figure \ref{fig:RUP} \subref{fig:RUP2d} and \subref{fig:RUP3d} show two examples of the production function with one-input and two-input, respectively. Both functions satisfy the RUP law and the S-shaped definition. 
The relationship between the RUP law and an S-shape function is characterized in the following lemma.

\begin{lemma}
\label{lmm:RUP}
     If a production function $g_0: \mathbb{R}_+^d \rightarrow \mathbb{R}_+$ is second-differentiable, monotonically increasing and satisfies the RUP law and there exists a single inflection point $\bm{x}^*$ where $\nabla_{\bm{v}}^2 g_0(\bm{x^*}) = 0$ for any ray from the origin defined by a direction $\bm{v}\in\mathbb{R}_+^d$, then $g_0$ is S-shaped. 
\end{lemma}
\nomenclature[$x$]{$\bm{x}^*$}{Inflection point}
\nomenclature[$v$]{$\bm{v}$}{Direction vector defining a ray in input space}

\begin{figure}
	\centering
	\subfloat[Homothetic]{\includegraphics[width=0.5\textwidth]{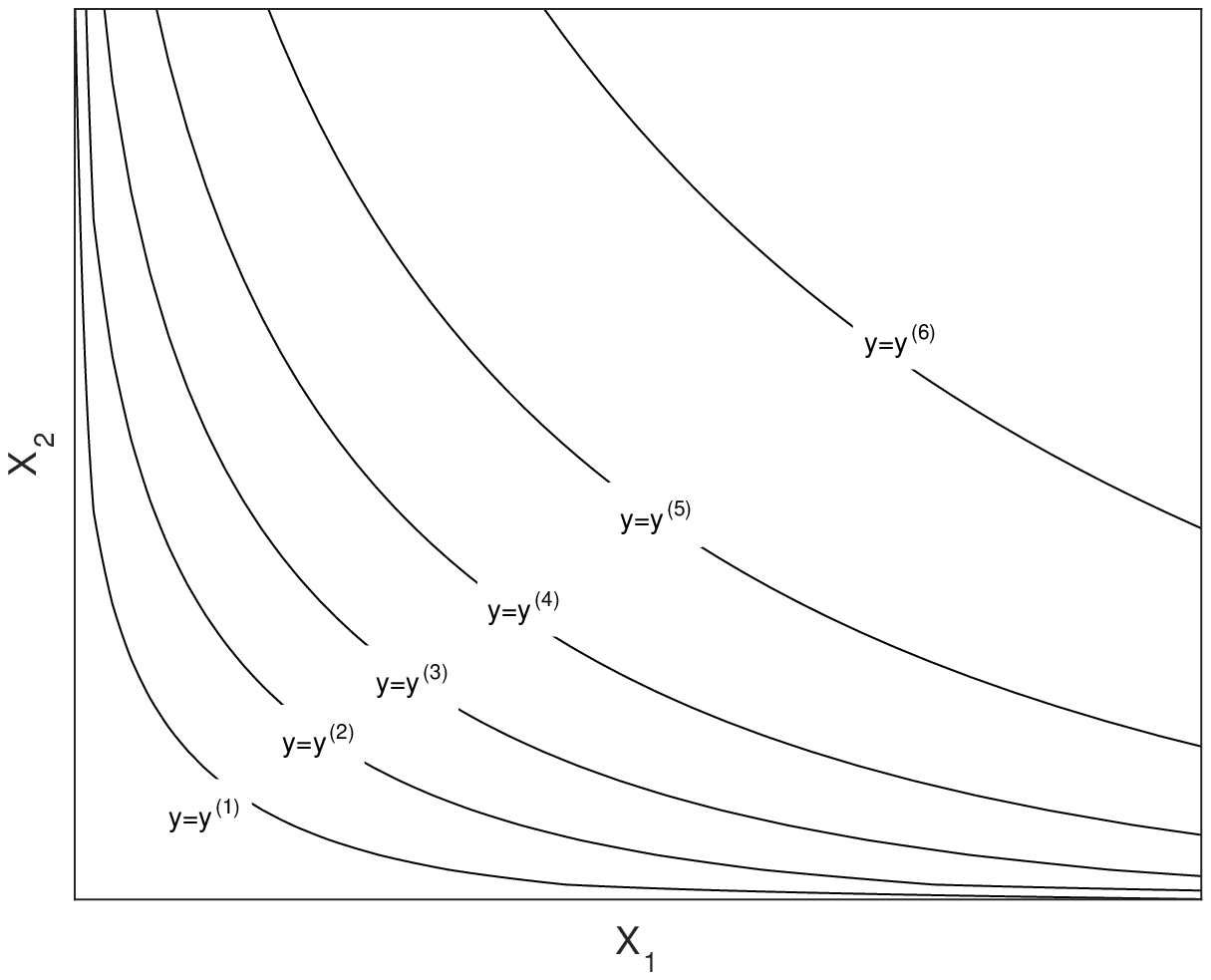}\label{fig:isoq_h}}
	\hfill
	\subfloat[Non-homothetic]{\includegraphics[width=0.5\textwidth]{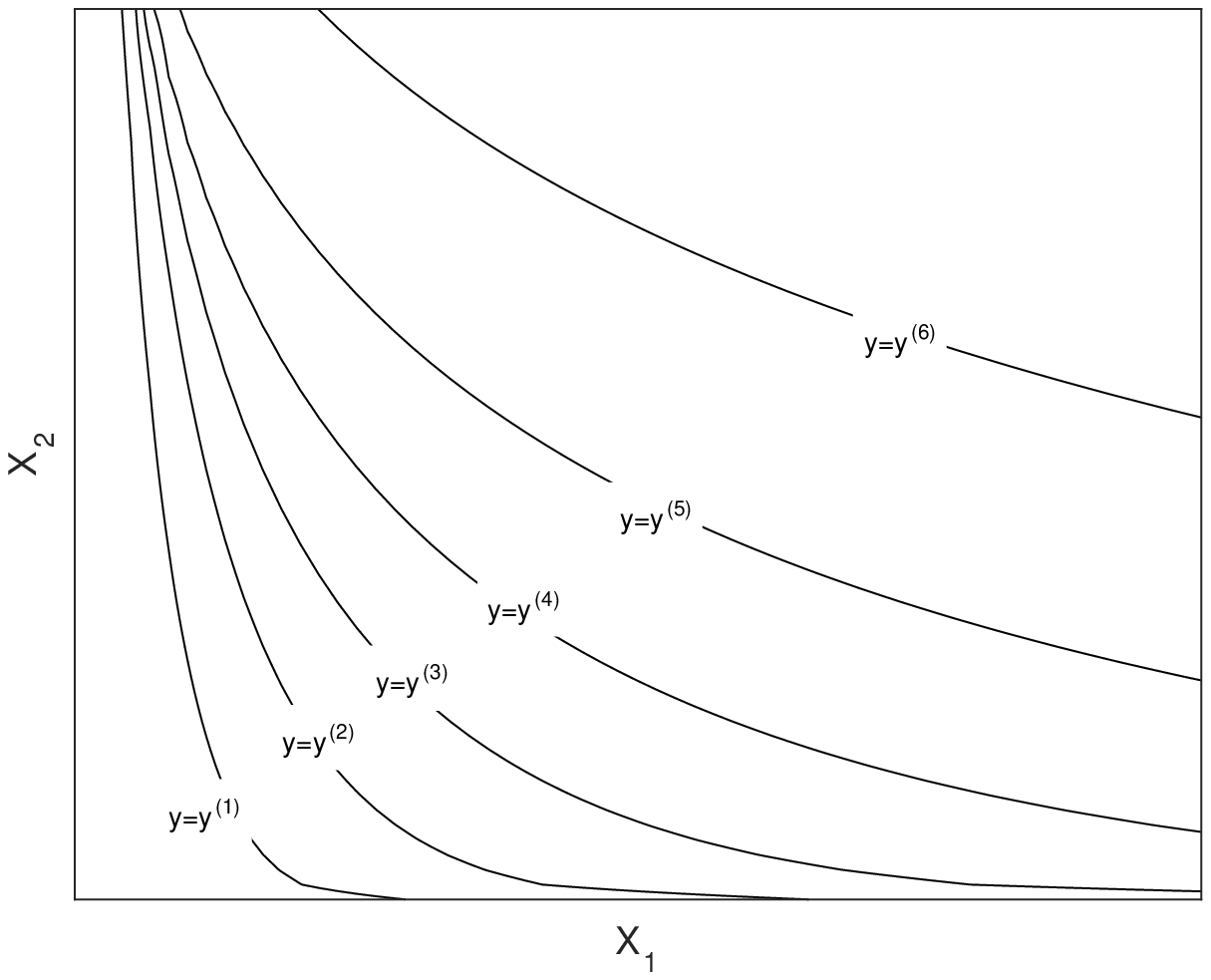}\label{fig:isoq_nh}}
	\caption{Input isoquants satisfying input convexity.}
	\label{fig:isoq}
\end{figure}

Another common assumption for production functions is homotheticity.
\begin{definition}
\label{def:homothetic}
    A production function $g_0$ is homothetic if for every $\bm{x}$, $\alpha > 0$ and $k=1,\ldots,d$, the implicit function $\mathscr{H}_{0,k}$ is homogeneous of degree one, i.e.
    \[
    \alpha x_k = \mathscr{H}_{0,k}(\alpha x_1,\ldots,\alpha x_{k-1},\alpha x_{k+1},\ldots,\alpha x_d;g_0(\alpha \bm{x})).
    \]  
\end{definition}
Input homotheticity is a strong assumption because it restricts input elasticity to be constant for a given input mix at all scales of production. However, by relaxing input homotheticity and assuming only input-convexity, each isoquant can have different shapes and curvatures at a given $y$-level. We refer to isoquants of this type as non-homothetic, convex input isoquants. Figure \ref{fig:isoq} shows examples of production functions with homothetic and non-homothetic isoquants with two-dimensional input.  

In the following, we prove that a homothetic production function which satisfies the S-shape definition for a single ray from the origin will also satisfy the S-shape definition for any expansion path. To achieve this, we require the following alternative characterization for a homothetic production function. 

\begin{definition}
    (Alternative definition of homothetic production function)
    A production function $g_0(\bm{x})=F_0(\mathscr{H}_0(\bm{x}))$ is homothetic if 
    \begin{enumerate}[(i)]
        \item{Scale function $F_0:\mathbb{R}\rightarrow\mathbb{R}$ is a strictly monotone increasing function, and}
        \item{Core function $\mathscr{H}_0: \mathbb{R}^d\rightarrow\mathbb{R}$ is a homogeneous of degree 1 function which implies $\mathscr{H}_0(\alpha\bm{x})=\alpha\mathscr{H}_0(\bm{x})$ for all $\alpha>0$,}
        \item with the identifiablility condition $\mathscr{H}_0((1,\ldots,1)^T) = 1$.
    \end{enumerate}
\end{definition}
\nomenclature[$f$]{$F_0(\cdot)$}{Scale function}

Note that the identifiability condition is necessary because otherwise, we could have set $F'(\cdot) = F(t\times \cdot)$ and $\mathscr{H}'(\cdot) = \mathscr{H}(\cdot)/t$ for any constant $t>0$ so that $F(\mathscr{H}(\cdot)) = F'(\mathscr{H}'(\cdot))$, so $F$ and $\mathscr{H}$ would not be identifiable. 

Define ${X}_{max,k}=\max {\bm{X}}_k$ for $\forall k=1,\ldots,d$ and $\bm{X}_{M}=({X}_{max,1},...,{X}_{max,k},...,{X}_{max,d})$. And also define $\bm{X}_0 =\bm{0}$. 
The value of the core function, $g$ when evaluating the input vector, $\bm{X}$, is referred to as aggregate input, specifically $x_A=\mathscr{H}_0(\bm{X})$.

\begin{definition}
 A rising curve (commonly referred to as an expansion path) is a series of $M+1$ input vectors, $\{\bm{X}_0,\ldots,\bm{X}_M\}$ such that $x_{A,m} < x_{A,m+1}$ for every $m=0,\ldots,M-1$, where $x_{A,m} = \mathscr{H}_0(\bm{X}_{m})$. The corresponding $\Big\{\big(\mathscr{H}_0(\bm{X}_0),g(\mathscr{H}_0(\bm{X}_0))\big),\ldots, \big(\mathscr{H}_0(\bm{X}_{M}),g(\mathscr{H}_0(\bm{X}_M))\big)\Big\}$ is called the aggregated input/output of that expansion path.

 %
\end{definition}
\nomenclature[$xc$]{$\bm{X}_{M}$}{Vector of max of $\bm{X}$ for each dimension $k$}
\nomenclature[$x$]{$x_{A,m}$}{Aggregate input at step $m$}
\nomenclature[$m$]{$m$}{Index of steps of expansion path}

If the production function is homothetic, \cite{forsund1975homothetic} shows the scale elasticity is constant on each isoquant. Here we build on these results to show that, given a function is homothetic, then for any ray from the origin, $\alpha \bm{v}$, the associated inflection points, $x_v^*$ lies on the same isoquant. This statement holds when inflection point is replaced by most productive scale size (point) where most productive scale size on a particular ray $\alpha \bm{v}$ is $\max_{\bm{X} \in \alpha \bm{v}} \frac{F(\mathscr{H}(\bm{X}))}{\mathscr{H}(\bm{X})}$.


\begin{theorem}
	\label{thm:1}
	Assume a production function is homothetic in inputs and the S-shape definition holds for a single ray from the origin, then the S-shape definition will hold for the aggregated input/output of any expansion path.  Furthermore, consider any pair of rays from the origin and define two 2-D sectionals of the production function. For both rays from the origin, the S-shape definition is satisfied and the inflection points lie on the same input isoquant with aggregate input level, $x_A^*$. 
\end{theorem}

\noindent If we interpret the expansion path as the growth in inputs from one period to the next. Then consider any two expansion paths $i$ and $j$, $\{\bm{X}_{0i},\ldots,\bm{X}_{Mi}\}$ and $\{\bm{X}_{0j},\ldots,\bm{X}_{Mj}\}$, such that $\mathscr{H}_0(\bm{X}_{mi}) = x_{A,m,i} = x_{A,m,j}=\mathscr{H}_0(\bm{X}_{mj})$ for all $m$, the previous results implies the two expansion paths cross the the inflection point isoquant during the same period $m$ in which $x_{A,m-1}=\mathscr{H}_0(\bm{X}_{m-1}) \leq x_A^* < \mathscr{H}_0(\bm{X}_{m}) = x_{A,m}$. Notice there is no restriction that expansion paths are radial.  In addition, the aggregated input/output of this non-radial expansion path is S-shaped. 

\section{Estimation Algorithm}
\label{sec:iter}
\subsection{Framework}
Given observations $\{\bm{X}_j,y_j\}_{j=1}^{n}$ satisfying $y_j = g_0(\bm{X}_j) + \epsilon_j$, where $\epsilon_j$ are i.i.d. noise with zero-mean and finite variance. Our goals include the following:
\begin{enumerate}
    \item For a given level $y$, estimate the isoquant function satisfying both the convex input and the monotone decreasing input assumptions (see Definition~\ref{def:inputconvex}).
    \item For a given direction $\bm{v} \in \mathbb{R}_+^d$, estimate the production curve along that direction, i.e. $g_0(\alpha \bm{v})$ for $\alpha>0$,  satisfying monotonicity and S-shaped assumptions (see Definition~\ref{def:s-shape}).
\end{enumerate}
 Our algorithm could also be used as intermediate steps to tackle more involved problems, such as  optimal resource allocation when giving the unit cost of each input as well as the total budget.
\nomenclature[$j$]{$j$}{Index of observations}
\nomenclature[$n$]{$n$}{Number of observations}
\nomenclature[$\epsilon$]{$\epsilon_j$}{Residual of observation $j$}

\subsection{Overview}
We propose an estimation algorithm for a production function satisfying both the S-shape definition and input convexity without any further structural assumptions. The algorithm combines two different shape constrained nonparametric estimation methods. Succinctly, the algorithm is constructed by two estimations: (1) Input isoquants for a set of $y$--levels, and (2) S-shape functions on a set of rays from the origin. Algorithm \ref{algo:basic} presents our basic algorithm which is composed of these two estimators.\footnote{The algorithm refers to CNLS-based and SCKLS-based estimators for a description of these methods see Appendix \ref{subsec:isoq} and Appendix \ref{subsubsec:SCKLS} respectively.} We reference a pilot estimate which can be any estimator that will provide an initial rough estimate of the function\footnote{In our particular application the use of the pilot estimator does not impact the estimation results. However, in other context, the use of a pilot estimator simplifies our theoretical analysis and may have significant computational benefits.}. The right-hand column of Algorithm \ref{algo:basic} reports the section numbers where the details of each step are described.

\begin{figure}
	\centering
	\subfloat[Functional estimates]{\includegraphics[width=0.5\textwidth]{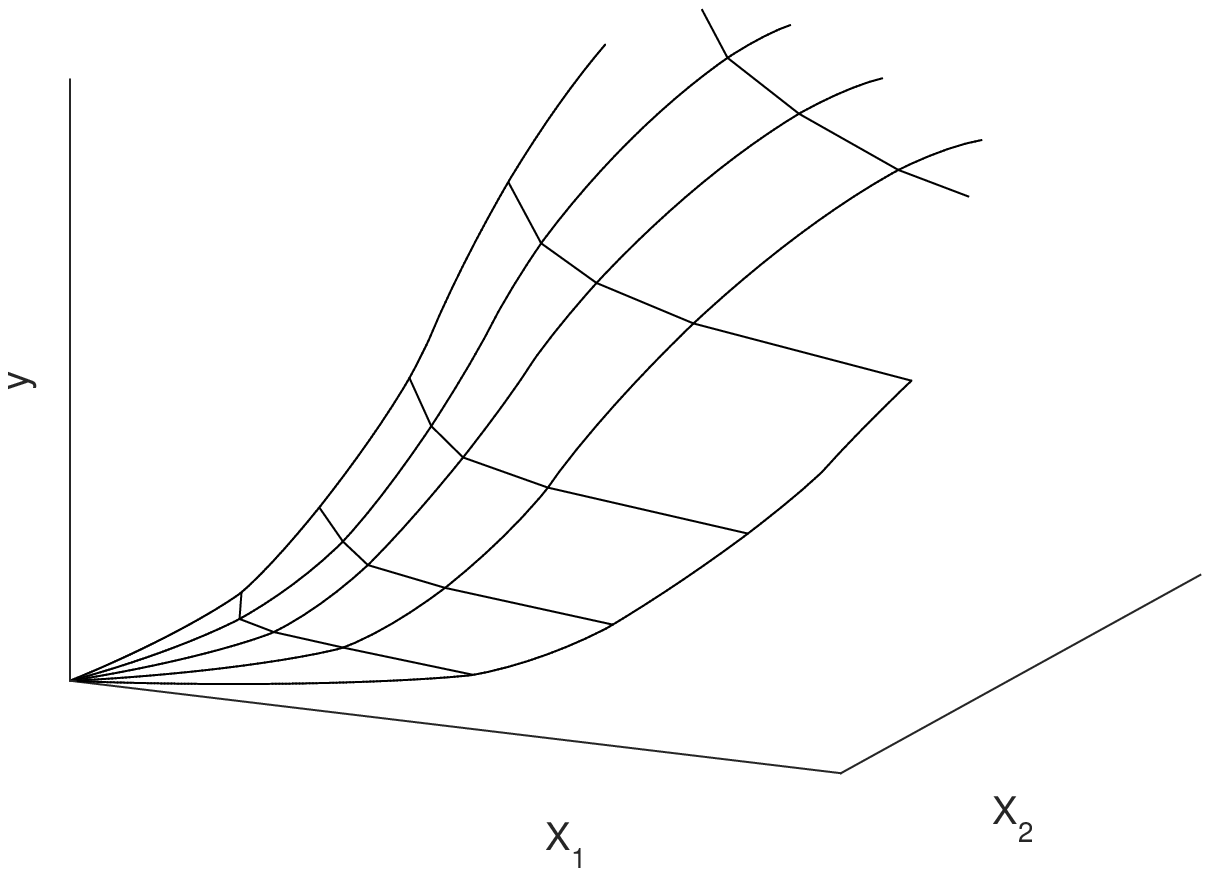}\label{fig:func_est_no_inter}}
	\hfill
	\subfloat[Interpolated functional estimates]{\includegraphics[width=0.5\textwidth]{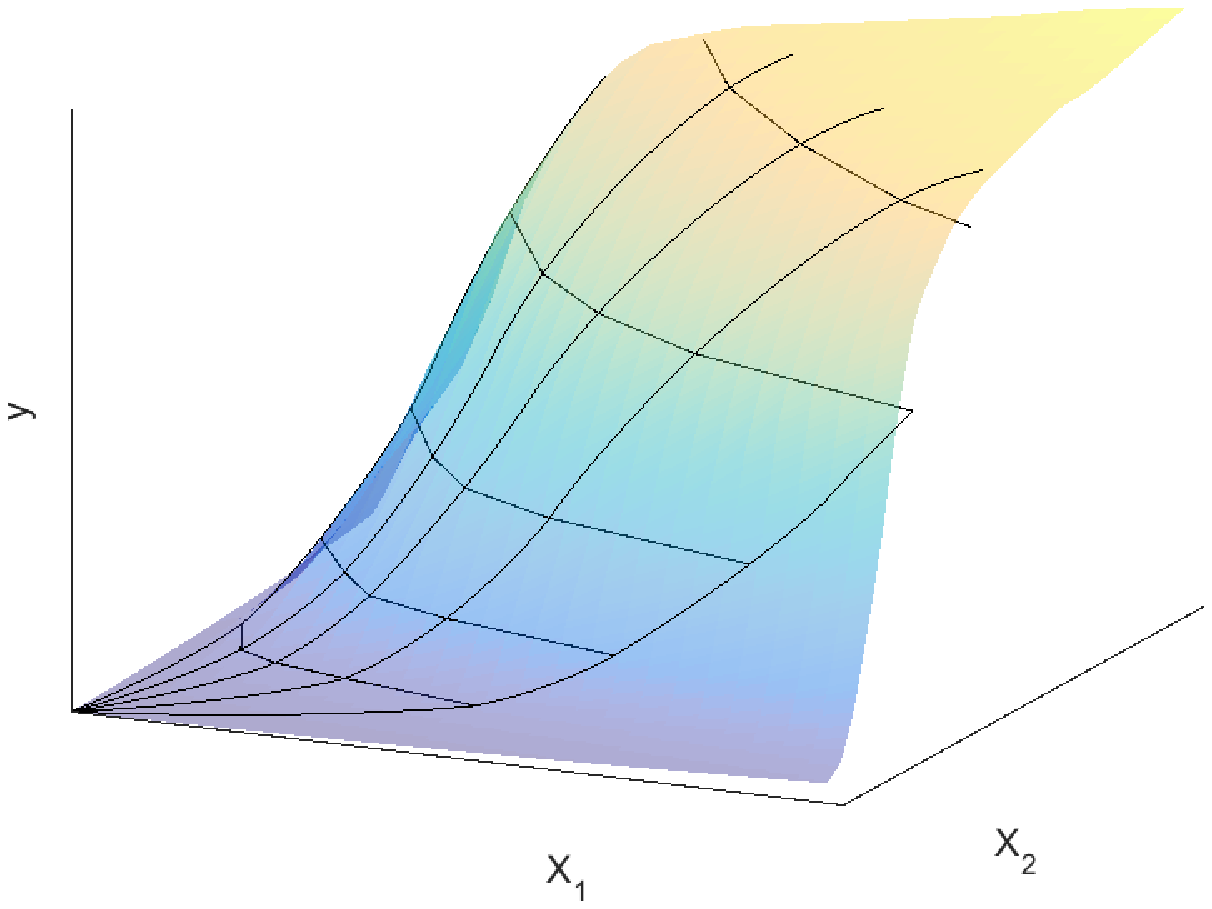}\label{fig:func_est_inter}}
	\caption{Illustration of functional estimates.}
	\label{fig:func_est}
\end{figure}

We approximate a production function $g_0$ with isoquant estimates for a set of output levels, and S-shape functional estimates for a set of rays from the origin as shown in Figure \ref{fig:func_est}\subref{fig:func_est_no_inter}. We also develop the interpolation procedure to obtain the functional estimates $\hat{g}_0(\bm{x})$ at any given input $\bm{x}$. Figure \ref{fig:func_est}\subref{fig:func_est_inter} shows the interpolated surface of the estimated production function.

\algrenewcommand\algorithmicindent{2.0em}%
\begin{algorithm}
    \caption{Basic estimation algorithm}\label{algo:basic}
    \begin{algorithmic}[1]
        \BState \textbf{Data: } \text{observations } $\{\bm{X}_j,y_j\}_{j=1}^{n}$
        \Procedure{}{}\hspace{\fill}(Section)
        \BState \emph{Initialization}:\hspace{\fill}(\ref{subsec:init})
            \State $I \gets$ Initialize number of isoquants
            \State $R \gets$ Initialize number of rays 
            \State $\{y^{(i)}\}_{i=1}^{I} \gets$ Initialize isoquant $y$-levels with $y^{(1)} < \cdots < y^{(I)}$.
            \State $\{\bm{\theta}^{(r)}\}_{r=1}^{R} \gets$ Initialize rays from origin
        \BState \emph{Estimation}:\hspace{\fill}(\ref{subsec:est})
            \State For $j=1,\ldots,n$, let $\tilde{y}_j = \tilde{g}_0(\bm{X}_j)$, where $\tilde{g}_0$ is the pilot estimator of $g_0$
            \State Project $\{\bm{X}_j,\tilde{y}_j\}_{j=1}^{n}$ to the isoquant level $y^{(i)}$
            \State Estimate convex isoquants by the CNLS-based estimation
            \State Project observations onto the ray $\bm{\theta}^{(r)}$ \State Estimate S-shape functions using the SCKLS-based estimator
        \BState \textbf{Return: } Estimated isoquants and S--shape functions
    \EndProcedure
    \end{algorithmic}
\end{algorithm}
\nomenclature[$i$]{$I$}{Number of isoquants}
\nomenclature[$i$]{$i$}{Index of isoquants}
\nomenclature[$r$]{$R$}{Number of rays}
\nomenclature[$r$]{$r$}{Index of rays}
\nomenclature[$y$]{$y^{(i)}$}{Isoquants level of isoquant $i$}
\nomenclature[$\theta$]{$\bm{\theta}^{(r)}$}{Angle of ray $r$}
\nomenclature[$g$]{$\tilde{g}_0(\bm{x})$}{Pilot estimator}
\nomenclature[$y$]{$\tilde{y}_j$}{Estimates by pilot estimator}

Since we estimate the S-shape function on rays from the origin, it is convenient to use a spherical coordinates system which is defined by the angle and distance (radius) of observed points to the origin. Therefore, our observed input vector $\bm{X}_j=(X_{j1},\ldots,X_{jd} )'$ in spherical coordinates system $(r_j,\bm{\phi}_j)=(r_j,\phi_{j,1},\ldots,\phi_{j,d-1})$ is defined as:
\begin{equation}
\label{eq:polar}
    \begin{aligned}
        r_j &= \sqrt{X_{j1}^2+\ldots+X_{jd}^2} \\
        \phi_{j,1} &= \arccos\frac{X_{j1}}{\sqrt{X_{j1}^2+\ldots+X_{jd}^2}}\\
        \phi_{j,2} &= \arccos\frac{X_{j2}}{\sqrt{X_{j2}^2+\ldots+X_{jd}^2}}\\
        &\vdots\\
    \phi_{j,d-1} &= \arccos\frac{X_{j,d-1}}{\sqrt{X_{j,d-1}^2+X_{jd}^2}},\\        
    \end{aligned}
\end{equation}
where $r_j$ is the radial distance from the origin, and $\{\phi_{j,1}\ldots\phi_{j,d-1}\}$ defines the angle of the observation. For notational conveinience, in the rest of the manuscript, we denote the angle of $\bm{v}$ for any $\bm{v} \in \mathbb{R}^d$ with $\|v\|_2=1$ as  ${\phi}(\bm{v})$ (and its inverse function as $\phi^{-1}(\cdot)$).
\nomenclature[$r$]{$r_j$}{Radius of observation $j$}
\nomenclature[$\phi$]{$\bm{\phi}_j$}{Angels of observation $j$}


\subsection{Initialization}
\label{subsec:init}

We initialize the parameters used in the estimation. The number of isoquants $I$ and the number of rays from the origin $R$ affect the flexibility of the estimated function (computation time increases with the number of isoquants and rays). We initialize isoquant $y$-levels, $\{y^{(i)}\}_{i=1}^{I}$, and rays from the origin, $\{\bm{\theta}^{(r)}\}_{r=1}^{R}$, based on the distribution of the observations. We propose three options: (1) Evenly spaced grid, (2) Equally spaced percentile grid, and (3) Centroid of $K$-means cluster of observations. To set notation, given  the number of isoquants, $I$, and rays, $R$, we set the grid as $y^{(i)}$ and $\bm{\theta}^{(r)}$, the locations of the isoquants and rays respectively. To overcome skewness in the empirical data in which there are many smaller firms and only a few large firms, we recommend an equally spaced percentile grid or $K$-means cluster.



\subsection{Two--step estimation}
\label{subsec:est}
During the estimation step, we approximate the production function by estimating the isoquants at a set of $y$-levels and estimating the S-shape functions on a set of rays from the origin. We calculate the estimates over different tuning parameters, compute the mean squared errors (MSE) against observations, and return the final estimates corresponding to the tuning parameters with the minimum MSE.

\subsubsection{Isoquant estimation}
\label{subsubsec:isoq_est}

\begin{figure}[ht]
	\centering
	\subfloat[Projected observations to each $y^{(i)}$]{\includegraphics[width=0.5\textwidth]{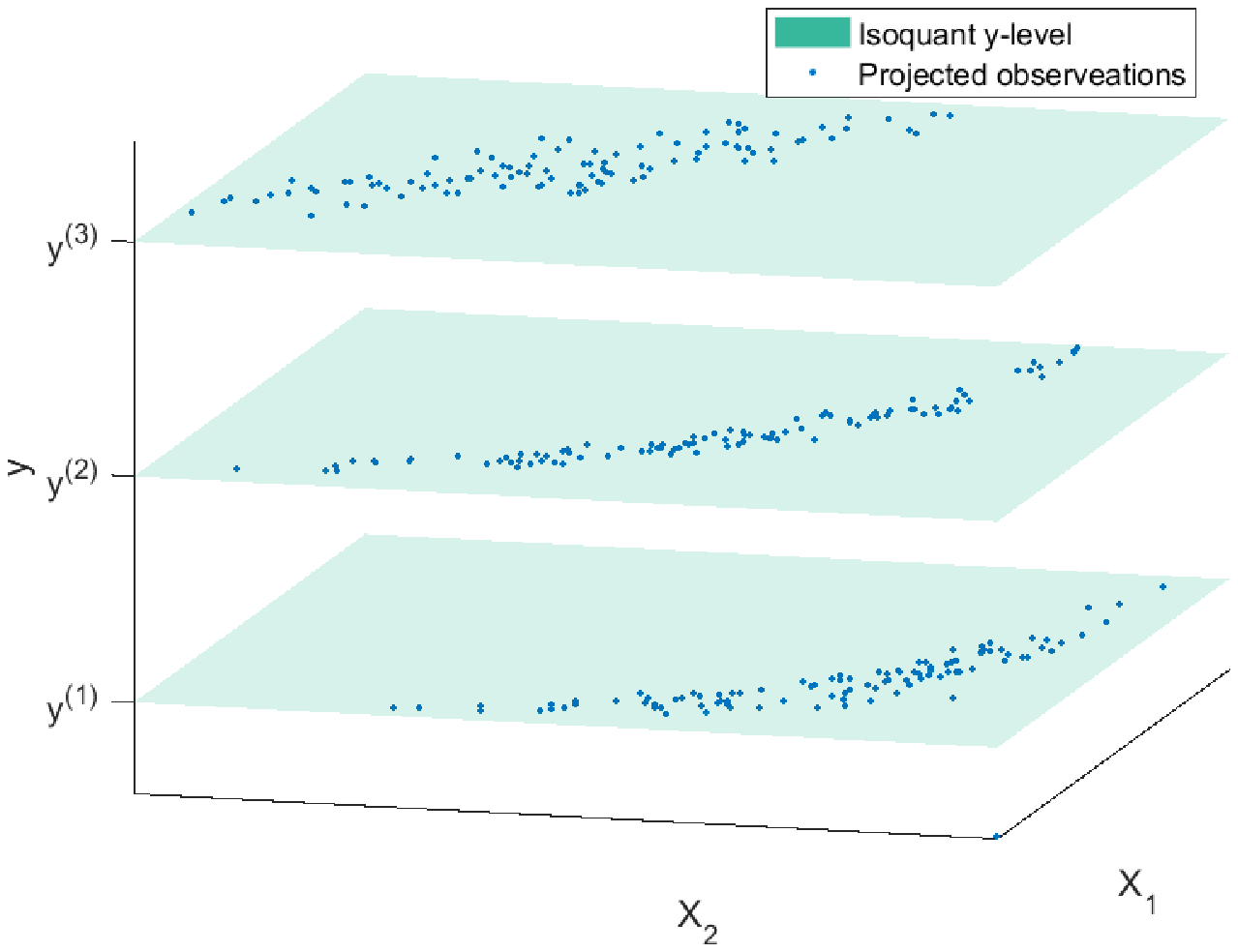}\label{fig:isoq_b}}
	\hfill
	\subfloat[Isoquant estimates on $y^{(i)}$]{\includegraphics[width=0.5\textwidth]{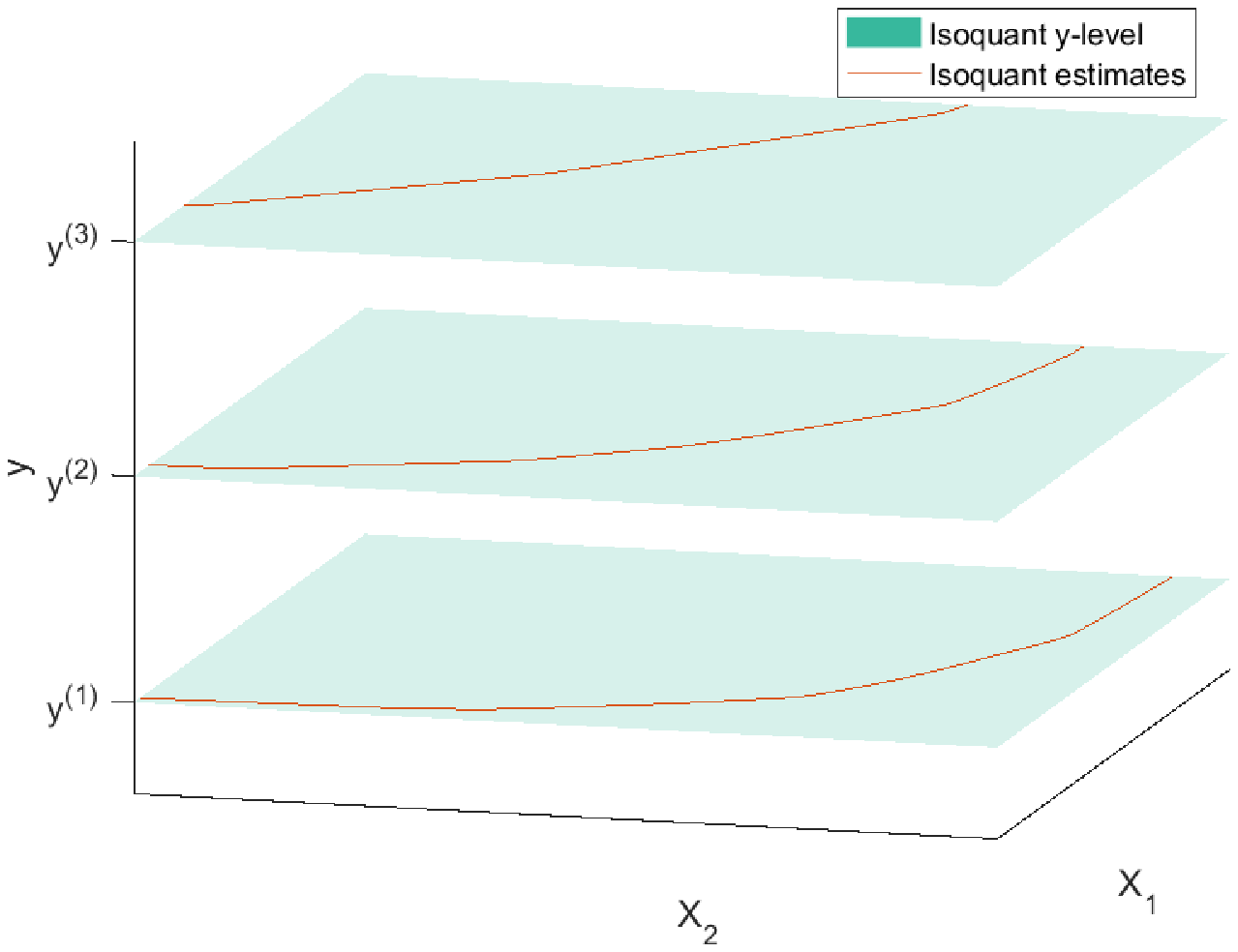}\label{fig:isoq_c}}
	\caption{Isoquant estimation}
	\label{fig:isoq_est}
\end{figure}

Before estimating the isoquants, we need to assign each observation $\{\bm{X}_j,y_j\}_{j=1}^{n}$ to an isoquant $y$-level, $y^{(i)}$ based on $\tilde{y}_j$ from a pilot estimator. The purpose of the pilot estimator is to improve the classification of observations to isoquant levels. Most well-known nonparametric estimators, such as local linear estimator could be used. We suggest simply assigning each observation to the closest isoquant $y$-level, which means
\begin{equation}
    i_j=\argmin_{i\in\{1,\ldots,I\}}\left(\tilde{y}_j-y^{(i)}\right)^{2} ~~~~~ \forall j=1,\ldots,n,
\end{equation}
\nomenclature[$i$]{$i_j$}{Index of isoquant where observation $j$ is assigned}
where $i_j$ indicates the isoquant index to which we assign observation $j$. Then, we define the projected observations for the $i^{th}$ isoquant as $\{\bm{X}_j,y^{(i_j)}\}_{\{j:i_j=i\}}$, where $y^{(i_j)}$ is the output level of the $i^{th}$ isoquant (ties are broken by assigning the observation to the a lower-level isoquant). Figure \ref{fig:isoq_est}\subref{fig:isoq_b} shows the projection of each observation to the corresponding isoquant $y$-level. We estimate a set of isoquants using the CNLS-based method which is a nonparametric estimation method imposing convexity for each $y^{(i)}$-level. Intuitively, we estimate the convex isoquant estimates nonparametrically without imposing any ex ante functional specification for each $y^{(i)}$-level. Figure \ref{fig:isoq_est}\subref{fig:isoq_c} shows the isoquant estimates obtained with projected observations $\{\bm{X}_j,y^{(i_j)}\}$. The mathematical formulation is described in Appendix \ref{subsec:isoq}. 

\subsubsection{S-shape estimation}
\label{subsubsec:s_est}

To estimate the S-shape functions on rays from the origin, we begin by project all observations $\{\bm{X}_j,y_j\}_{j=1}^n$ to each ray from the origin $\bm{\theta}^{(r)}$. We can either project the observations directly onto the rays, or use the estimated isoquants from the previous step to project the observations. For the second approach, in short, we find the level of an isoquant to which $\bm{X}_j$ belongs. 
Below we also provide an alternative way of thinking about this step. Considering the observations input level, $\bm{X}_j$, we select the two closest isoquants associated with a larger and smaller aggregate inputs. Here the definition of larger and smaller vectors are in terms of a proportional expansion or contraction of the input vector, $\lambda\bm{X}_j$ where $0 \leq \lambda < \infty$ with $\lambda \geq 1$ indicating expansion and $\lambda \leq 1$ indicating contraction. We will refer the two closest isoquants as ``sandwiching" the input vector of interest. Then, we assign weights to these two isoquants based on the distance to the observed input $\bm{X}_j$ along a ray from the origin through the observed points. 
\nomenclature[$\lambda$]{$\lambda$}{Scaling factor for input vector}
Finally, we project the observation with the weighted average of the two isoquant estimates. Figure \ref{fig:s_shape_est}\subref{fig:s_shape_b} shows the projection of our observations. Details are described in Appendix \ref{subsubsec:project}.

Next, we use the SCKLS-based method to estimate the S-shape function on each ray from the origin. Note that this estimation assigns two different kernel weights to each observation. The first weight is a function of the angle(s) formed by a ray from the origin through the observation and a ray from the origin through the current evaluation point. The angle will be a vector if there are more than two regressors. The second weight is a function of the distance measured along the ray between the projected observation and the evaluation point.  

SCKLS-based estimation requires the selection of a smoothing parameter which we refer to as the bandwidth. Intuitively, a smaller bandwidth will lead to over--fitting the data, and a larger bandwidth will lead to over--smoothing. Thus, it is crucial to select the optimal bandwidth by balancing the bias--variance tradeoff of the estimator. In our algorithm, the bandwidth of the kernel weights for angles, $\bm{\omega}$, is optimized via a grid search, and the bandwidth of the kernel weights for distance along the ray, $h^{(r)}$, is optimized by leave-one-out cross-validation, given kernel weights for angles. 
We adapt the SCKLS estimator by introducing an inflection point, below this point the function is convex and after this point the function is concave. The estimation is preformed for each ray, thus inputs are aggregated to a single univariate regressor. Therefore, the number of constraints used are on order of evaluation points. We search over a large set of potential inflection points similar to the estimator studied in \cite{liao2017change}. Figure \ref{fig:s_shape_est}\subref{fig:s_shape_c} shows the S-shape estimates obtained with projected observations. The mathematical details are described in Appendix \ref{subsubsec:SCKLS}.
\nomenclature[$\omega$]{$\bm{\omega}$}{Bandwidth to smooth over angles}
\nomenclature[$h$]{$h^{(r)}$}{Bandwidth to smooth over radius}

\begin{figure}[ht]
	\centering
	\subfloat[Projected observations to each $\bm{\theta}^{(r)}$]{\includegraphics[width=0.5\textwidth]{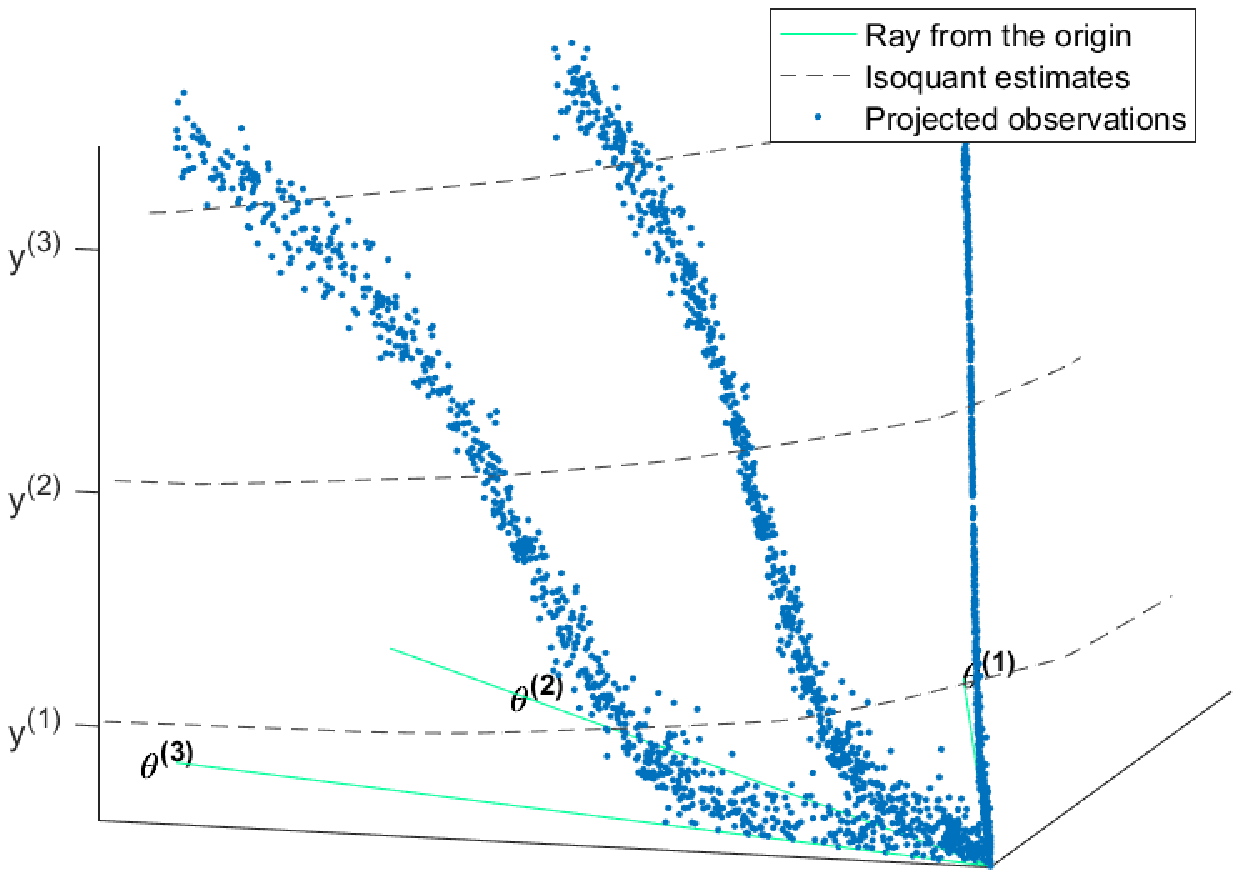}\label{fig:s_shape_b}}
	\hfill
	\subfloat[S-shape estimates on $\bm{\theta}^{(r)}$]{\includegraphics[width=0.5\textwidth]{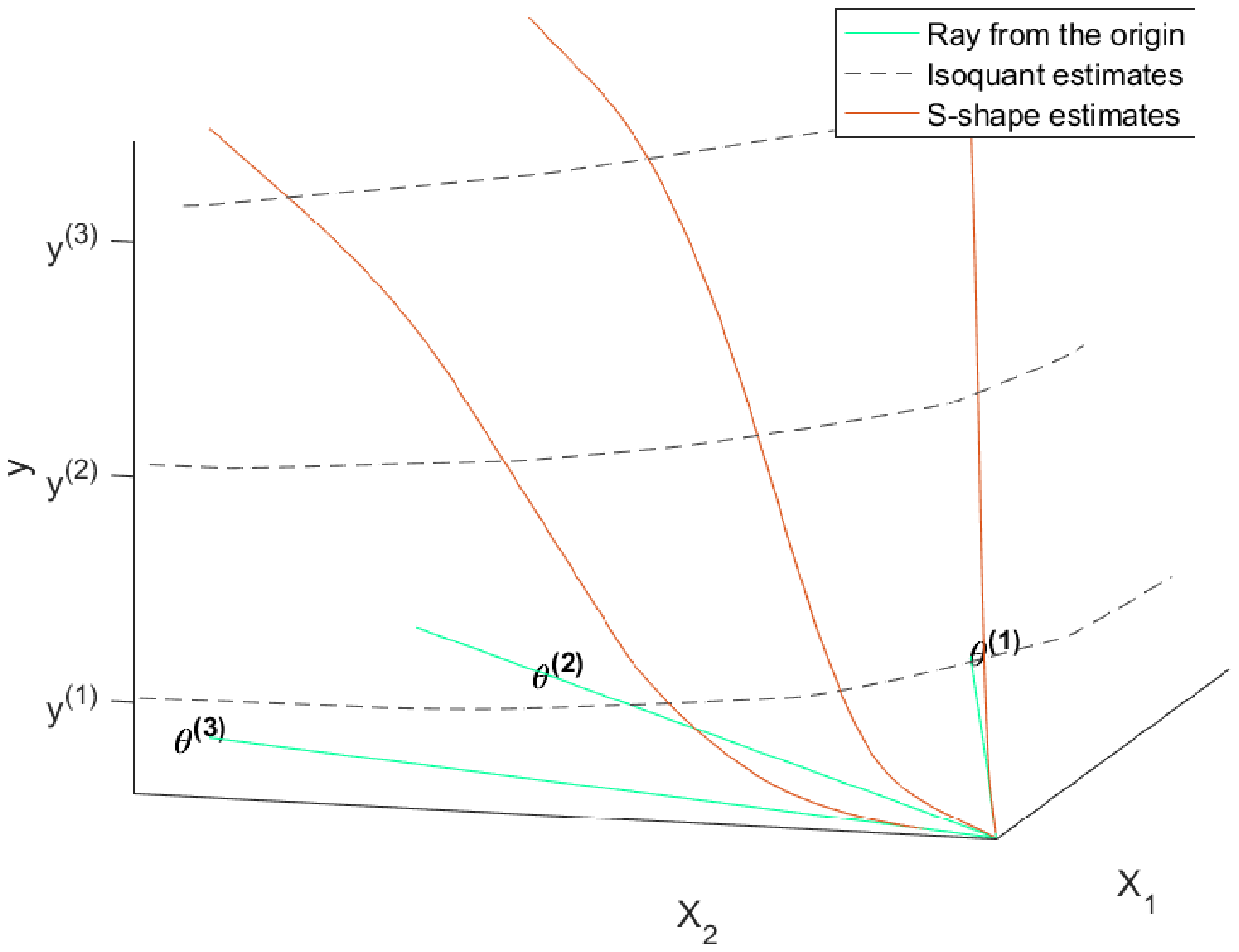}\label{fig:s_shape_c}}
	\caption{S-shape estimation}
	\label{fig:s_shape_est}
\end{figure}

\subsubsection{Computing functional estimates at a given input vector}
\label{subsubsec:funest}

The last step of Algorithm \ref{algo:basic} obtains the functional estimates $\hat{g}(\bm{x})$ at any given value of input vector $\bm{x}$, and computes the MSE against observations $\{\bm{X}_j,y_j\}_{j=1}^{n}$.
\nomenclature[$g$]{$\hat{g}(\bm{x})$}{S--shape functional estimates}

First we compute the weighted average of the two closest isoquants which sandwich the observed input $\bm{X}_j$. The details are given in Appendix $\ref{subsubsec:project}$. Second, we assign weights to each S-shape estimate based on the angle between a given input vector $\bm{x}$ and each ray from the origin $\bm{\theta}^{(r)}$ on which we have estimated the S-shape functions, followed by scaling and computing the weighted average of the S-shape estimates and obtaining the final functional estimates on a given input $\bm{x}$, $\hat{g}(\bm{x})$. Figure \ref{fig:func_est}\subref{fig:func_est_inter} shows the interpolated functional estimates. The details are given in Appendix \ref{subsubsec:estimates_on_obs}.

\begin{figure}[ht]
    \includegraphics[width=0.7\textwidth]{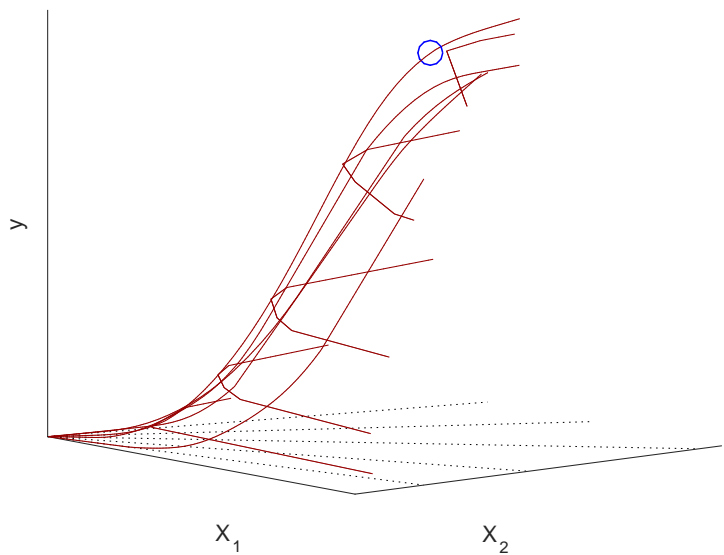}
    \centering
    \caption{Gap between convex isoquant and S-shape estimates}
    \label{fig:gap}
\end{figure}

Note that there may be a gap between the convex isoquant estimates and the S-shape estimates on rays from the origin. Specifically, if the S-shape estimates do not all lie on the input isoquant for each evaluated output level $y^{(i)}$, then the S-shape estimates will not match the isoquant estimates at some isoquant $y$-level as indicated by the blue circle in Figure \ref{fig:gap}. The gap tends to be larger when the data are noisier. However, the gaps can be assured to be zero if we impose homotheticity. In the non-homothetic case, we can always reduce the gap to zero by using fewer rays for estimation, although at the cost of a rougher functional estimate.\footnote{When the gaps are significant, selecting the value for tuning parameters becomes a multi-criteria problem in which we want to minimize both the largest gap and Mean Squared Error (MSE). We do this by setting a threshold on the largest acceptable gap level and picking the tuning parameter value with the smallest MSE. For details of the implementation see Appendix \ref{subsec:update}.}


\subsection{Other variants}

\subsubsection{Homothetic isoquants}
\label{subsec:homo1}
If we know that the isoquants are homothetic, then isoquants at different levels would have the same shape. This means that we could estimate the isoquant at any given $y$-level (say, $y^{(\lfloor I/2\rfloor)}$), and scale it to other $y$-level accordingly. Alternatively, we could estimate the isoquants at different levels jointly via the following procedure. Insert the following steps between Line~11 and Line~12 of Algorithm~\ref{algo:basic}.
\begin{enumerate}
    \item For $i = 1,\ldots,I$, let $\mathcal{I}_i \subset \{1,\ldots,n\}$ be the index set with $\{\bm{X}_j,\tilde{y}_j\}_{j \in \mathcal{I}_i}$ projected to the isoquant level $y^{(i)}$. After we estimate isoquants at different $y$-levels, we let $\hat{\lambda}_i$ be the scalar such that $\hat{\lambda}_i(1,\ldots,1)^T$ is on the estimated isoquant at level $y^{(i)}$. 
    \item For some pre-defined $\delta \in (0,1/2)$, apply the CNLS-based estimator on $$\bigcup_{\{i=\lceil \delta I\rceil, \ldots, \lfloor (1-\delta) I\rfloor\}}\{\hat{\lambda}_i^{-1} \bm{X}_j\}_{j \in \mathcal{I}_i}$$ and denote the curve by $\hat{\mathscr{H}}_{0,k}(\bm{x}_{-k}; F_0(1))$.
    \item Re-estimate the isoquant at $y^{(i)}$ level by 
    $\hat{\mathscr{H}}_{0,k}(\bm{x}_{-k}; y^{(i)}) \equiv \lambda_i \hat{\mathscr{H}}_{0,k}(\hat{\lambda}_i^{-1}\bm{x}_{-k}, F_0(1))$.
\end{enumerate}
\nomenclature[$i$]{$\mathcal{I}_i$}{Index set of observations projected to isoquant $i$}
\nomenclature[$\lambda$]{$\hat{\lambda}_i$}{Scaling factor to the estimated isoquant at level $y^{(i)}$}
Note that in the second step above, we do not make use of the estimated isoquant at the top and bottom quantiles of the $y$-levels. This is due to the fact that isoquant estimation at extreme levels could be inconsistent.

Due to homotheticity, given the estimated isoquants, instead of estimating the S-shape function along different rays, we could concentrate on estimation along a single ray. Without loss of generality, we could project all observations to the ray $\alpha(1,\ldots,1)^T$ (with $\alpha > 0$) along the isoquants, and then perform SCKLS. 

\subsubsection{Parametric and homothetic isoquants}
\label{subsec:homo2}
Recall that $g_0(\bm{x})=F_0(\mathscr{H}_0(x))$ in the homothetic setting, with $F_0$ following the S-shape. Given the parametric form of $H$ and for each possible parameter value, we could derive the profile log-likelihood use the CNLS-type approach. As such, we can obtain the estimates by directly  solving a semi-parametric optimization problem (without the need of a pilot estimator). As an illustration, two concrete examples are given below. Here we denote $\mathcal{F}$ as the class of increasing and S-shaped functions from $[0,\infty) \rightarrow \mathbb{R}$. 
\nomenclature[$f$]{$\mathcal{F}$}{Class of increasing S--shape functions}

\begin{enumerate}
\item \textbf{Linear isoquants} 

$\mathscr{H}_0(\bm{x}) = \boldsymbol{\beta}_0^T \bm{x}$ with $\boldsymbol{\beta}_0> \mathbf{0}$ and $\|\boldsymbol{\beta}_0\|_1 = 1$ (so that $\mathscr{H}_0((1,\ldots,1)^T)=1$). 

We estimate $\boldsymbol{\beta}_0$ by 
\[
\hat{\boldsymbol{\beta}}_0 \in  \argmin_{ \|\boldsymbol{\beta}\|_1 = 1, \boldsymbol{\beta} \ge \mathbf{0}} \min_{F \in \mathcal{F}} \sum_{j=1}^n \big(Y_j - F(\boldsymbol{\beta}^T \bm{X}_j)\big)^2
\]
and $F_0$ by
\[
\hat{F}_0 \in \argmin_{F \in \mathcal{F}} \sum_{j=1}^n \big(Y_j - F(\hat{\boldsymbol{\beta}}_0^T \bm{X}_j)\big)^2.
\]

\nomenclature[$\beta$]{$\boldsymbol{\beta}_0$}{True parameters of isoquants}

\item \textbf{Power isoquants} 

We consider $\mathscr{H}_0(\bm{x})=\bm{x}^{\boldsymbol{\beta}_0} \equiv \prod_{i=1}^d x_i^{\beta_{0,i}}$, where $\boldsymbol{\beta}_0 = (\beta_{0,1}, \ldots, \beta_{0,d})^T$, which is a Cobb-Douglas type of isoquant. Here  $\boldsymbol{\beta}_0> \mathbf{0}$ and $\|\boldsymbol{\beta}_0\|_1 = 1$ (so that $\mathscr{H}_0$ is homothetic). Also note that there is no extra coefficient in front of $\bm{x}^{\boldsymbol{\beta}_0}$ in $\mathscr{H}_0$ since we require $\mathscr{H}_0((1,\ldots,1)^T)=1$.

We estimate $\boldsymbol{\beta}_0$ by 
\[
\hat{\boldsymbol{\alpha}}_0 \in  \argmin_{ \|\boldsymbol{\beta}\|_1 = 1, \boldsymbol{\beta} \ge \mathbf{0}} \min_{F \in \mathcal{F}} \sum_{j=1}^n \big(Y_j - F(\bm{X}_j^{\boldsymbol{\beta}})\big)^2
\]
and $F_0$ by
\[
\hat{F}_0 \in \argmin_{F \in \mathcal{F}} \sum_{j=1}^n \big(Y_j - F(\bm{X}_j^{\hat{\boldsymbol{\beta}}_0})\big)^2.
\]

\end{enumerate}

Finally, we note that one could also use the SCKLS-type instead of CNLS-type approach in the above estimation procedures, see Appendix \ref{app:algo} for details of the two types of estimators.

\subsubsection{Parametric isoquants}
Suppose we know the parametric (but not necessarily homothetic) form of the isoquants, then we could replace the CNLS-based method in Line~11, Algorithm~\ref{algo:basic} by the ordinary least-squares-based method.

\subsection{Further extensions to the estimation algorithm}
\label{subsec:adv}
Note that in the homothetic cases in Section~\ref{subsec:homo1} and Section~\ref{subsec:homo2}, our estimator provides estimates for convex isoquants and S-shape curves with no gap. However, as stated above, Algorithm \ref{algo:basic} may result in a production function estimate with a gap between the convex isoquant estimates and the S-shape estimates in the non-homothetic setting. To address this issue, we develop several extensions, which allow us to estimate a production function by iterating between the estimations of isoquants and S-shape functions to reduce the size and number of gaps that may exist.

\newcounter{mycount}
\setcounter{mycount}{1}
\counterwithin{algorithm}{mycount}
\refstepcounter{mycount}
\renewcommand\thealgorithm{\arabic{mycount}\Alph{algorithm}}

\algrenewcommand\algorithmicindent{2.0em}%
\begin{algorithm}
    \caption{Concise summary of the advanced estimation algorithm}\label{algo:advanced}
    \begin{algorithmic}[1]
        \BState \textbf{Data: } \text{observations } $\{\bm{X}_j,y_j\}_{j=1}^{n}$
    \Procedure{}{}
            \State Initialize the parameters\hspace{\fill}(\ref{subsec:init2})
            \State Estimate each convex isoquant by the CNLS-based method\hspace{\fill}(\ref{subsec:isoq})
            \State Estimate S-shape curve along the rays by the SCKLS-based method\hspace{\fill}(\ref{subsec:s-shape})
            \State Compute a gap between estimates \hspace{\fill}(\ref{subsec:update})
            \State Iterate previous steps until the parameters stabilize
        
        \BState \textbf{Return: }
            \State Estimated function with minimum Mean Squared Errors and a gap smaller than threshold
    \EndProcedure
    \end{algorithmic}
\end{algorithm}
    
Algorithm \ref{algo:advanced} is a concise summary of our algorithm. The mathematical details and an extended description is available in Appendix \ref{app:algo} and is labeled, Algorithm  \ref{algo:advanced2}. We use Algorithm  \ref{algo:advanced2} in the following simulation and application sections.

\subsection{Quantifying uncertainty of the estimator}
\label{subsec:uncertainty}

In addition to estimating the conditional mean, understanding uncertainty of the estimator is critical for practitioners to make actual managerial decisions. However our estimator is piece-wise linear and thus require non-standard analysis to derive asymptotic properties. \cite{yagi2018shape} develop the bootstrapping procedure to validate the shape constraints imposed. We can also use the same wild bootstrap procedure to resample the response variable. Then we can use boostrap samples to emprically compute uncertainty of the estimator.

We can also use bootstrapping to validate the RUP law and input convex isoquants similar to \cite{yagi2018shape}. The test statistic is defined as a difference between shape constrained and unconstrained estimates. Intuitively, when shape constraints are correctly specified, then both estimates should have similar shape, and a test statistic becomes small. We describe the detailed procedure of bootstrapping in Appendix \ref{app:uncertainty}.

\section{Theoretical properties of the estimator}
\label{sec:property}
\subsection{The non-homothetic case}
We show the consistency of Algorithm~\ref{algo:basic}. We make the following assumptions: 
	\begin{assumption}
		\leavevmode
		\label{ass:2}
		\begin{enumerate}[(i)]
			\item \label{ass:2.1} $\{\bm{X}_j,y_j\}_{j=1}^n$ are a sequence of i.i.d. random variables with  $y_j = g_0(\bm{X}_j)+ \epsilon_j$.
			\item \label{ass:2.2} $g_0:\bm{S}\rightarrow \mathbb{R}$ satisfy Assumption~\ref{ass:1}, Definition~\ref{def:inputconvex} (convex input insoquants) and Definition~\ref{def:s-shape} (S-shaped). For simplicity, we also assume that $\bm{S}=[0,c]^d$ for some $c > 1$.
			\item \label{ass:2.3} $\bm{X}_j$ follows a distribution with continuous density function $f$ and support $\bm{S}$. 
			Moreover, $\min_{\bm{x} \in \bm{S}} f(\bm{x}) > 0$.	
			\item \label{ass:2.4} The conditional probability density function of $\epsilon_j$, given $\bm{X}_j$, denoted as $p(e|\bm{x})$, is continuous with respect to both $e$ and $\bm{x}$, with the mean function \[\mu(\cdot) = E(\epsilon_j|\bm{X}_j=\cdot) = 0\] and the variance function \[\sigma^2(\cdot) = \mathrm{Var}(\epsilon_j|\bm{X}_j=\cdot)\] being continuous over $\bm{S}$. Moreover, $\sup_{\bm{x} \in \bm{S}}E\Big(\epsilon_j^4\Big|\bm{X}_j = \bm{x}\Big) < \infty$.
		\end{enumerate}
	\end{assumption}
	\nomenclature[$c$]{$c$}{Boundary of an input support}
	\nomenclature[$s$]{$\bm{S}$}{Support of inputs}
	\nomenclature[$f$]{$f(\bm{x}) $}{Density function of inputs $\bm{x}$}
	\nomenclature[$p$]{$p(e|\bm{x})$}{Conditional distribution of residual on inputs $\bm{x}$}
	\nomenclature[$\mu$]{$\mu(\bm{x})$}{Mean function of residual at $\bm{x}$}
	\nomenclature[$\sigma$]{$\sigma^2(\bm{x})$}{Variance function of residual at $\bm{x}$}

Most parts of Assumption~\ref{ass:2} are typical in the nonparametric regression setting. Here (\ref{ass:2.1}) states that the data are i.i.d.; (\ref{ass:2.2}) says that the constraints we impose are satisfied by the true function; (\ref{ass:2.3}) makes a further assumption on the distribution of the covariates; and (\ref{ass:2.4}) states that the noise can be heteroscedastic in certain ways, but requires the change in the variance to be smooth. 
	
To simplify our theoretical development, below we impose some more specific assumptions regarding the construction of our estimator. 
	
	\begin{assumption}
		\leavevmode
		\label{ass:3}
		\begin{enumerate}[(i)]
		\item For the pilot estimator, we use the local linear estimator with the sphereically symmetric Epanechnikov kernel and bandwidth $h' \asymp  n^{-1/(4+d)}$ as $n \rightarrow \infty$.
		\item  $I = o(n^{2/(4+d)}/\log n)$, with $I \rightarrow \infty$ as $n \rightarrow \infty$. Moreover, let $y_\circ =  \inf_{\bm{S}} g_0(\bm{x})$ and $y^\circ = \sup_{\bm{S}} g_0(\bm{x})$.  The initial $y$-values of the isoquants are set as
		\[y^{(i)} = y_\circ + \frac{i}{I+1}(y^\circ-y_\circ)\]
		for $i=1,\ldots,I$.
		\item We use the spherically symmetric Epanechnikov kernel, with bandwidths, $\bm{\omega}=(\omega,\ldots,\omega)'$ and $h$. For simplicity, we take $\omega = h$ and $h \asymp n^{-1/(4+d)}$ as $n\rightarrow \infty$.
		\item The number of rays $R \rightarrow \infty$ as $n\rightarrow \infty$. Moreover, the empirical distribution of $\{\bm{\theta}^{(r)}\}_{r=1}^{R}$ converges to the uniform distribution on $[0,\pi/2]^{d-1}$.
		
		\item For any $\bm{v} \in \mathbb{R}_+^d$, define $c_{\bm{v}} = \sup \{t > 0: \; t\bm{v} \in \bm{S}\}$. For the SCKLS estimator along each ray of direction $\bm{\theta}^{(r)}$, evaluation points are equally spaced over $[0,c_{\phi^{-1}(\bm{\theta}^{(r)})}]$, where $\phi^{-1}$ is the inverse angle function. The number of evaluation points, $m$, goes to $\infty$, as $n\rightarrow \infty$. 
	
		\end{enumerate}
	\end{assumption}
	\nomenclature[$h$]{$h'$}{Bandwidth for pilot estimator}
The following theorems establish the consistency for isoquant estimation and  estimation along the rays. Without loss of generality, we focus on isoquants expressed as a function of the first $d-1$ coordinates (i.e. the truth is $\mathscr{H}_{0,d}$, with its estimator $\hat{\mathscr{H}}_{0,d}$).
\begin{theorem}
\label{thm:nonhomo_consistency1}
    Under Assumptions \ref{ass:1}--\ref{ass:3}, for any $y \in (y_\circ,y^\circ)$,  suppose that $\mathscr{H}_{0,d}(\cdot;y)$ has domain $\bm{C}_y \subset [0,c]^{d-1}$, and let $\hat{\mathscr{H}}_{0,d}(\cdot;y)$ be the estimated isoquant. Then, $\hat{\mathscr{H}}_{0,d}(\cdot;y)$ satisfies the input-convexity constraint. Moreover, for any compact set $\bm{C}'$ that belongs to the interior of $\bm{C}_y$, as $n \rightarrow \infty$,
    \[
    \sup_{\bm{x}_{-d} \in \bm{C}'} |\hat{\mathscr{H}}_{0,d}(\bm{x}_{-d};y)-{\mathscr{H}}_{0,d}(\bm{x}_{-d};y)|\stackrel{p}{\rightarrow} 0.
    \]
\end{theorem}
\nomenclature[$c$]{$\bm{C}_y$}{Domain of isoquants}
\nomenclature[$y$]{$y_\circ$}{Lower bound of output $y$}
\nomenclature[$y$]{$y^\circ$}{Upper bound of output $y$}

\begin{theorem}
\label{thm:nonhomo_consistency2}
    Under Assumptions \ref{ass:1}--\ref{ass:3}, for any direction $\bm{v}=(v_1,\ldots,v_d)^T$ with $\min_i v_i > 0$ and $\|\bm{v}\|_2=1$, we have that $\hat{g}_0(\alpha \bm{v})$ obeys the S-shape along $\bm{v}$. For any $\delta \in (0,c_{\bm{v}}/2)$, as $n \rightarrow \infty$, 
    \[
    \sup_{\alpha \in [\delta,c_{\bm{v}}-\delta]} |\hat{g}_0(\alpha\bm{v})-g_0(\alpha\bm{v})|\stackrel{p}{\rightarrow} 0.
    \] 
\end{theorem}

\subsection{The homothetic cases}
We also show consistency on the variants of our algorithm on the estimation of the isoquants in the homothetic settings. 

\subsubsection{Nonparametric isoquants}
\begin{theorem}
\label{thm:homo_consistency1}
	Under Assumptions \ref{ass:1}--\ref{ass:3} and suppose that $g_0(\bm{x})=F_0(\mathscr{H}_0(\bm{x}))$ is homothetic. Then, $\hat{\mathscr{H}}_{0,d}(\cdot;F_0(1))$ satisfies the input-convexity constraint. Moreover, for any compact set $\bm{C}'$ that belongs to the interior of $\bm{C}_{F_0(1)}$,  where $\bm{C}_{F_0(1)} \subset [0,c]^{d-1}$ is the domain of  $\mathscr{H}_{0,d}(\cdot;F_0(1))$, we have that, as $n \rightarrow \infty$,
	\[
	\sup_{\bm{x}_{-d} \in \bm{C}'} |\hat{\mathscr{H}}_{0,d}(\bm{x}_{-d};F_0(1))-{\mathscr{H}}_{0,d}(\bm{x}_{-d};F_0(1))|\stackrel{p}{\rightarrow} 0.
	\]
\end{theorem}

\subsubsection{Parametric isoquants}
Here for the brevity of our presentation, we focus on the case of linear and power isoquants. Similar consistency result could also be established under other parametric settings.
\begin{theorem}
\label{thm:homo_consistency2}
	Suppose that Assumptions \ref{ass:1}--\ref{ass:3} hold. Furthermore, assume that $g_0(\bm{x}) = F_0(\mathscr{H}_0(\bm{x}))$ is homothetic, with $\mathscr{H}_0(\bm{x}) = \boldsymbol{\beta}_0^T \bm{x}$ or $\mathscr{H}_0(\bm{x}) =  \bm{x}^{\boldsymbol{\beta}}$ , $\boldsymbol{\beta}_0> \mathbf{0}$ and $\|\boldsymbol{\beta}_0\| = 1$. Then, $\hat{\boldsymbol{\beta}}_0 \stackrel{p}{\rightarrow} \boldsymbol{\beta}_0$, as $n \rightarrow \infty$.
\end{theorem}

\section{Simulation study}
\label{sec:simulation}
We use Monte Carlo simulations to evaluate the finite sample performance of the proposed estimator with datasets generated by the different data generation process (DGP). We consider different models for estimating isoquants: parametric homothetic, nonparametric homothetic and nonparametric non-homothetic.

\subsection{The setup}
\label{sec:setup}
In our simulation, we compare the performance of the proposed estimator with a Local Linear estimator (LL), which is an unconstrained nonparametric estimation method using kernel weights. We run simulations using the built-in quadratic programming solver, \texttt{quadprog}, in \texttt{MATLAB}. We define three DGPs to compare different models for estimating isoquants: parametric homothetic, nonparametric homothetic and nonparametric non-homothetic input isoquants. For each case, we run experiments varying the sample size and the size of noise. For a testing set drawn from the true DGP, we measure the Root Mean Squared Errors (RMSE) against the true function.

\subsection{Parametric homothetic isoquants}
\label{subsec:para}
 Here we compute the performance of S--shape estimator in case that we correctly specify the parametric expression of the input isoquant. The true production function used in the simulation is defined by the following scale and core function:
\begin{equation}
\label{eq:scale}
F_0(z)=\frac{15}{1+\exp(-5\log{z})}
\end{equation}
\begin{equation}
\label{eq:core_para}
\mathscr{H}_0(X_1,X_2)= X_1^\beta X_2^{\left(1-\beta\right)},
\end{equation}
where the intensity of the first input, $X_1$, is $\beta=0.50$. We generate samples from
\begin{equation}
\label{eq:dgp_prod_para}
y_j=F_0\left(\mathscr{H}_0(X_{1j},X_{2j})\right)+\epsilon_j,
\end{equation}
with an additive noise term generated as $\epsilon_j\sim N(0,\sigma_v)$, where $\sigma_v$ is the standard deviation of the additive noise. We radially generate inputs to the production function, $(X_1,X_2)$, as
\begin{equation}
\label{eq:input_dist}
\bm{X}=(X_1,X_2)=(\psi\cos\eta,\psi\sin\eta),
\end{equation}
\noindent with the modulus, $\psi$, generated as $\psi\sim U(0,2.5)$ and angles, $\eta$, generated as $\eta\sim U(0.05,\frac{\pi}{2}-0.05)$. Note this DGP specifies that inputs are generated radially and noise is additively contained in the output. 

We consider 9 scenarios varying the training set sample size $(100,500, 1000)$ and the standard deviations of the noise term, $\sigma_v\in(1.0, 2.0, 3.0)$. We compare our proposed estimator to the LL estimator. For the S--shape estimation, we use the procedure proposed in Section \ref{subsec:homo2} which uses parametric estimation for isoquants. Specifically, we search for the optimal value of $\beta$ which minimizes the residuals of S--shape estimation at a ray from the origin. For the S--shape estimation of our algorithm, we implement the SCKLS estimator.\footnote{We also implement the CNLS estimator for the S--shape estimation. The results are not significantly different from the one with the SCKLS estimator.} To compute the bandwidths for both the LL estimators and the SCKLS estimator for the S--shape part of our algorithm, we use Leave-one-out cross-validation (LOOCV) with the LL estimator. LOOCV is a data-driven bandwidth selection method that has been shown to perform well for unconstrained and constrained kernel estimators, respectively; see \cite{stone1977consistent} and \cite{yagi2018shape}.

\begin{figure}[p]
    \includegraphics[width=\textwidth]{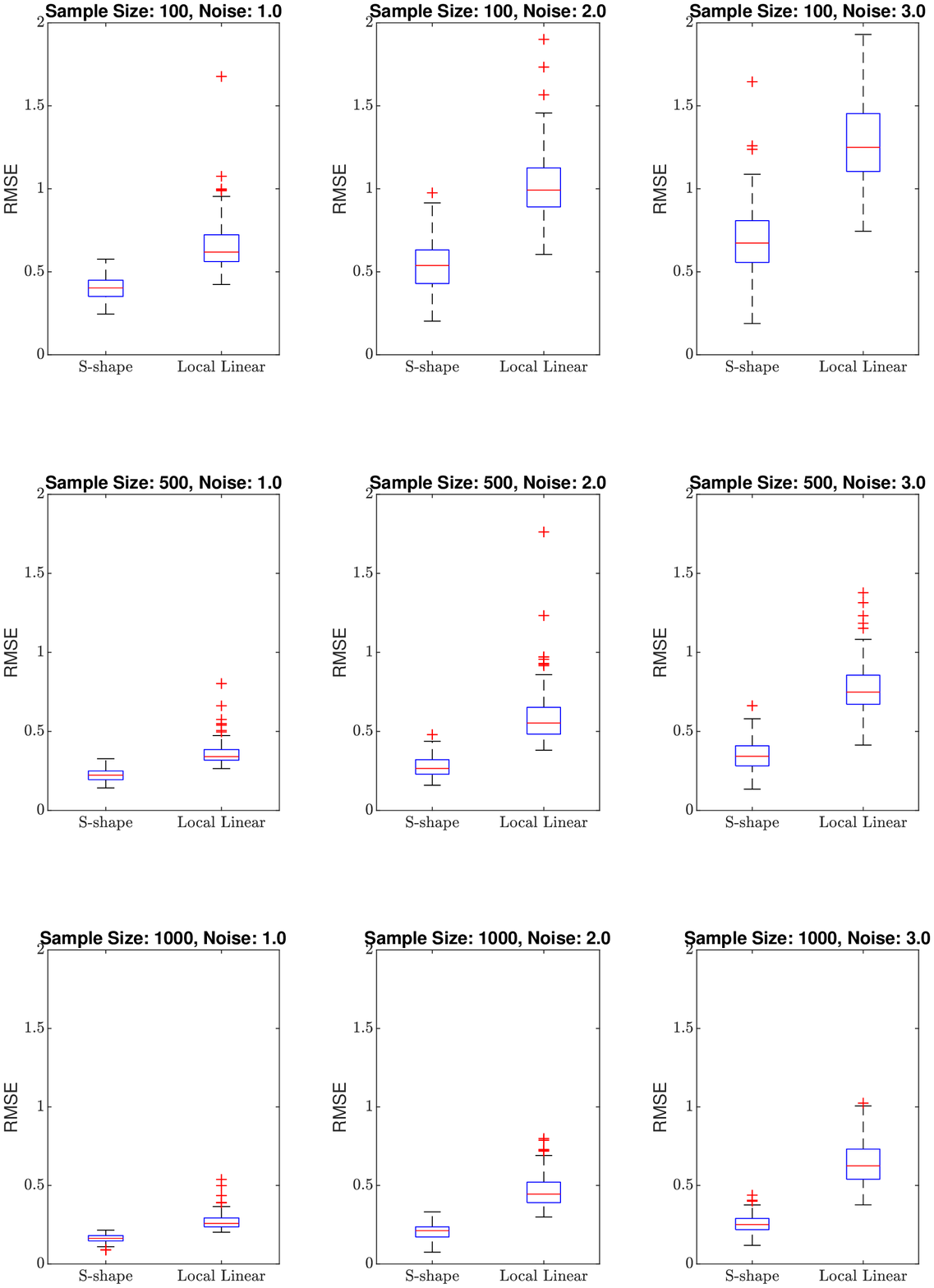}
    \centering
    \caption{Estimation results on the testing sets with the parametric homothetic isoquants}
    \label{fig:expParatest}
\end{figure}

We generate 100 training-testing set pairs for each scenario, and draw box plots\footnote{We define a maximum whisker length of a box plot as $\lbrack q_1 - 1.5(q_3 - q_1), q_3 + 1.5(q_3 - q_1)\rbrack$, where $q_1$ and $q_3$ denote the 25 and 75 percentiles, respectively.} of RMSE against the true function for both estimators shown in Figure \ref{fig:expParatest}. The size of the testing set is $1000$, and it is randomly drawn from the same distribution as the training set.

We find that the S--shape estimator performs significantly better than the LL estimator for all scenarios. This is because our estimator correctly specifies and imposes the parametric input isoquants. Due to the slower rate of convergence of the nonparametric estimator, the difference between the S--shape and LL estimator is large even with a larger sample size. Further, the variance in the S--shape estimator is smaller than that of LL estimator because the shape constraints and parametric structure reduce the estimator's variance.

\subsection{Nonparametric homothetic isoquants}
\label{subsec:homo}
The DGP we use has the same scale function (\ref{eq:scale}) and the following core function, which is used by \cite{olesen2014maintaining}:
\begin{equation}
\label{eq:core}
\mathscr{H}_0(X_1,X_2;y)=\left(\beta(y) X_1^\frac{\sigma-1}{\sigma}+(1-\beta(y)) X_2^\frac{\sigma-1}{\sigma}\right)^\frac{\sigma}{\sigma-1},
\end{equation}
where the elasticity of substitution is $\sigma=1.51$ and the intensity of the first input, $X_1$, is $\beta(y)=0.45$. For the homothetic case, the value of $\beta(y)$ is independent of output level $y$. We generate samples from
\begin{equation}
\label{eq:dgp_prod}
y_j=F_0\left(\mathscr{H}_0(X_{1j},X_{2j};y^*_j)\right)+\epsilon_j,
\end{equation}
where $y^*_j$ indicates a true functional value at $(X_{1j},X_{2j})$ satisfying 
\begin{equation}
\label{eq:output}
y^*_j=F_0\left(\mathscr{H}_0(X_{1j},X_{2j};y^*_j)\right) 
\end{equation}
with an additive noise term generated as $\epsilon_j\sim N(0,\sigma_v)$, where $\sigma_v$ is the standard deviation of the additive noise. This DGP generates homothetic input isoquants because the core function, $\mathscr{H}(\cdot)$, is independent of the output level, $y$. Input is radially generated as in the previous experiment and defined in (\ref{eq:input_dist}).

\begin{figure}[p]
    \includegraphics[width=\textwidth]{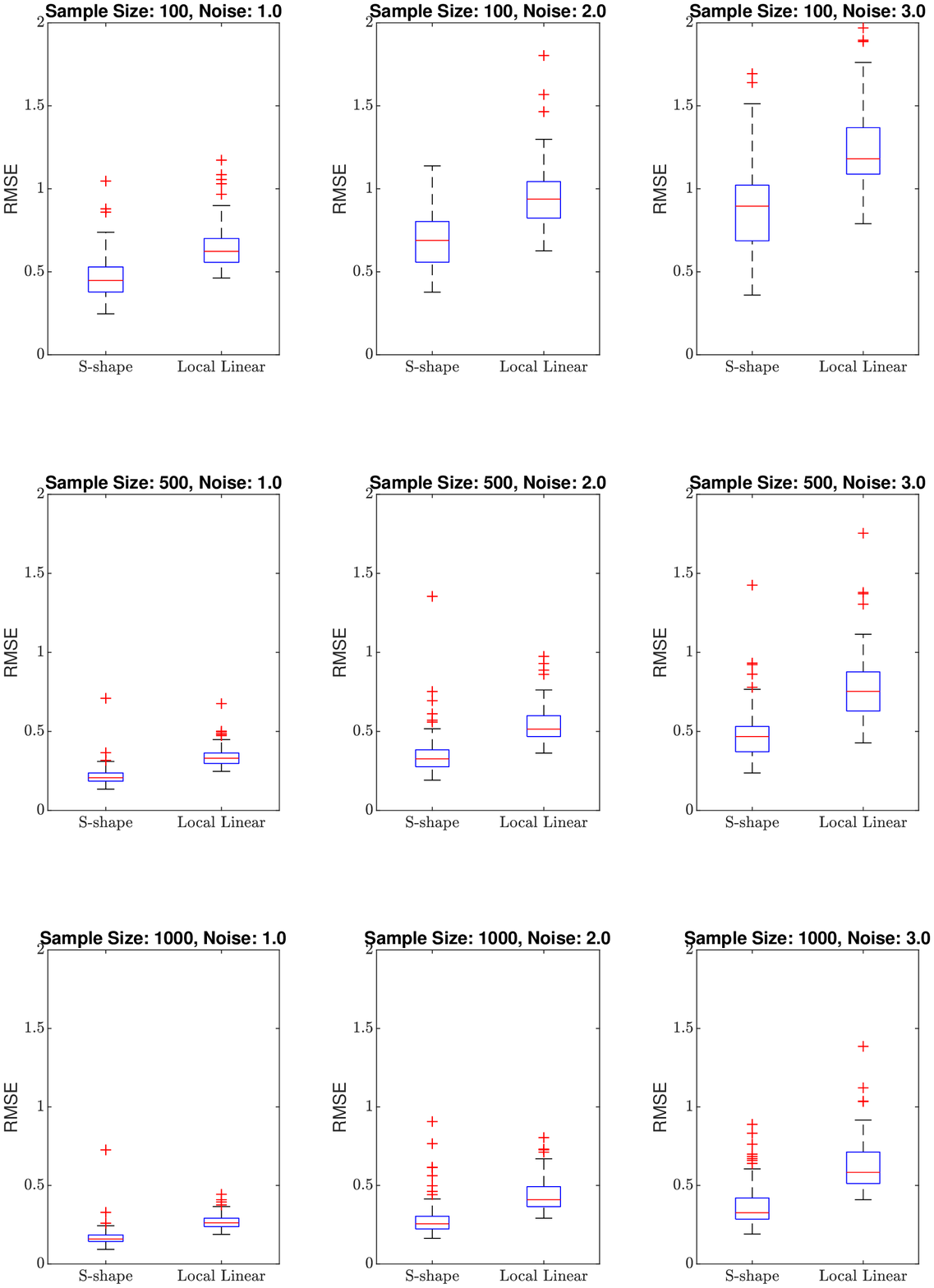}
    \centering
    \caption{Estimation results on the testing sets with the nonparametric homothetic isoquants}
    \label{fig:exp1test}
\end{figure}

We use the S--shape estimator with nonparametric homothetic input isoquants which is described in Section \ref{subsec:homo1}. We use the LL estimator as the pilot estimator of our S--shape model. We run simulations with same settings described in Section \ref{subsec:para}, and draw box plots of RMSE values against the true function for each estimator on testing set shown in Figure \ref{fig:exp1test}.

We find that the S--shape estimator performs  better than the LL estimator in all scenarios. Specifically, our S--shape estimator has better out-of-sample performance because the shape constraints add structures to the estimator, which helps to avoid over-fitting the observations. The difference in performance becomes larger as the noise increases because the flexible nature of the LL estimator. We find that the shape constraints in our S--shape estimator make it robust to noisy data. 

\subsection{Nonparametric non-homothetic isoquants}
\label{subsec:nhces}
We consider the same scale function (\ref{eq:scale}) and core function (\ref{eq:core}) as defined in Section \ref{subsec:homo}. We make the function non-homothetic by redefining the $\beta$ value as
\begin{equation}
\label{eq:nonhomo}
    \beta(y)=0.25 + \frac{y}{15} \times0.30,
\end{equation}
where $\beta(y)\in[0.25,0.55]$ depends on the output level $y\in[0,15]$. We generate the observations by solving equation (\ref{eq:output}) for a given $(X_{1j},X_{2j})$. This function is non-homothetic because the core function $g(\cdot)$ is dependent on an output level $y$. 

\begin{figure}[p]
    \includegraphics[width=\textwidth]{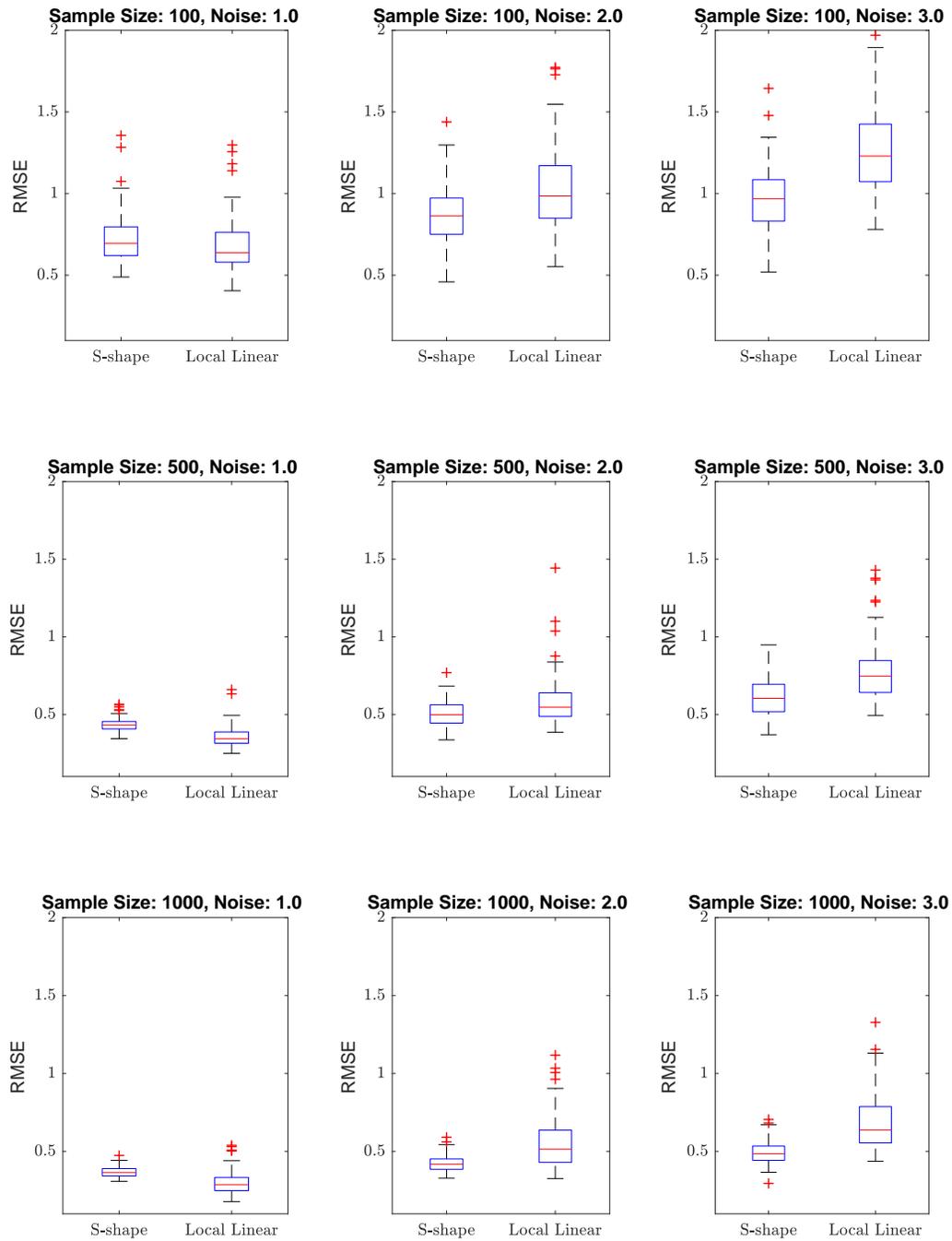}
    \centering
    \caption{Estimation results on the testing sets with the nonparametric non-homothetic isoquants}
    \label{fig:exp2test}
\end{figure}

We use the S--shape estimator with nonparametric non-homothetic input isoquants. We use Algorithm \ref{algo:advanced} to implement our estimator. We specify the number of isoquants and rays as $I=5$ and $R=5$, and compute equally spaced percentiles to set the location of the isoquant-level, $\{y^{(i)}\}_{i=1}^{I}$, and rays, $\{\bm{\theta}^{(r)}\}_{r=1}^{R}$, respectively. We use the average directional CNLS estimates for the isoquant estimation; the details are in Appendix \ref{subsubsec:ADCNLS}. 
We initialize the bandwidth between angles, $\bm{\omega}$, as $\omega_{1}=0.20$, and increment it by $\Delta\omega=0.25$. We iterate the procedure 20 times, increasing $\omega$ by $\Delta\omega$ in each iteration. After 20 iterations, we select the solution with the smallest sum of squared residuals as our final estimate. We allow the estimator to have a 1\% of gap between the convex isoquant estimates and the S-shape estimates. We run simulations with same settings described in Section \ref{subsec:para}, and draw box plots of RMSE values against the true function for each estimator on testing set shown in Figure \ref{fig:exp2test}.
\nomenclature[$\omega$]{$\Delta\omega$}{Inclement of bandwidth between angles}

We find that the LL estimator performs slightly better than our proposed estimator when the noise is very small, this is likely because our estimator optimizes the fit of the estimated function only on a limited set of grid points. Again, the LL estimator has a larger RMSE variance than our estimator for medium and high noise settings. However, both estimators have larger RMSE variance in the non-homothetic scenarios, particularly in very noisy instances. Our estimator still performs well in terms of RMSE, which indicates its robustness to different assumptions about the production function.

\section{Application}
\label{sec:application}
In this section, we estimate the production function using firm-level industry data from Japan's \emph{Census of Manufactures} provided by METI from 1997 to 2007, when demand for cardboard was relatively constant. Although some researchers have used the same dataset to estimate production functions (\cite{ichimura2011econometric}), they rely on strong parametric functional assumptions, whereas we relax them and estimate a production function nonparametrically under the RUP law and input convexity. We focus on economic insights related to the cardboard firms' productivity and scale of production.

\subsection{Census of Manufactures, Japan}
\label{subsec:census}

The annual \emph{Census of Manufactures} covers all establishments with four or more employees and is conducted by METI under the Japanese Statistics Act. We use establishment-level data with 30 or more employees since the establishment with less than 30 employees do not report capital stock values. We use the same definition of the variables for production functions as \cite{ichimura2011econometric}:
\begin{itemize}
    \item $L$ = (sum of total regular employees\footnote{Regular employees include full-time, part-time, and dispatched workers who work 18 days or more per month.} at the end of each month)
    \item $K$ = (starting amount of tangible assets\footnote{Tangible assets include machines, buildings, and vehicles.})
    \item $y$ = (total amount shipped) + ({ending inventory of finished and work-in-progress products}) - (starting inventory of finished and work-in-progress products) - (cost for intermediate inputs\footnote{Intermediate inputs include raw materials, fuel and electricity.}) 
\end{itemize}

where $L$,$K$ and $y$ indicate the labor, capital and value added, respectively, and the production function is modeled as $y = g_0(L,K)$.

\begin{figure}[p]
    \includegraphics[width=0.7\textwidth]{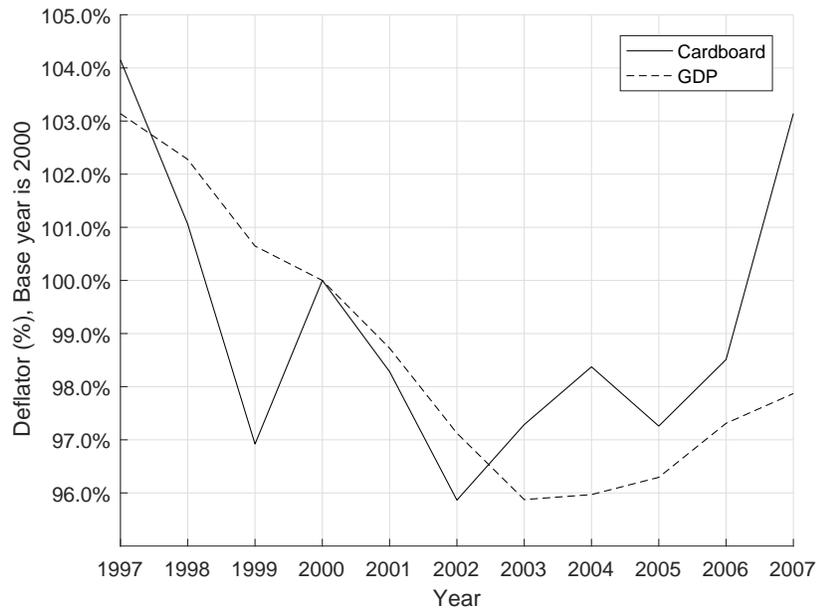}
    \centering
    \caption{Price deflator (Base year = 2000)}
    \label{fig:deflator}
\end{figure}

We use industry-level deflators obtained from the Japan Industrial Productivity Database (JIP)\footnote{The JIP database is publicly available at \emph{Research Institute of Economy, Trade and Industry} (REITI) (\url{https://www.rieti.go.jp/en/database/jip.html})} to convert into year 2000 values. Figure \ref{fig:deflator} shows the price deflator of the cardboard industry and the deflator for Japan's GDP. Note that the price deflator of the cardboard industry is larger than that of GDP after 2003. This finding is consistent with larger firms shrinking their production capacity, which led to higher cardboard prices after 2003, \cite{Iguchi2015}.

We convert establishment-level data into firm-level data by summing up the establishment-level data which belong to the same firm. We use firm-level data because expansion decisions are typically made at the firm-level by investing capital, labor, or merging with other firms.

The sample size of the panel data set is $n=4316$, and there are approximately 400 observations in each year.  We normalize each variable by dividing by the standard deviation for data confidentiality. 
Positive skewness of both the input and output variables implies the existence of many small and a few large firms. Table \ref{tab:summaryStat} reports the summary statistics.

\renewcommand{\arraystretch}{0.65}

\begin{table}[htbp]
  \centering
  \caption{Summary Statistics of the corrugated cardboard industry (1453)}
    \begin{tabular}{C{1in}C{1in}C{1in}C{1in}}
    \toprule
          & \multicolumn{1}{c}{Labor} & \multicolumn{1}{c}{Capital} & \multicolumn{1}{c}{Value added} \\
    \midrule
    Mean  & 0.554 & 0.283 & 0.340 \\
    Skewness & 10.28 & 11.87 & 11.86 \\
    10-percentile & 0.217 & 0.024 & 0.059 \\
    25-percentile & 0.253 & 0.047 & 0.093 \\
    50-percentile & 0.334 & 0.100 & 0.158 \\
    75-percentile & 0.539 & 0.231 & 0.298 \\
    90-percentile & 0.861 & 0.519 & 0.567 \\
    \bottomrule
    \end{tabular}%
  \label{tab:summaryStat}%
\end{table}%

\begin{figure}[p]
    \includegraphics[width=0.7\textwidth]{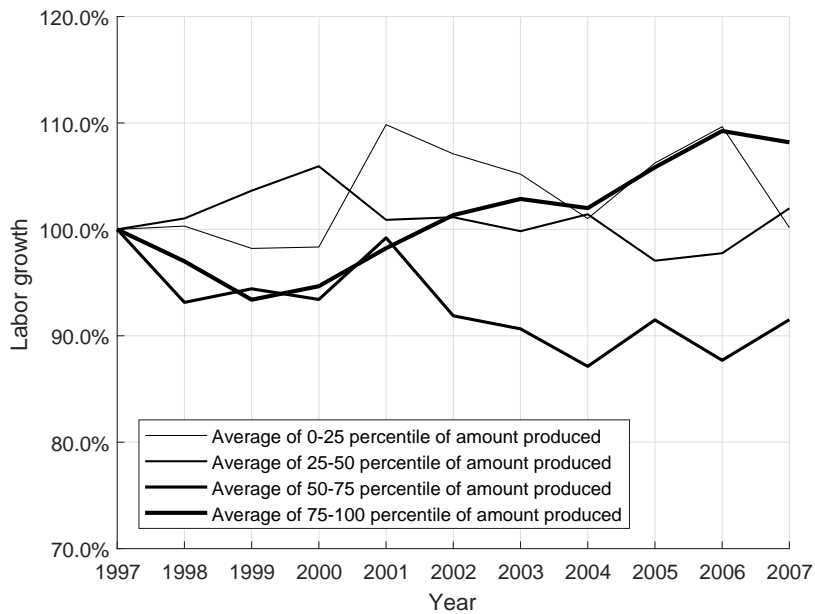}
    \centering
    \captionsetup{justification=centering}
    \caption{Percentage change of quartile mean of labor\\(by amount produced, base year = 1997)}
    \label{fig:labor_panel}
\end{figure}

\begin{figure}[p]
    \includegraphics[width=0.7\textwidth]{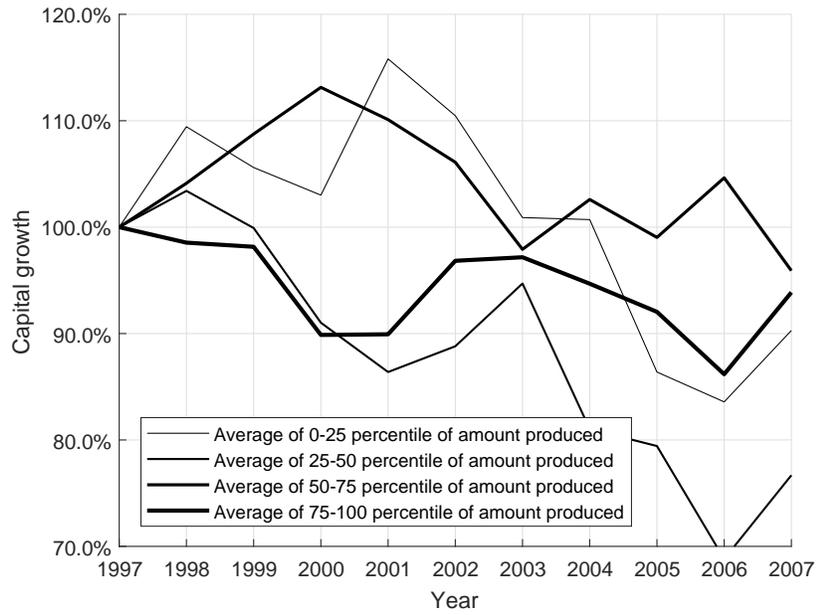}
    \centering
    \captionsetup{justification=centering}
    \caption{Percentage change of quartile mean of capital\\(by amount produced, base year = 1997)}
    \label{fig:capital_panel}
\end{figure}

\begin{figure}[p]
    \includegraphics[width=0.7\textwidth]{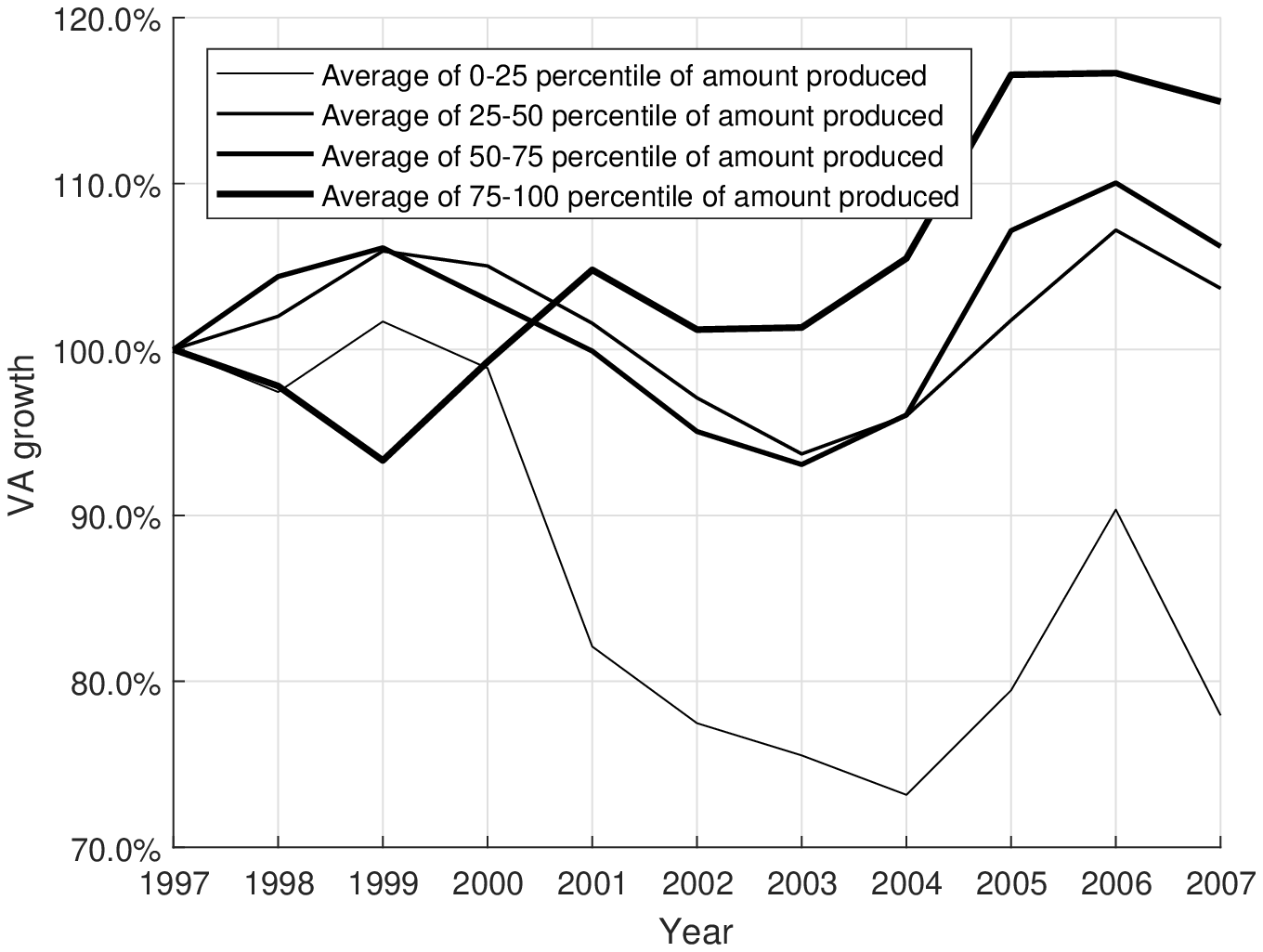}
    \centering
    \captionsetup{justification=centering}
    \caption{Percentage change quartile mean of value added\\(by amount produced, base year = 1997)}
    \label{fig:va_panel}
\end{figure}

Figure \ref{fig:labor_panel}, \ref{fig:capital_panel} and \ref{fig:va_panel} show the evolution of each variable across the panel periods by plotting the percentage change of each variable's quartile mean for each year compared with 1997. Here, we compute the quartiles by total amount produced, i.e. firms in the 75\%-100\% bin have the highest total amount produced, while firms in the lower percentile bin have lower total amount produced. We define total amount produced as:
\begin{itemize}
    \item (total amount produced) = (total amount shipped) + ({ending inventory of finished and work-in-progress products}) - (starting inventory of finished and work-in-progress products)
\end{itemize}
Intuitively, we use the total amount produced as an indicator of a firm's scale size.

The four lines indicate from thinnest to thickest, the 0--25 percentile mean, 25--50 percentile mean, 50--75 percentile mean, and 75--100 percentile mean, respectively. During the time period, firms did not need to adjust their labor levels significantly while most firms reduce their capital levels between 2004 and 2006. We can interpret this as firms in the cardboard industry realized their over-investment in capital and readjusted for more efficient resource use. We observe that the larger firms in our panel dataset expanded value added while reducing their capital levels.

\subsection{Initialization}
\label{subsec:appsetup}
Before using our iterative algorithm, we specify (1) Number and location of the rays and (2) Number and location ($y$-levels) of the isoquants. Table \ref{tab:summaryStat} reports significant skewness of our dataset, i.e. many small firms and only a few large firms. An equally spaced percentile grid will not work well because it may fail to define the rays and isoquant $y$-levels corresponding to the large firms. Therefore, we use the $K$-means clustering method to cluster the data into $\mathscr{K}$ groups. 
\nomenclature[$k$]{$\mathscr{K}$}{Number of clusters}

\begin{figure}[p]
    \includegraphics[width=0.7\textwidth]{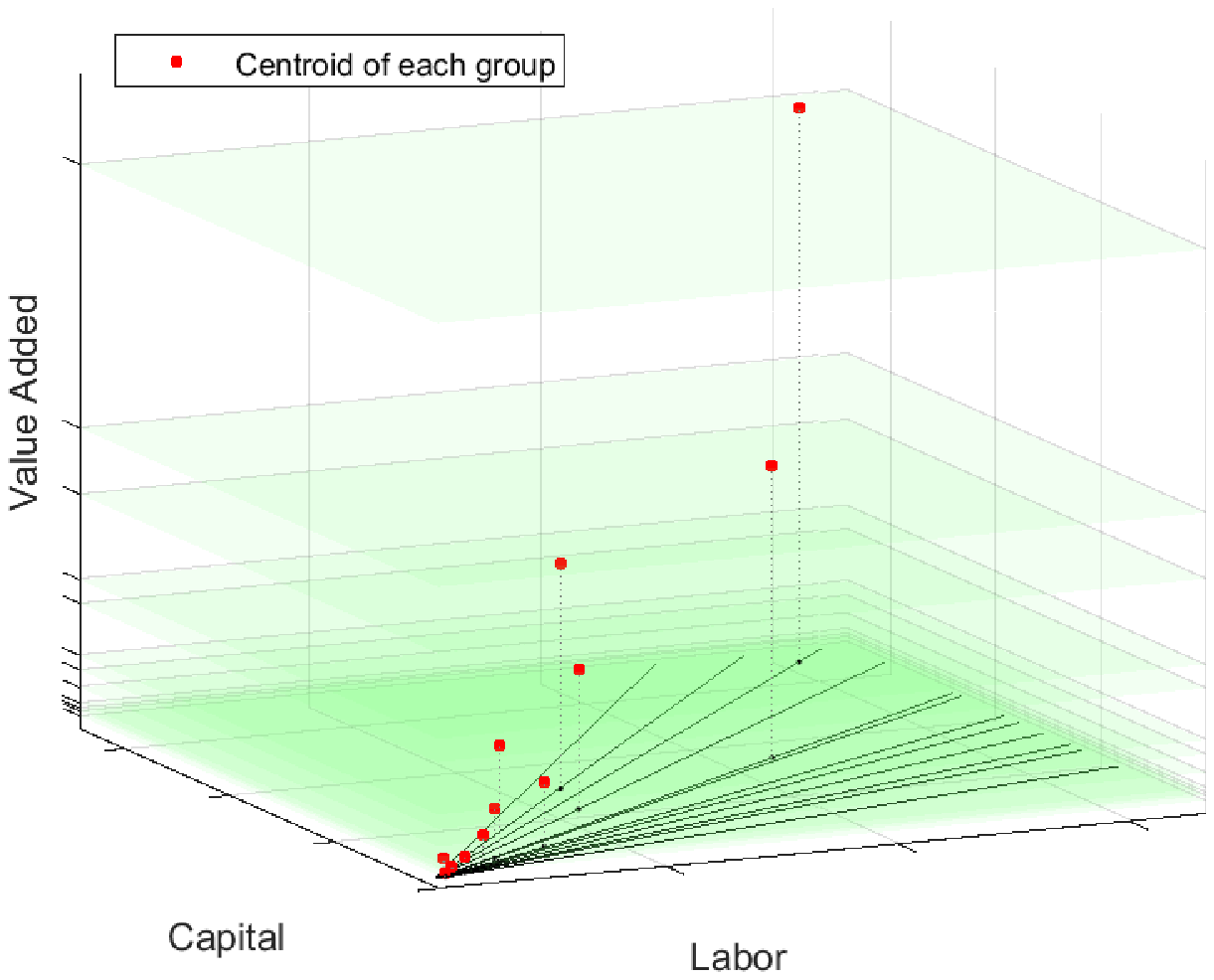}
    \centering
    \captionsetup{justification=centering}
    \caption{Centroid of each group estimated by $K$-means clustering}
    \label{fig:cluster}
\end{figure}

However, since $K$-means clustering requires pre-defining parameter $\mathscr{K}$ which is the number of clusters, we use Bayesian Information Criteria (BIC) to balance the model complexity and explanatory power and avoid over-fitting. We iterate the algorithm 100 times over different $K$, and find that $\mathscr{K}=12$ provides the lowest BIC value for our dataset. We define the rays and isoquant $y$-levels as the centroid of each cluster. Figure \ref{fig:cluster} shows the rays and isoquant $y$-levels defined by $K$-means clustering. 
There are many clusters defined for small scale firms and labor intensive firms and there are also a few clusters defined for large firms and capital intensive firms. 


We initialize the bandwidth between angles as $\omega_{1}=0.20$, and increase it by $\Delta\omega=0.20$ for each iteration. We iterate the procedure 50 times until $\omega$ becomes large enough that the functional estimates are stable between iterations.\footnote{When we use a very large bandwidth between angles $\omega$, the shape of function on each ray will be almost linear and violations of the convex-concave function definition, as defined by \cite{ginsberg1974multiplant}, are more common. In such cases, we cannot define the most productive scale size. When MSE is used as the criteria for selecting among alternative estimated, this problem typically does not arise, but if alternative objectives are used, such as smoothness of the estimator, this could be a potential issue.} From the 50 estimates, we select the solution with the smallest sum of squared residuals in our solution set as our final estimate\footnote{We allow five-percent gap between isoquant and S--shape estimates as a threshold.}.

\subsection{Estimated production function and interpretation}
\label{subsec:appresult}

\begin{figure}[htbp]
	\centering
	\subfloat[Estimated input isoquants]{\includegraphics[width=0.5\textwidth]{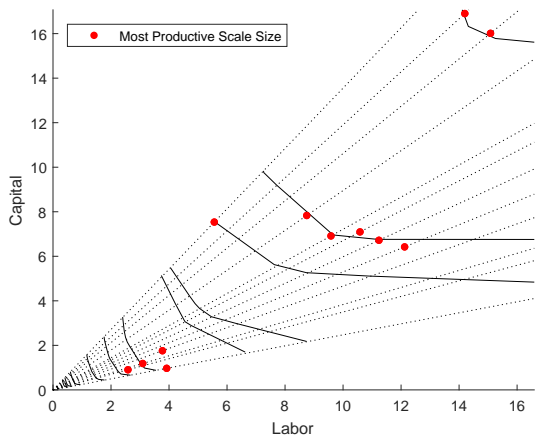}\label{fig:isoq_card}}
	\hfill
	\subfloat[Estimated production function]{\includegraphics[width=0.5\textwidth]{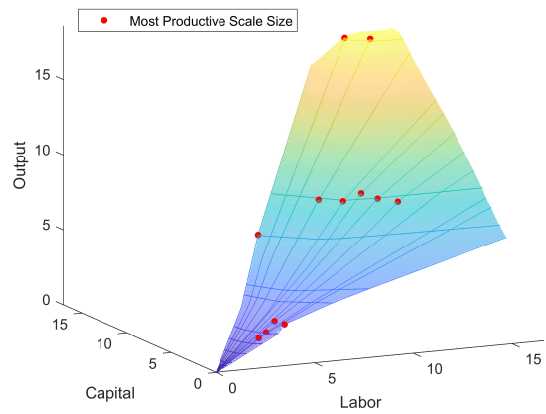}\label{fig:prod_card}}
	\caption{Estimated results of the corrugated cardboard industry}
	\label{fig:cardboard_result}
\end{figure}

Figure \ref{fig:cardboard_result} shows graphs of: \subref{fig:isoq_card} the estimated input isoquants, and \subref{fig:prod_card} the estimated S-shape production function on each ray. The black lines indicate the estimates on the centroid of each cluster defined by $K$-means clustering, and the red points indicate the most productive scale size on each ray from the origin. Figure \ref{fig:cardboard_result} \subref{fig:isoq_card} shows that the marginal rate of technical substitution (MRTS) of labor for capital is high when the scale of production is smaller. This indicates that labor is a more important input factor for firms operating at a smaller scale. In contrast, the isoquant becomes flat as the scale of production increases, i.e. the MRTS is low for large firms. These isoquants imply that capital is a more important input factor for larger firms because labor levels need to increase significantly to offset a small reduction in capital.

Figure \ref{fig:cardboard_result} \subref{fig:prod_card} shows that labor intensive firms have a much smaller most productive scale size than capital intensive firms. This finding coincides with the production economics theory stating that firms become more capital intensive as they grow larger by automating processes with capital equipment and using less labor. Note that the most capital intensive ray has a smaller most productive scale size. This is likely a result of over-investment in capital. Therefore, these capital intensive firms could reduce their capital intensity in order to increase their productivity and scale of operations.

\subsection{Analysis on productivity measure}
\label{subsec:appproductivity}
Our production function estimator makes a new decomposition of productivity possible and allows further investigation of productivity variation. Productivity is the ratio of observed output $y_{jt}$ to aggregate input $\mathscr{H}_{0}(L_{jt},K_{jt})$. Intuitively, if firms have higher productivity, they can produce larger value added with a given amount of input factors. Total factor productivity, the residual in a growth accounting exercise, can measure the firms' deviation of output (value added) which cannot be explained by the input factors. \cite{syverson2011determines} enumerates the primary causes of productivity dispersion as managerial practices, quality of input factors, R\&D, learning by doing, product innovation, firms' structure decisions, or other external drivers. 

We measure unexplained productivity residual defined as follows: 
\begin{equation}
\label{eq:TFP}
TFP_{jt} = \frac{y_{jt}}{\hat{g}_0(L_{jt},K_{jt})}~~~~~\forall j=1,\ldots,n_t\mbox{ and }\forall t= 1,\ldots,T,
\end{equation}
where $n_t$ is a sample size for each time period $t$, $T$ denotes the panel periods, and $\hat{g}_0$ is a S-shape estimator of the production function used to aggregate inputs. 

\begin{figure}[t]
    \includegraphics[width=0.7\textwidth]{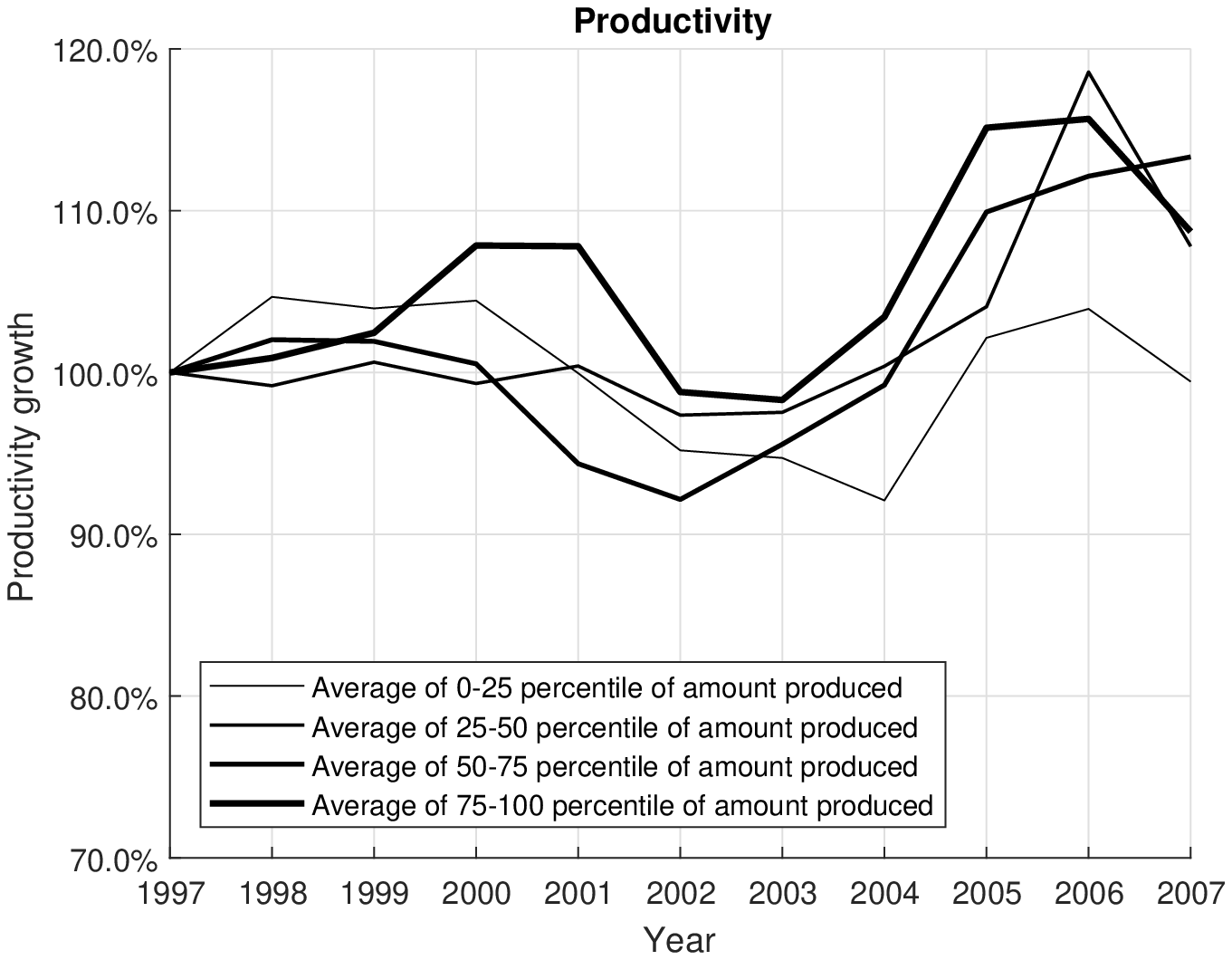}
    \centering
    \captionsetup{justification=centering}
    \caption{Percentage change of quartile mean of productivity\\(by amount produced, base year = 1997)}
    \label{fig:prod_trans_scale}
\end{figure}

\begin{figure}[t]
    \includegraphics[width=0.7\textwidth]{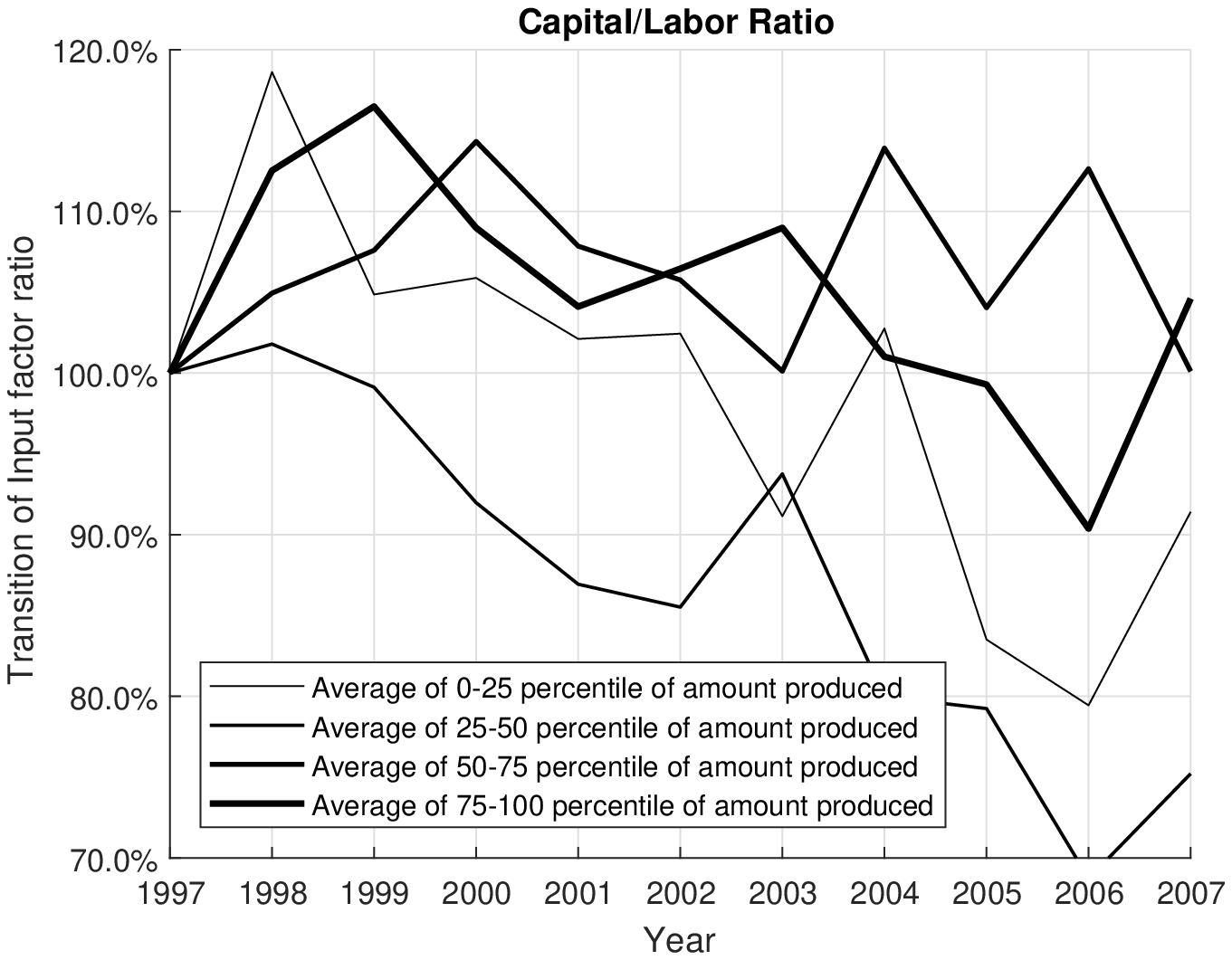}
    \centering
    \captionsetup{justification=centering}
    \caption{Percentage change of quartile mean of input ratio\\(by amount produced, base year = 1997)}
    \label{fig:input_ratio_panel}
\end{figure}

First, we investigate how the productivity for the cardboard industry is changing over time. Figure \ref{fig:prod_trans_scale} and \ref{fig:input_ratio_panel} plot the percentile change of quartile mean of productivity and capital-to-labor input factor ratio for each year compared with 1997, respectively.

 Figure \ref{fig:prod_trans_scale} shows that the medium and large firms have significant productivity growth after 2004, whereas small firms have more stable productivity transition. In contrast, Figure \ref{fig:input_ratio_panel} describes that smaller firms tend to shrink capital-to-labor ratio after 2004. 
 Since the productivity of the cardboard industry is heavily dependent on the amount of capital investment, small firms had difficulty to improve their productivity endogenously over the 11 years. 

We now turn our attention to the decomposition of the productivity to investigate the cause of productivity deviation. We will use three methods to calculate the production function: Cobb--Douglas with Constant Returns to Scale (CRS), homothetic S--shape and non-homothetic S--shape\footnote{We find significant differences between the productivity variation measures when TFP is calculated using growth accounting measures and when a Cobb-Douglas production function is estimated. We believe these differences are driven by the fact that there are significant fixed costs to capital and therefore setting the cost share of capital equal to its marginal product is a weak assumption. Further, under Constant returns-to-scale the coefficients of the input factors are restricted to sum to 1, thus the distortion on the capital coefficient is transmitted to the other input variables. We describe this in detail in Appendix \ref{app:prod_disp}.}. The first method is the most restrictive model since the scale function at any rays from the origin is linear. The second method can explain the benefit of increasing the scale size since the scale function follows the S--shape axiom. However, since the model assumes homothetic isoquants, it cannot explain the benefits to changing the input factor ratio. The last model is the most flexible model, and characterizes the benefit of changing input factor ratio and scale.

The productivity defined by Cobb--Douglas with CRS is decomposed into following three terms:
\begin{equation}
    \frac{y_{jt}}{\hat{g}_0^{CRS}(L_{jt},K_{jt})} = \frac{\hat{g}_0^{H}(L_{jt},K_{jt})}{\hat{g}_0^{CRS}(L_{jt},K_{jt})} \cdot \frac{\hat{g}_0^{NH}(L_{jt},K_{jt})}{\hat{g}_0^{H}(L_{jt},K_{jt})} \cdot \frac{y_{jt}}{\hat{g}_0^{NH}(L_{jt},K_{jt})}
\end{equation}
where $\hat{g}_0^{CRS}$, $\hat{g}_0^{H}$, and $\hat{g}_0^{NH}$ denote the estimated production function with Cobb--Douglas CRS, homothetic S--shape, and non-homothetic S--shape respectively. The productivity estimated with the CRS model can be decomposed into: (1) scale productivity which is the ratio of Cobb--Douglas CRS and homothetic S--shape, (2) input mix productivity which is the ratio of non-homothetic and homothetic S--shape, and (3) unexplained productivity by non-homothetic S--shape.
\nomenclature[$g$]{$\hat{g}_0^{CRS}(\bm{x})$}{Cobb--Douglas CRS estimates}
\nomenclature[$g$]{$\hat{g}_0^{H}(\bm{x})$}{Homothetic S--shape Estimates}
\nomenclature[$g$]{$\hat{g}_0^{NH}(\bm{x})$}{Non-homothetic S--shape estimates}

\begin{figure}[t]
    \includegraphics[width=0.99\textwidth]{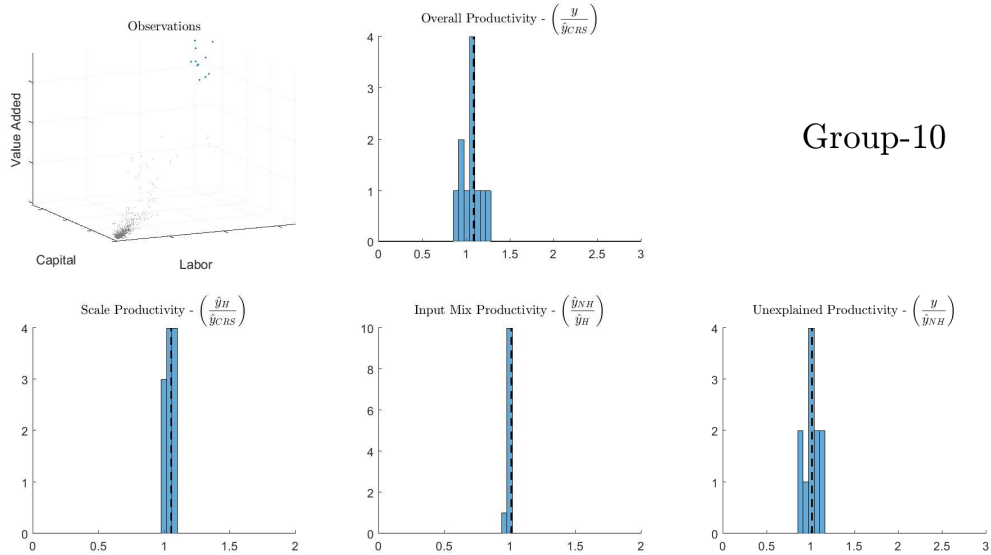}
    \centering
    \captionsetup{justification=centering}
    \caption{Productivity decomposition of firms belong to the group of the largest firms (Group--10)}
    \label{fig:prod_decom_largest}
\end{figure}

Here, we compute a productivity decomposition for each group defined by the $K$--means clustering. The group number is arranged in the ascending order of a capital intensity: Group--1 is the most labor intensive and Group--12 is the most capital intensive group. Figure \ref{fig:prod_decom_largest} shows the histogram of each decomposed productivity for the group of the largest firms (Group--10) which is highlighted in the left-top figure. The dash line in each histogram indicates the median productivity level within the group. Since these firms in this group are operating near the scale close to the most productive scale size, they have both high scale and input mix productivity. This indicates that the firms are productive with current scale size and input mix, and they can produce at relatively lower costs than other firms operated lower productivity level.

\begin{figure}[t]
    \includegraphics[width=0.99\textwidth]{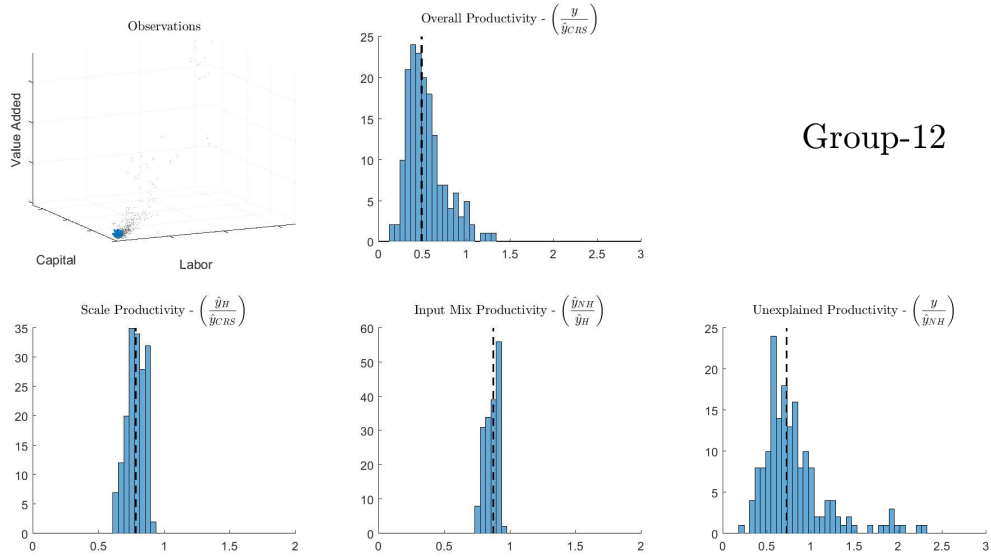}
    \centering
    \captionsetup{justification=centering}
    \caption{Productivity decomposition of firms belong to the group of the capital intensive small firms (Group--12)}
    \label{fig:prod_decom_small_capital}
\end{figure}

Figure \ref{fig:prod_decom_small_capital} shows the same histogram for the group of smaller and capital intensive firms (Group--12). While these firms are capital intensive, the scale size is much smaller than the most productive scale size. Thus, we can observe that both their scale productivity and input mix productivity is low. This indicates that these firms should increase their scale size to improve their scale productivity, or they should change the input mix to become more labor intensive to improve their input mix productivity.

Appendix \ref{app:prod_decomp} contains the histogram of decomposed productivity for all 12 groups defined by K--means clustering. In summary, the productivity decomposition provides the source of productivity-level of each firm. Furthermore, it also provides the critical managerial insights for the expansion of firms to make them more productive and increase their survival probability. 

\section{Conclusion}
\label{sec:conclusion}
This paper develops an approach to estimate a general production function imposing economic axioms, both the RUP law and the input convexity. The axioms can be stated as shape constraints and the proposed estimator is implemented as a non-parametric shape constrained regression. This approach allows considerable more flexibility than the widely-used parametric methods. 

We use this newly-proposed approach to analyze a panel dataset of Japan's cardboard industry from 1997 to 2007. We observe a capacity contraction after 2004 across most of the larger firms in the industry. The contraction's timing corresponds to an increase in the price index for cardboard productions, indicating increasing market power of firms in the industry. We estimate the production function and compute the most productive scale size and the productivity of each firm. We find most productive scale size is significantly influence by the capital-labor ratio of the firm. In particular firms with higher capital-to-labor ratios have a larger most productive scale size than firms with lower capital-to-labor ratios. We also decompose the productivity into the scale and input mix productivity to analyze the cause of productivity level. While large capital intensive firms benefit from both their scale size and input mix, we find that the small capital intensive firms need an improvement by either expanding their scale or if the firm cannot expand production, then adjusting their input mix. 


We plan to extend our analysis to other industries in Japan which have roughly homogeneous outputs such as bread, coffee, concrete, plywood, and sugar. Census of Manufacturing data are self reported by firms and are notoriously noisy. 
Thus, estimators that take advantage of additional axiomatic information are beneficial in this setting. We will study the patterns across industries to identify which factors (scale, input mix, etc.) consistently influencing productivity. 

As managers strategically plan the expansion of their firm, estimates of the most productive scale size, the trade-offs between manual and automated operations, and the potential outputs gains to expansion provide critical insights to the benefit-cost analysis. The proposed axiomatic approach imposes a minimum set of axioms that still allows for the standard interpretation of the production function allowing managers to be better informed when taking critical planning decisions for the firm.


\nomenclature[$x$]{$\bm{X}^{(i)}$}{Subset of observations used for estimating isoqaunt $i$}
\nomenclature[$n$]{$n_i$}{Number of observations used for estimating isoquant $i$}

\nomenclature[$e$]{$e_j$}{Residual for observation $j$ for isoquants estimates}
\nomenclature[$\alpha$]{$\alpha_j^{(i)}$}{Intercept estimates of isoquant estimates}
\nomenclature[$\beta$]{$\bm{\beta}_j^{(i)}$}{Slope estimates of isoquant estimates}

\nomenclature[$g$]{$\bm{g}^{X_{-d}}$}{Direction vector for $d-1$ input components}
\nomenclature[$g$]{$g^{X_d}$}{Direction scalar for $d$-th input}
\nomenclature[$\gamma$]{$\gamma_j^{(i)}$}{Slope coefficient for the $d$-th input in CNLSD}


\nomenclature[$r$]{$R_j^{(r)}$}{Radial distance to observation $j$ mesuared along ray $r$}
\nomenclature[$r$]{$R_j^{(below)}$}{Radial distance to isoquant below observation $j$ along ray $r$}
\nomenclature[$r$]{$R_j^{(above)}$}{Radial distance to isoquant above observation $j$ along ray $r$}
\nomenclature[$\rho$]{$\rho_j$}{Weight for interpolating isoquants}

\nomenclature[$r$]{$r_g^{(r)}$}{Radius of evaluation point $g$ on ray $r$}
\nomenclature[$g$]{$g$}{Index of evaluation points on ray}
\nomenclature[$g$]{$G$}{Number of evaluation points on ray}

\nomenclature[$d$]{$D(\cdot, \cdot)$}{Euclidian distance function}

\nomenclature[$a$]{$a_g^{(r)}$}{Functional estimate on ray $r$ at evaluation point $g$}
\nomenclature[$b$]{$b_g^{(r)}$}{Slope estimate on ray $r$ at evaluation point $g$}
\nomenclature[$g$]{$g_*^{(r)}$}{Index of inflection point on ray $r$}
\nomenclature[$k$]{$k(\cdot)$}{Kernel function}
\nomenclature[$k$]{$K(\cdot)$}{Product kernel}

\nomenclature[$r$]{$r^{(i)(r)}$}{Radial distance to isoquant $i$ along ray $r$}
\nomenclature[$g$]{$G'$}{Union of $G$ and $I$}
\nomenclature[$g$]{$g'$}{Index of union of $G$ and $I$}

\nomenclature[$a$]{$\tilde{a}_g^{(r)}$}{Revised functional estimate on ray $r$ at evaluation point $g$}
\nomenclature[$w$]{$w^{S}$}{Weight for S--shape estimates}
\nomenclature[$w$]{$w^{I}$}{Weight for isoquant estimates}

\nomenclature[$r$]{$r_{\bm{x}}$}{Radius of given input $\bm{x}$}
\nomenclature[$\phi$]{$\bm{\phi_{x}}$}{Angles of given input $\bm{x}$}

\nomenclature[$r$]{$r_{\bm{x}}^{(r)}$}{Radial distance to $\bm{x}$ along ray $r$}

\nomenclature[$r$]{$r^*$}{Index of ray closest to angle $\bm{\phi_{x}}$}
\nomenclature[$a$]{$\tilde{a}_{\bm{x}}^{(r)}$}{Functional estimates of given input $\bm{x}$ evaluated on ray $r$}
\nomenclature[$p$]{$p^{(r)}$}{Inverse distance weight on ray $r$}

\nomenclature[$t$]{$T$}{Number of iterations allowing errors over tolerance}
\nomenclature[$\delta$]{$\delta$}{Tolerance of errors between S--shape and isoquant}

\nomenclature[$h$]{$\mathscr{H}(\cdot)$}{Isoquant function}


\nomenclature[$x$]{$x_A^*$}{Inflection point with aggregate input $x_A$}
\nomenclature[$x$]{$x_A$}{Aggregate input}
\nomenclature[$\delta$]{$\Delta_n$}{Difference of isoquant levels}
\nomenclature[$c$]{$\bm{C}'$}{Compact set belongs to the interior of $\bm{C}_y$}
\nomenclature[$g$]{$\mathcal{G}_{d-1}$}{Class of convex and decreasing functions}
\nomenclature[$b$]{$B(x,r)$}{Closed ball centered at $x$ of radius $r$}


\nomenclature[$\lambda$]{$\lambda_y$}{Scaling factor to the true isoquant of the level $y$}

\nomenclature[$f$]{$\mathcal{F}'$}{Class of increasing functions}
\nomenclature[$c$]{$\bm{C}$}{Compact set belongs to the interior of $\bm{S}$}
\nomenclature[$g$]{$G_0(\cdot)$}{Transformed scale function $F_0(\cdot)$}
\nomenclature[$g$]{$\mathcal{G}$}{Sub-class of increasing functions}




\newtheorem{example}{Example}

    \newpage
    \appendix

\noindent{\huge\textbf{Appendix}}
	
    \vspace{1cm}

\bigskip
This appendix includes:
\begin{itemize}
    \item List of symbols (Appendix \ref{app:symblos})
	\item Detailed algorithm and estimation procedure (Appendix \ref{app:algo}),
	\item Comparison of different isoquant estimators (Appendix \ref{App:compCNLS}),
	\item Technical proofs of the theoretical results (Appendix \ref{app:proof})
	\item Comparison between S--shape and the RUP Law (Appendix \ref{app:ce_rup})
    \item Quantifying uncertainty of our estimator (Appendix \ref{app:uncertainty})
	\item Productivity dispersion among different models (Appendix \ref{app:prod_disp})
	\item Comprehensive results of productivity decomposition (Appendix \ref{app:prod_decomp})
\end{itemize}

\noindent%

\newpage
\spacingset{1.5} 

\appendix
	
\section{List of symbols}
\label{app:symblos}
\xpatchcmd{\thenomenclature}{%
  \section*{\nomname}
}{
}{\typeout{Success}}{\typeout{Failure}}\setlength{\nomitemsep}{-\parsep}
\begin{multicols}{2}
\printnomenclature
\end{multicols}

\section{Detailed algorithm and estimation procedure}
\label{app:algo}
	
In this section, we described the detailed estimation algorithm and mathematical formulations. The algorithm consists of two estimation steps: (1) input isoquants estimation for a set of $y$--levels using Convex Nonparametric Least Squares (CNLS) type estimator, and (2) S-shape functions on a set of rays from the origin using Shape Constrained Kernel Least Squares (SCKLS). Algorithm \ref{algo:advanced2} presents the details of our algorithm which is composed of three steps: Initialization, Iteration and Updating parameters. The section numbers, where the details of each step are described, are displayed in the right column of the table.

\newcounter{mycount1}
\setcounter{mycount1}{1}
\counterwithin{algorithm}{mycount1}
\refstepcounter{mycount1}
\setcounter{algorithm}{1}
\renewcommand\thealgorithm{\arabic{mycount1}\Alph{algorithm}}

\algrenewcommand\algorithmicindent{2.0em}%
\begin{algorithm}
    \caption{Details of the advanced estimation algorithm}\label{algo:advanced2}
    \begin{algorithmic}[1]
        \BState \textbf{Data: } \text{observations } $\{\bm{X}_j,y_j\}_{j=1}^{n}$
        \Procedure{}{}\hspace{\fill}(Section)
        \BState \emph{Initialization}:\hspace{\fill}(\ref{subsec:init})
            \State $I \gets$ Initialize number of isoquants
            \State $R \gets$ Initialize number of rays 
            \State $\{y^{(i)}\}_{i=1}^{I} \gets$ Initialize isoquant $y$-levels
            \State $\{\bm{\theta}^{(r)}\}_{r=1}^{R} \gets$ Initialize rays from origin
            \State $\bm{\omega}\gets$ Initialize smoothing parameter between rays
            \State Project observations $\{\bm{X}_j,y_j\}_{j=1}^{n}$ to the isoquant level $y^{(i)}$
            \State Estimate initial isoquants by the CNLS-based estimation \hspace{\fill}(\ref{subsec:isoq})
        \BState \emph{Iteration}:
            \While{Termination condition not reached}
                \State Project observations onto the ray $\bm{\theta}^{(r)}$ \hspace{\fill}(\ref{subsubsec:project})
                \State Update S-shape estimates using the SCKLS-based estimator\hspace{\fill}(\ref{subsubsec:SCKLS})
                \State Update isoquant estimates by the CNLS-based estimator\hspace{\fill}(\ref{subsubsec:input_info})
                \State Minimize the gap between S-shape and isoquant estimates\hspace{\fill}(\ref{subsubsec:min_gap})
                \State Compute Mean Squared Errors against observations\hspace{\fill}(\ref{subsubsec:estimates_on_obs})
        \BState \emph{Updating parameters}:\hspace{\fill}(\ref{subsec:update})
                \State $I \gets$ Update number of isoquants
                \State $R \gets$ Update number of rays from the origin
                \State $\{y^{(i)}\}_{i=1}^{I} \gets$ Update isoquant $y$-levels
                \State $\{\bm{\theta}^{(r)}\}_{r=1}^{R} \gets$ Update rays from origin
                \State $\bm{\omega} \gets$ Update smoothing parameter between rays
            \EndWhile
        \State \textbf{end}
        \BState \textbf{Return: } 
        \State Estimated function with minimum Mean Squared Errors and gap smaller than threshold
    \EndProcedure
    \end{algorithmic}
\end{algorithm}


\subsection{Initialization}
\label{subsec:init2}
The number of isoquants $I$, the number of rays from the origin $R$, isoquant $y$-levels, $y^{(i)}$, and rays from the origin, $\bm{\theta}^{(r)}$ can be initialized in the same way as what we discussed in section \ref{subsec:init}. 

In the estimation of S-shape function on rays from the origin, we need to specify the smoothing parameter (bandwidth) between rays, $\bm{\omega}$, which determines the weights on each observation based on the angle between the observation and the ray from the origin on which we are currently estimating. Instead of optimizing bandwidth between rays, $\bm{\omega}$, by grid search in Algorithm \ref{algo:basic}, we try to find the optimal bandwidth between rays by increasing $\bm{\omega}$ by some increments, $\Delta\bm{\omega}$, with updating both isoquants and S-shape estimates in each iteration of Algorithm \ref{algo:advanced2}. Based on our numerical experiments, we recommend to start from a small value and increase $\bm{\omega}$ by small increment $\Delta\bm{\omega}$ in each iteration. We will generate a set of estimates and select from the set. Intuitively, the S-shape function, estimated along the ray, only gives significant weight to observations close to the ray in the first iteration. As our algorithm progresses, the S-shape estimation step gives weight to observations more distance from the ray. For more details of the S-shape estimation and the smoothing parameters, see Appendix \ref{subsubsec:SCKLS}.

\subsection{Estimate convex isoquants}
\label{subsec:isoq}
We are interested in estimating the isoquant function $\mathscr{H}$ in (\ref{eq:isoq}) at a given level of output. Assume that a set of output levels for isoquant estimation is given by
\begin{equation}
    \label{eq:y_iso}
    y^{(i)},~i=1,\ldots,I
\end{equation}
where $I$ is the number of isoquants to be estimated. Also assume that the  input data used to estimate the isoquant at $y^{(i)}$ is given by
\begin{equation}
    \label{eq:x_iso}
    \bm{X}^{(i)},~i=1,\ldots,I
\end{equation}
where $\bm{X}^{(i)}$ is subset of observations of input used for the estimation of isoquant at level $y^{(i)}$. $\bm{X}^{(i)}$ is $n_i\times d$ matrix and $n_i$ denotes the number of observations used for estimation of isoquant $i$ at level $y^{(i)}$. We have already described the procedure for specifying isoquant level $y^{(i)}$ in section \ref{subsec:init} and how to obtain the input data $\bm{X}^{(i)}$ associated with the isoquant level $y^{(i)}$ in section \ref{subsubsec:isoq_est}.

We first propose to use the existing nonparametric estimation method called Convex Nonparametric Least Squares (CNLS) to estimate isoquants.  We also propose two modifications to the CNLS estimator which improve the performance of the isoquant estimation.

\subsubsection{Convex Nonparametric Least Squares (CNLS)}
\label{subsubsec:CNLS}
\cite{kuosmanen2008representation} extends Hildreth's least squares approach to the multivariate setting with a multivariate $\bm{x}$, and coins the term Convex Nonparametric Least Squares (CNLS). CNLS builds upon the assumption that the true but unknown function belongs to the set of continuous, monotonic increasing/decreasing and globally concave/convex functions. We describe the isoquant function $\mathscr{H}$ at $y^{(i)}$ as
\begin{equation}
	\begin{aligned}
        \label{eq:isoq3}
        X_{j,d}^{(i)} = \mathscr{H}\left(\bm{X}_{j,-d}^{(i)};y^{(i)}\right)+e_j = \alpha_j^{(i)} + \bm{\beta}_j^{(i)}{'}\bm{X}_{j,-d}^{(i)}+e_j, && \forall j=1,\ldots,n_i.\\
    \end{aligned}
\end{equation}
where $e_j$ is the random small error, $\alpha_j^{(i)}$ and $\bm{\beta}_j^{(i)}$ define the intercept and slope parameters that characterize the estimated set of hyperplanes.

For each $i=1,\ldots,I$, we compute the CNLS estimator using $\Big\{\big(\,\bm{X}_{j,-d}^{(i)},X_{j,d}^{(i)}\, \big)\Big\}_{j=1}^{n_i}$, and obtain the isoquant estimates $\hat{\mathscr{H}}(\bm{x};y^{(i)})=\max_{j=1,\ldots,n_i}\Big\{\hat{\alpha}^{(i)}_j + \hat{\bm{\beta}}_j^{(i)}{'}(\bm{x}-\bm{X}_{j,-d}^{(i)})\Big\}$ at each isoquant level $y^{(i)}$. Here the CNLS estimator can be computed by solving the quadratic programming problem:
\begin{equation}
	\begin{aligned}
    	\label{eq:CNLS}
    	& \min_{\alpha,\bm{\beta}}
    	& & \sum_{j=1}^{n_i}\left(X_{j,d}^{(i)}-\left(\alpha_j^{(i)}+\bm{\beta}_j^{(i)}{'}\bm{X}_{j,-d}^{(i)}\right)\right)^2\\
    	& \mbox{subject to}
    	& & \alpha_j^{(i)}+\bm{\beta}_j^{(i)}{'} \bm{X}_{j,-d}^{(i)} \geq\alpha_l^{(i)}+\bm{\beta}_l^{(i)}{'} \bm{X}_{j,-d}^{(i)}, \; & \forall j,l=1,\ldots,n_i\\
    	&
    	& & \bm{\beta}_j^{(i)}\leq 0, \; & \forall j=1,\ldots,n_i
    \end{aligned}
\end{equation}
The first set of inequality constraints in (\ref{eq:CNLS}) can be interpreted as a system of Afriat inequalities that imposes convexity. See \cite{afriat1972efficiency} and \cite{varian1984nonparametric}. The second set of inequality constraints imposes monotonicity. We note that the functional estimates resulting from (\ref{eq:CNLS}) is unique only for the observed data points. \cite{seijo2011nonparametric} and \cite{lim2012consistency} proved the consistency of the CNLS estimator. Also \cite{chen2016convex} proves that the CNLS estimator attains $n^{-1/2}$ pointwise rate of convergence in the univariate setting when the true function is piece-wise linear.

\subsubsection{Directional CNLS}
\label{subsubsec:DCNLS}
The CNLS estimator in the previous section assumes that the input data contains errors only in the $d$-th input direction while all input variables are typically measured with error. \cite{kuosmanen2017} introduces the CNLS estimator within the directional distance function (DDF) framework. The DDF indicates the distance from a given sample vector to the estimated function in some pre-assigned direction. In our isoquant estimation, we can write the DDF function as follows:
\begin{equation}
    \vec{D}(\bm{X}_{j,-d}^{(i)},X_{j,d}^{(i)},\bm{g}^{X_{-d}},g^{X_d})=e_j, \:\:\: \forall j=1,\ldots,n_i
\end{equation}
where $(\bm{g}^{X_{-d}},g^{X_d})\in\mathbb{R}^d$ is the pre-assigned error direction. We can choose the error direction $(\bm{g}^{X_{-d}},g^{X_d})$ empirically from the density of the input data. Specifically we select the 50th percentile capital to labor ratio as the direction for the estimator. We also normalize input data $\{\bm{X}_{j}^{(i)}\}_{j=1}^{n_i}$ by dividing the inputs by their corresponding sample standard deviations, so that they all have unit sample variance. Here normalizing inputs avoids the situation where one input, measured on a large scale, dominates other inputs, measured on smaller scales.

Similar to the CNLS estimator, for $i=1,\ldots,I$, we compute the directional CNLS estimator with $\Big\{\big(\,\bm{X}_{j,-d}^{(i)},X_{j,d}^{(i)}\, \big)\Big\}_{j=1}^{n_i}$, and obtain the isoquant estimation at each isoquant level $y^{(i)}$. The directional CNLS estimator is computed by solving the quadratic programming problem:
\begin{equation}
	\begin{aligned}
    	\label{eq:DCNLS}
    	& \min_{\alpha,\bm{\beta},\gamma}
    	& & \sum_{j=1}^{n_i}\left(\gamma_j^{(i)} X_{j,d}^{(i)}-\left(\alpha_j^{(i)}+\bm{\beta}_j^{(i)}{'}\bm{X}_{j,-d}^{(i)}\right)\right)^2\\
    	& \mbox{subject to}
    	& & 
    	\scalebox{0.95}{$\displaystyle
    	\alpha_j^{(i)}+\bm{\beta}_j^{(i)}{'} \bm{X}_{j,-d}^{(i)} - \gamma_j^{(i)} X_{j,d}^{(i)} \geq\alpha_l^{(i)}+\bm{\beta}_l^{(i)}{'} \bm{X}_{j,-d}^{(i)} - \gamma_l^{(i)} X_{j,d}^{(i)}, 
    	$}
    	& \forall j,l=1,\ldots,n_i\\
    	&
    	& & \bm{\beta}_j^{(i)}\leq 0, \; & \forall j=1,\ldots,n_i\\
    	&
		& & \gamma_j^{(i)}\geq 0, \; & \forall j=1,\ldots,n_i\\
    	&
    	& & \gamma_j^{(i)} g^{X_d} + \bm{\beta}_j^{(i)}{'} \bm{g}^{X_{-d}} = 1, \; & \forall j=1,\ldots,n_i
    \end{aligned}
\end{equation}
This formulation introduces new coefficients $\gamma_j^{(i)}$ that represents marginal effects of the $d$-th input to the DDF. Similar to the CNLS estimator (\ref{eq:CNLS}), first three constraints impose convexity and monotonicity in all input directions respectively. The last constraints are normalization constraints that ensure the translation property \citep{chambers1998profit}.

\subsubsection{Averaging directional CNLS}
\label{subsubsec:ADCNLS}
\begin{figure}[ht]
    \includegraphics[width=5in]{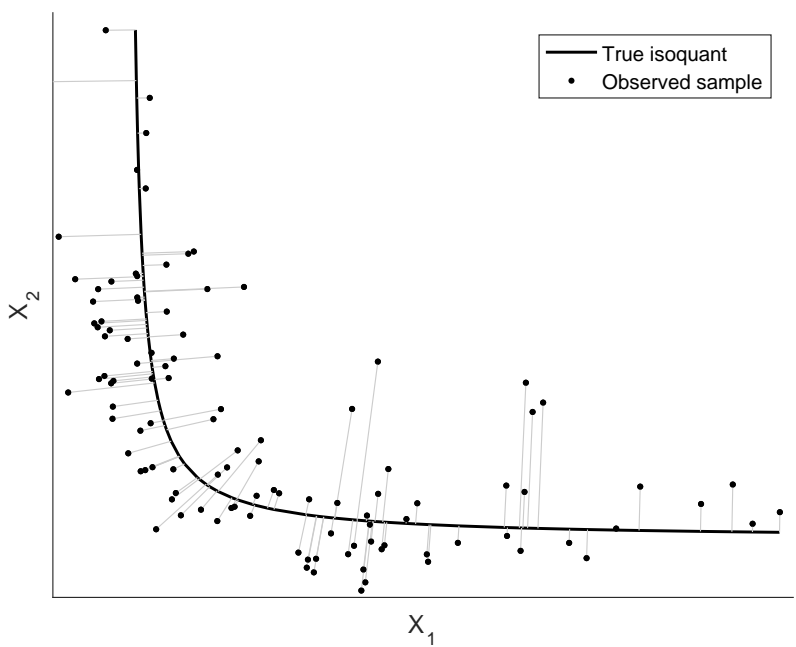}
    \centering
    \caption{Noise which is orthogonal to the true isoquant}
    \label{fig:error_isoq}
\end{figure}

The directional CNLS estimator in previous section assumes that the input data contains errors in potentially all variables, but in fixed ratios such that the over all error direction is $(\bm{g}^{X_{-d}},g^{X_d})$. However, in observed production data, the errors in different components of the input vector, $\bm{X}_j^{(i)}$, may vary in length randomly. Particularly when estimating input isoquants, observations can be projected to the function orthogonally as shown in Figure \ref{fig:error_isoq}. 
Noise here is mainly caused by the projection of observations to particular isoquant level $y^{(i)}$. This issue is discussed in Section \ref{app:algo}\ref{subsec:init}. 

If we misspecified the error direction, the estimated isoquants will be biased, and the bias will increase as the specified error direction moves further from the true error direction. We propose a simple algorithm to average out a bias from the misspecification of the error direction. We define the set of error directions $\left\{\left(\bm{g}_m^{X_{-d}},g_m^{X_d}\right)\right\}_{m=1}^M$ from the distribution of the input data $\bm{X}^{(i)}$ where $M$ is the number of error directions considered.\footnote{Based on our numerical experiments, we recommend to use $M=10$ and define error directions by the equally spaced percentile of the input ratio.} For each isoquant level $y^{(i)}$, we compute the directional CNLS estimator (\ref{eq:DCNLS}) with each error direction $\left\{\left(\bm{g}_m^{X_{-d}},g_m^{X_d}\right)\right\}_{m=1}^M$, and averaging them to obtain the final isoquant estimates. The final isoquant estimates still satisfied conditions for an isoquant in Assumption \ref{ass:isoq} since the average of convex monotone decreasing functions is a convex monotone decreasing function.

\begin{figure}[ht]
	\centering
	\subfloat[CNLS]{\includegraphics[width=0.33\textwidth]{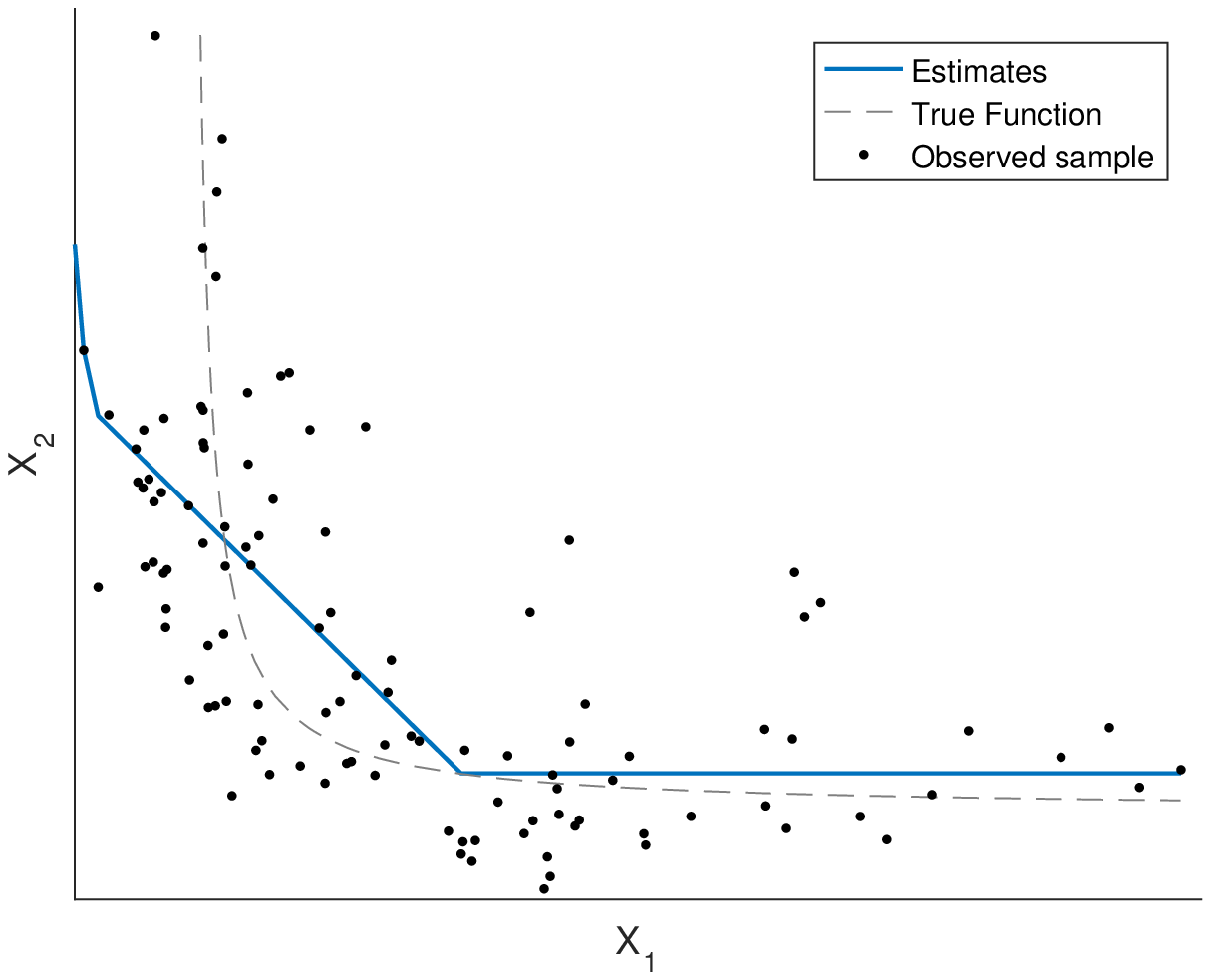}\label{fig:CNLS}}
	\hfill
	\subfloat[Directional CNLS]{\includegraphics[width=0.33\textwidth]{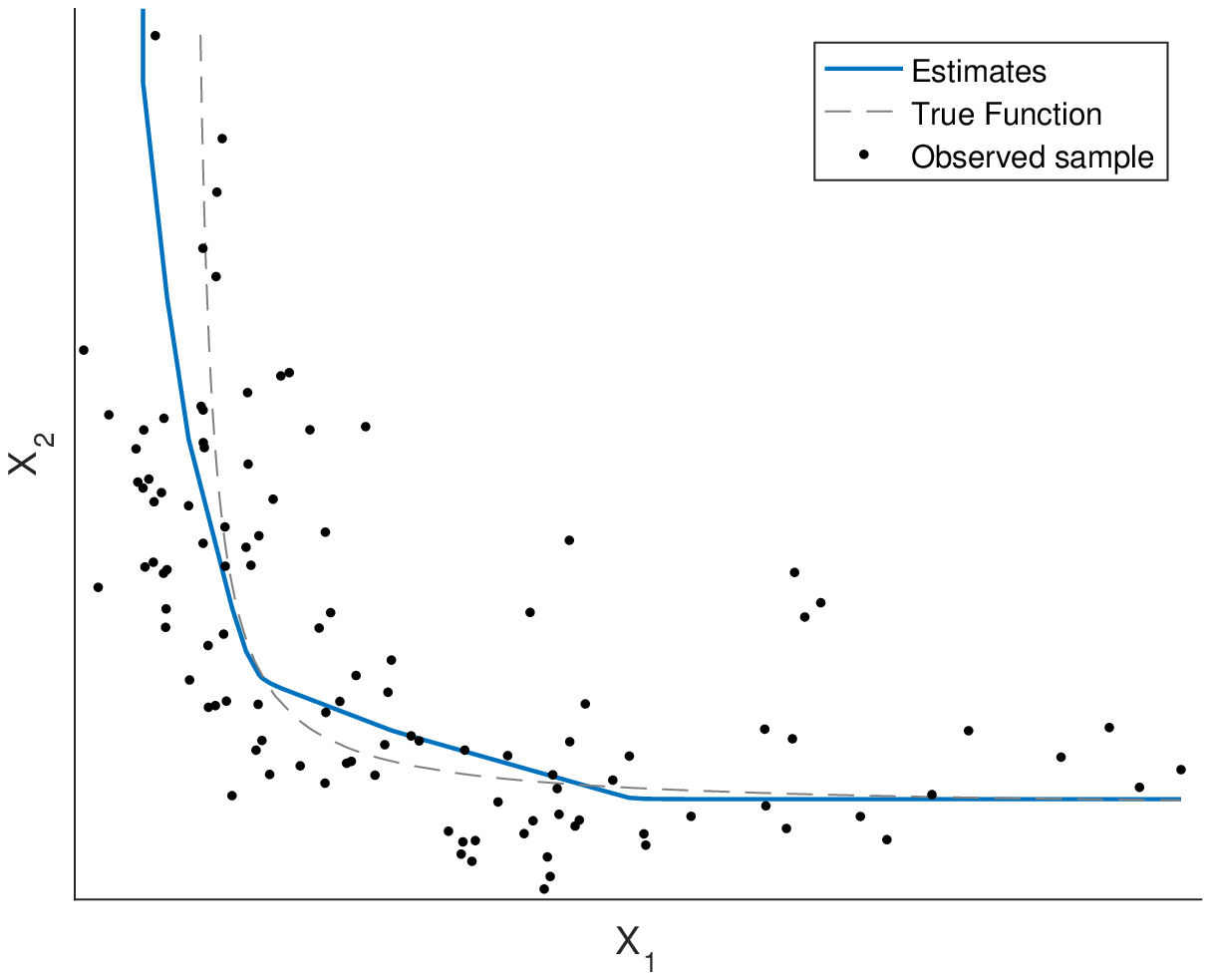}\label{fig:DCNLS}}
	\hfill
	\subfloat[Averaging directional CNLS]{\includegraphics[width=0.33\textwidth]{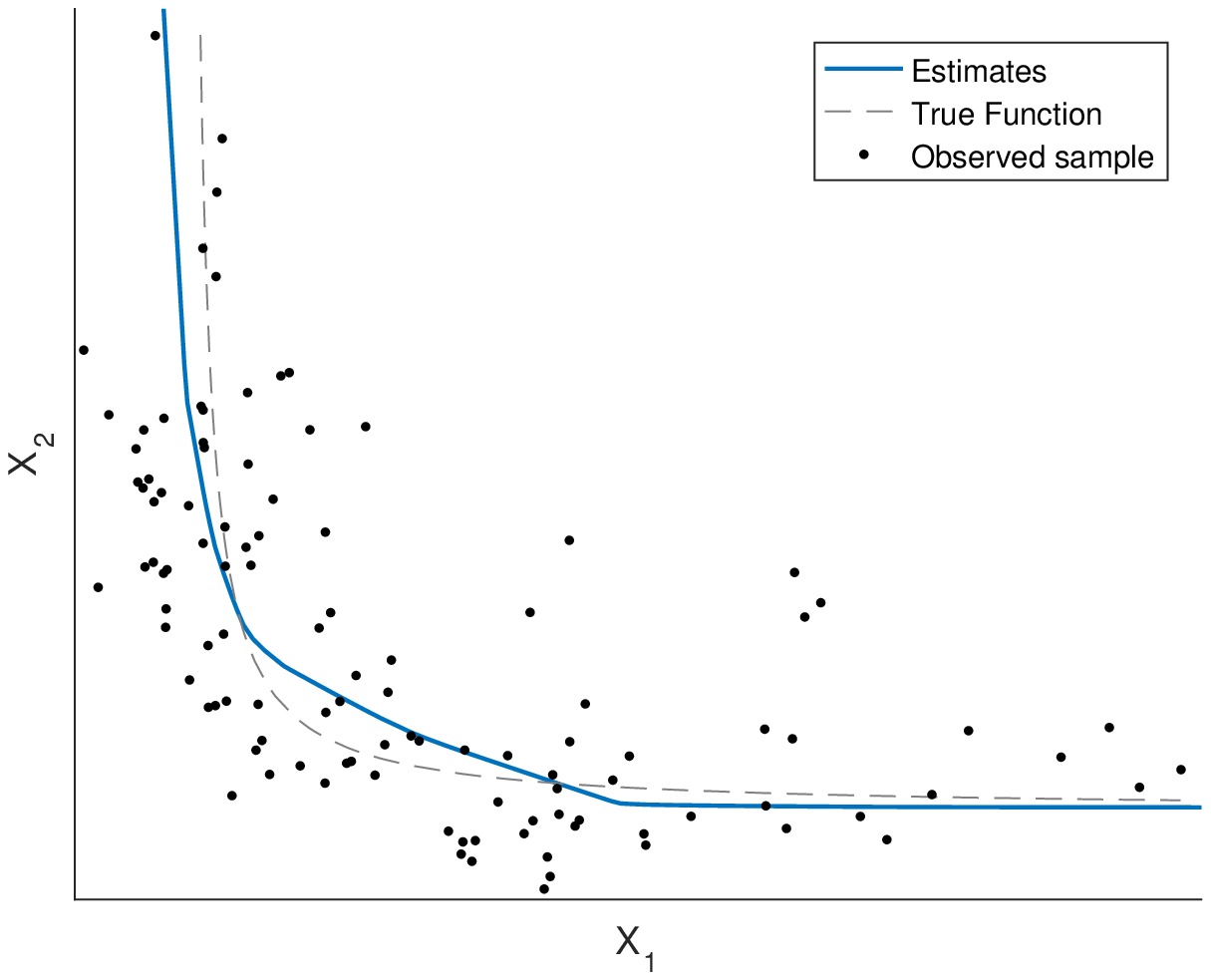}\label{fig:ADCNLS}}
	\caption{Estimated isoquant by CNLS, Directional CNLS and Averaging directional CNLS}
	\label{fig:est_isoq}
\end{figure}

Figure \ref{fig:est_isoq} \subref{fig:CNLS}, \subref{fig:DCNLS} and \subref{fig:ADCNLS} show the estimation results with CNLS, Direction CNLS and Averaging direction CNLS respectively with samples generated by radial errors. The CNLS estimator has noticeable bias for the observations close to the boundary. Directional CNLS and averaging multiple estimates of directional CNLS with different directions performs better than the CNLS estimator because these methods allow for errors in all input dimensions (instead of just in the $d$-th dimension, as implied in the original CNLS). Our experience suggests that both extensions perform well even for small sample size, as shown in Appendix \ref{App:compCNLS}.

\subsection{S-shape function}
\label{subsec:s-shape}
We are interested in estimating the S-shape function on rays from the origin as a component of our estimation procedure. This step is composed of two sub-steps: First, we project each observation to each ray from the origin by projecting along an estimated isoquant. Second, we estimate the S-shape function on each ray from the origin. We describe the procedure how to obtain the rays from the origin $\bm{\theta}^{(r)}$ in section \ref{subsec:init}.

\subsubsection{Projecting observations onto rays}
\label{subsubsec:project}
Before estimating S-shape functions, we project the observations $\{\bm{X}_j,y_j\}_{j=1}^n$ to each ray from the origin $\bm{\theta}^{(r)}$. Two approaches are described below.
\paragraph{Distance-based approach}
We perform the projection purely based on the covariates, i.e.
\[
R_j^{(r)}= \langle \bm{X}_j,\bm{\theta}^{(r)}\rangle /  \|\bm{\theta}^{(r)}\|.
\]

\begin{figure}[ht]
	\centering
	\subfloat[Estimated isoquants and predefined rays from the origin]{\includegraphics[width=0.33\textwidth]{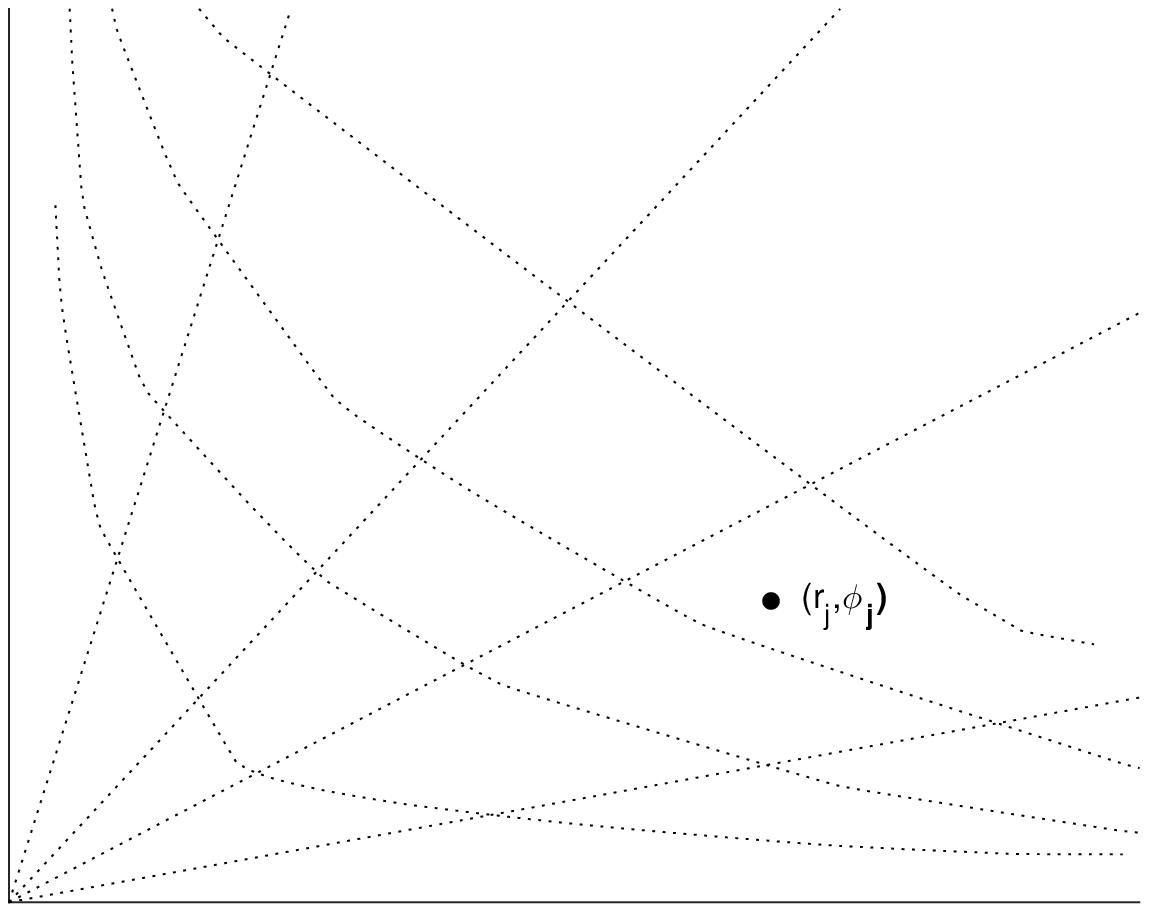}\label{fig:isoqray}}
	\hfill
	\subfloat[Two isoquants sandwich the observation]{\includegraphics[width=0.33\textwidth]{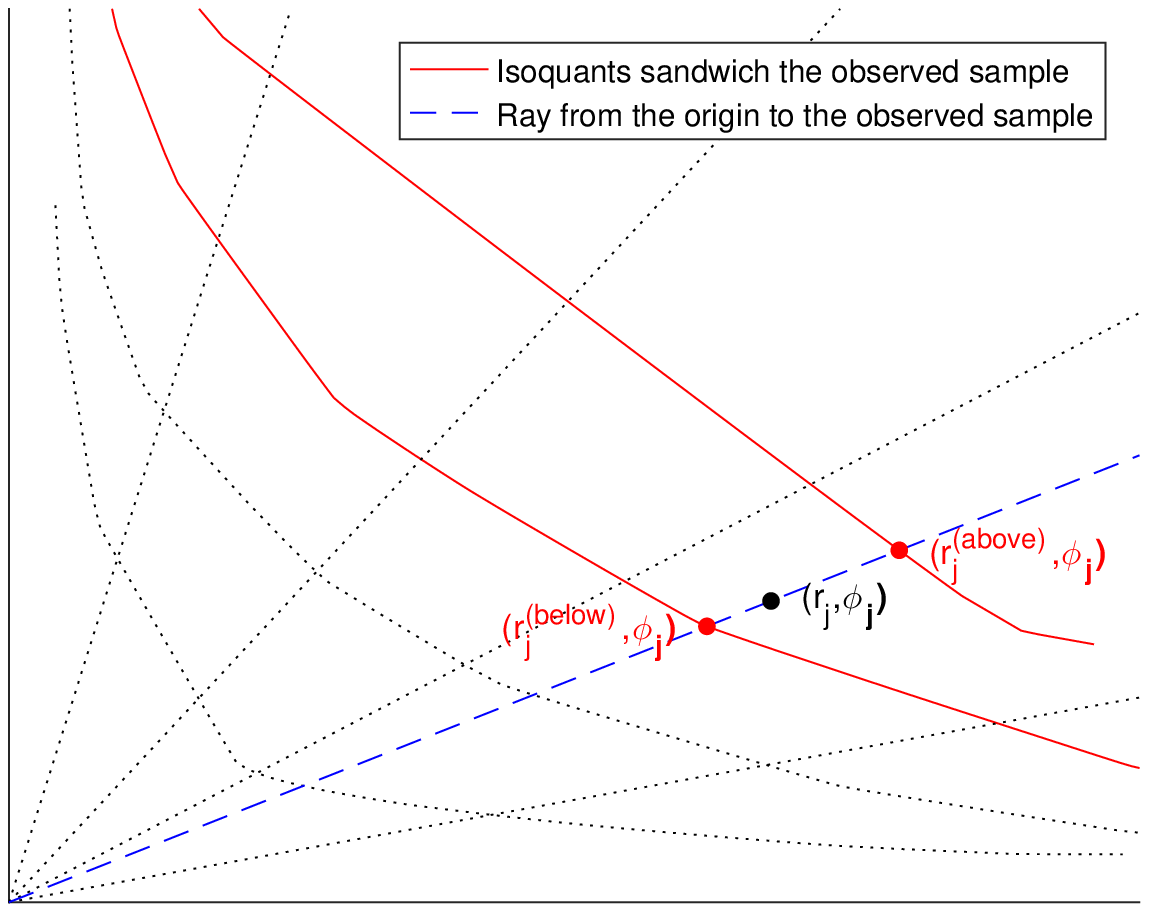}\label{fig:extisoq}}
	\hfill
	\subfloat[Projection of observation to each predefined rays from the origin]{\includegraphics[width=0.33\textwidth]{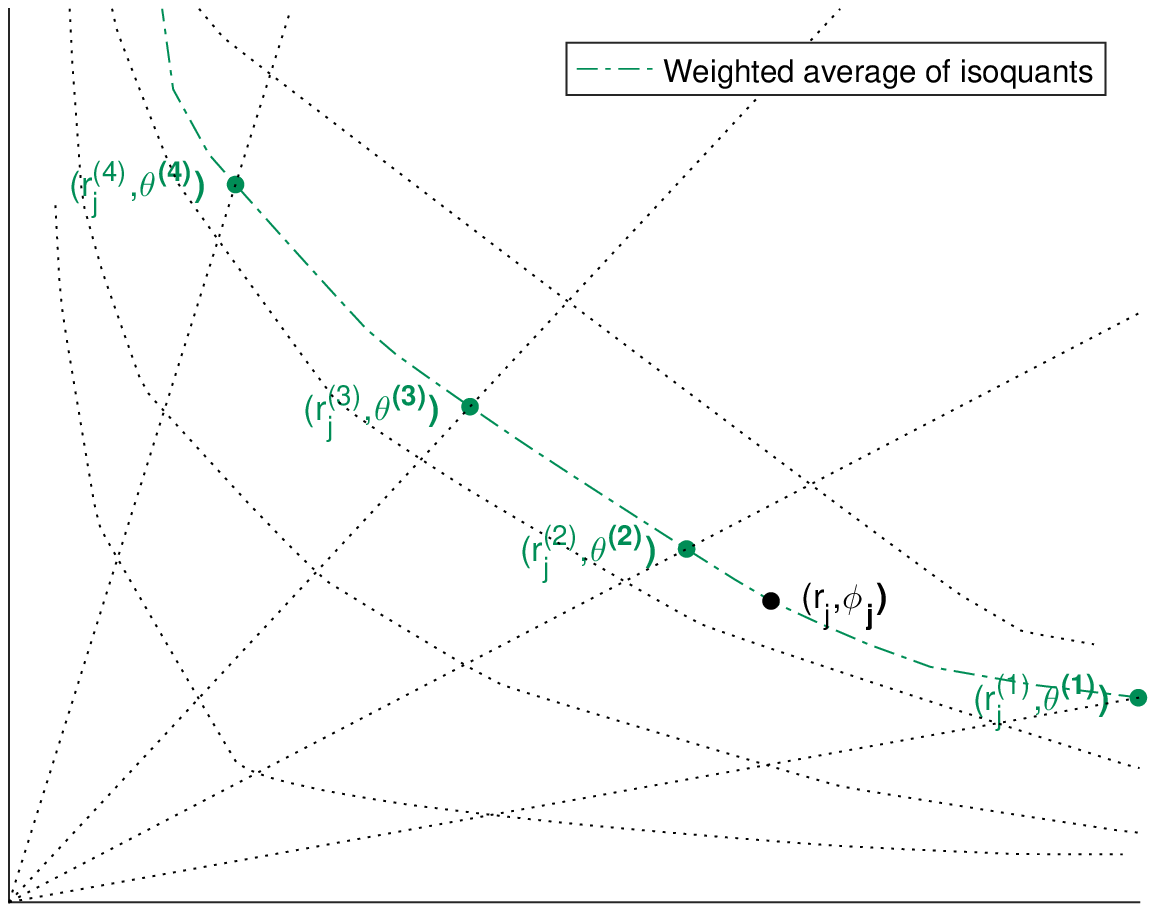}\label{fig:aveisoq}}
	\caption{Procedures of the projection of the observation in input space}
	\label{fig:projection}
\end{figure}

\paragraph{Using information from the estimated isoquants}
We can also use the estimated isoquants described in Section \ref{subsec:isoq} to project the observations. First, for each observation, we extract the two estimated isoquants which sandwich the observation in input space. Figure \ref{fig:projection}\subref{fig:extisoq} shows the example that two isoquants sandwiching the observation $\bm{X}_j$. 

Second, we compute the intersection of the extracted isoquants and the ray from the origin to the observation, and define distances to the isoquants below and above as $R_j^{(below)}$ and $R_j^{(above)}$ respectively. Then we can compute the weights $\rho_j$ which is defined as
\begin{equation}
    \rho_j=\frac{R_j-R_j^{(below)}}{R_j^{(above)}-R_j^{(below)}},~j=1,\ldots,n
\end{equation}
where $0\leq\rho_j\leq 1$, and $\rho_j$ approaches $1$ as $R_j$ is closer to $R_j^{(above)}$. Intuitively, we aim to use more information from the isoquant above when the observation is closer to the isoquant above. Figure \ref{fig:projection}\subref{fig:extisoq} also shows the definition of $R_j^{(below)}$ and $R_j^{(above)}$. In case that the observation is below or above the minimum or maximum isoquant, we define $R_j^{(below)}=0$ and $R_j^{(above)}=R_j$ respectively.

Finally, we compute the intersection of extracted isoquants and each ray from the origin, and define distances to the intersection with isoquants below and above as $R_j^{(below)(r)}$ and $R_j^{(above)(r)}$ respectively for $r=1,\ldots,R$. Then we obtain the projected observation $\{R_j^{(r)},\bm{\theta}^{(r)}\}_{r=1}^R$ as follows:
\begin{equation}
\label{eq:proj_point}
    R_j^{(r)}=\rho_j R_j^{(above)(r)} +\left(1-\rho_j\right) R_j^{(below)(r)} ~~~~~ \forall r=1,\ldots,R.
\end{equation}
Figure \ref{fig:projection}\subref{fig:aveisoq} shows the example of projection to each ray from the origin $\bm{\theta}^{(r)}$. Intuitively, we compute the inverse distance weighted average of two isoquants which sandwich the observation. 

\subsubsection{Shape Constrained Kernel Least Squares (SCKLS)}
\label{subsubsec:SCKLS}
\cite{yagi2018shape} proposed the Shape Constrained Kernel-weighted Least Squares (SCKLS) which is a kernel-based nonparametric shape constrained estimator. The SCKLS estimator is an extension of Local Polynomial estimator (\cite{stone1977consistent} and \cite{cleveland1979robust}) which imposes some constraints on parameters which characterize the estimated function such as intercept and slope. The SCKLS estimator introduces a set of $G$ evaluation points to impose shape constraints on each evaluation point. We are now interested in estimating S-shape function on each ray from the origin. Define the evaluation points on a ray from the origin $\bm{\theta}^{(r)}$ as follows
\begin{equation}
    r_g^{(r)}\in\{r_1^{(r)},\ldots,r_G^{(r)}\} ~~~~~ \forall r=1,\ldots,R.
\end{equation}
Note that evaluation points in input space on ray $r$ are defined by the scalar value $r_g^{(r)}$ which is a distance from the origin on the $r$-th ray.

\begin{figure}[ht]
    \includegraphics[width=5in]{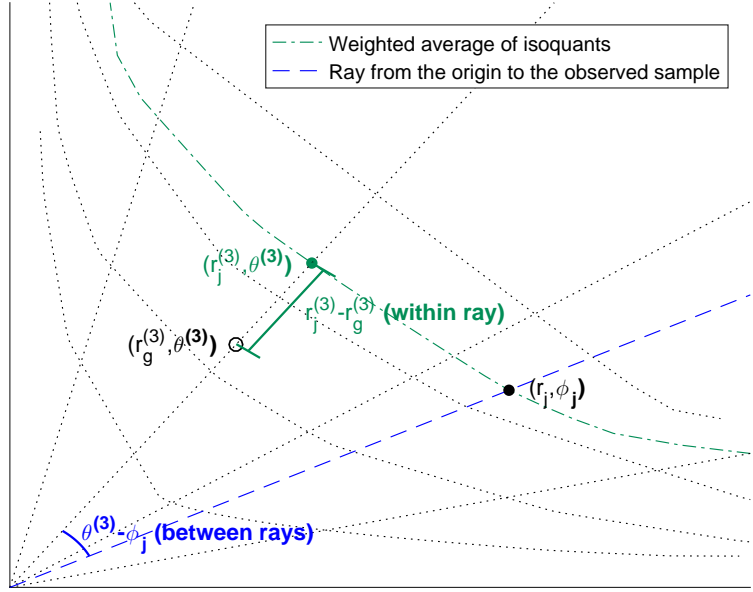}
    \centering
    \caption{Kernel weight in the S-shape estimation}
    \label{fig:weight}
\end{figure}

The objective function of the SCKLS estimator uses kernel weights, so more weight is given to the observations that are closer to the evaluation point. In our S-shape estimation, there exist two different weights to be considered: 1) the angle between the observation and the ray from the origin for which we are currently estimating, 2) the distance measured along the ray after the sample is projected using the estimated isoquant. Figure \ref{fig:weight} shows two different kernel weights imposed in our S-shape estimator.

Here  we define a distance measure in angles by their $L_2$ distance (in the $d-1$ Euclidean space), i.e.
$D(\bm{\phi}_1, \bm{\phi}_2) = \|\bm{\phi}_1-\bm{\phi}_2\|_2$.


For each ray from the origin $\bm{\theta}^{(r)}$, we solve the following 
quadratic programming problem:
\begin{equation}
\label{eq:SCKLS}
    \begin{aligned}
    	& \min_{\bm{a},\bm{b},g_{*}^{(r)}}
    	& & \sum_{g=1}^{G}\sum_{j=1}^{n}\left(\tilde{y}_j-\left(a_g^{(r)}+b_g^{(r)}\left(R_j^{(r)}-r_g^{(r)}\right)\right)\right)^2 
    	\scalebox{1}{$
    	K\left(\frac{D\left(\bm{\phi}_j,\bm{\theta}^{(r)}\right)}{\bm{\omega}}\right) k\left(\frac{R_j^{(r)}-r_g^{(r)}}{h^{(r)}}\right)
    	$}\\
    	& \mbox{subject to}
    	& &
    	a_g^{(r)}-a_l^{(r)}\leq b_g^{(r)}\left(r_g^{(r)}-r_l^{(r)}\right) 
    	~~~~~
    	\forall g,l=1,\ldots,g_*^{(r)}-1\\
    	&
    	& &
    	a_g^{(r)}-a_l^{(r)}\geq b_g^{(r)}\left(r_g^{(r)}-r_l^{(r)}\right) 
    	~~~~~
    	\forall g,l=g_*^{(r)},\ldots,G\\
    	&
    	& &
    	b_g^{(r)}\geq 0
    	~~~~~
    	\forall g,l=1,\ldots,G\\
    	&
    	& & g_*^{(r)} \in \{1,\ldots,G\}
    \end{aligned}
\end{equation}
where $a_g^{(r)}$ is a functional estimate, $b_g^{(r)}$ is an estimate of the slope of the function at $r_g^{(r)}$, the $g$-th evaluation point on the $r$-th ray. $k(\cdot)$ and $K(\cdot)$ denote the kernel and the product kernel function respectively. In fact, one could also replace $\tilde{y}_j$ (i.e. the pilot estimator)  by $y_j$ (observed response) in the above objective function of the minimization problem without affecting the correctness of our theory in consistency, and without having a noticeable difference in finite-sample performance. Here the observations which are closer to the evaluation points as measured by the angular deviation, and along the projected ray get more weights in the estimation procedure. $\bm{\omega}$ and $h^{(r)}$ are tuning parameters for the corresponding kernels which we will refer to as bandwidths. The first and second constraints in (\ref{eq:SCKLS}) are the convexity and concavity constraints respectively. We also need to estimate an index of an inflection point $g_*^{(r)}$ which is the point at which the S-shape function switches from convex to concave. We solve the quadratic programming problem $G$-times, once for each value $g_*^{(r)}\in\{1,\ldots,G\}$, and obtain a S-shape estimation by selecting the solution which has the minimum objective value among these $G$ solutions.

\subsubsection{Update isoquant estimates}
\label{subsubsec:input_info}

\begin{figure}[ht]
    \includegraphics[width=5in]{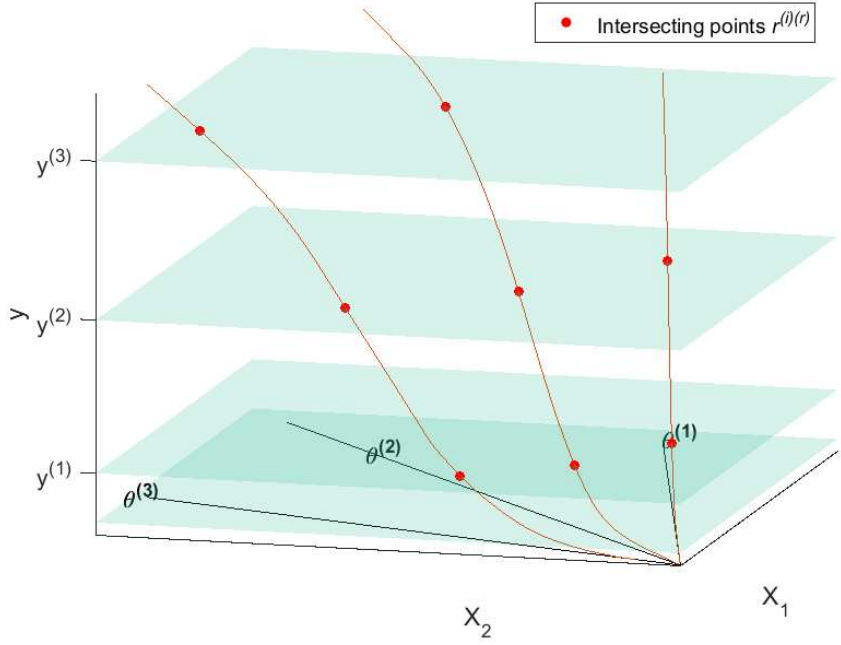}
    \centering
    \caption{How to obtain intersecting points $r^{(i)(r)}$}
    \label{fig:cut}
\end{figure}

After estimating the S-shape function along each ray, we need to verify whether the estimated S-shape functions satisfy the input convexity assumption. For this purpose, we cut the S-shape estimates at each isoquant level, $y^{(i)}$, and obtain intersecting points defined by radial coordinates as $\{r^{(i)(r)},\bm{\theta}^{(r)}\}$. Figure \ref{fig:cut} shows how we obtain the intersecting points $\{r^{(i)(r)},\bm{\theta}^{(r)}\}$ with 2-input example. We now re--estimate the isoquants by applying the CNLS-based method to the intersections for each isoquant $\{r^{(i)(r)},\bm{\theta}^{(r)}\}_{r=1}^{R}$. Note that we can convert this into Cartesian coordinate system through the inverse of the equations shown in (\ref{eq:polar}), and apply the CNLS-based method explained in Appendix \ref{subsec:isoq}.

\subsubsection{Minimizing the gap between estimates}

\begin{figure}[ht]
	\centering
	\subfloat[Before modification]{\includegraphics[width=0.5\textwidth]{modification_1.eps}\label{fig:modification_a}}
	\hfill
	\subfloat[After modification]{\includegraphics[width=0.5\textwidth]{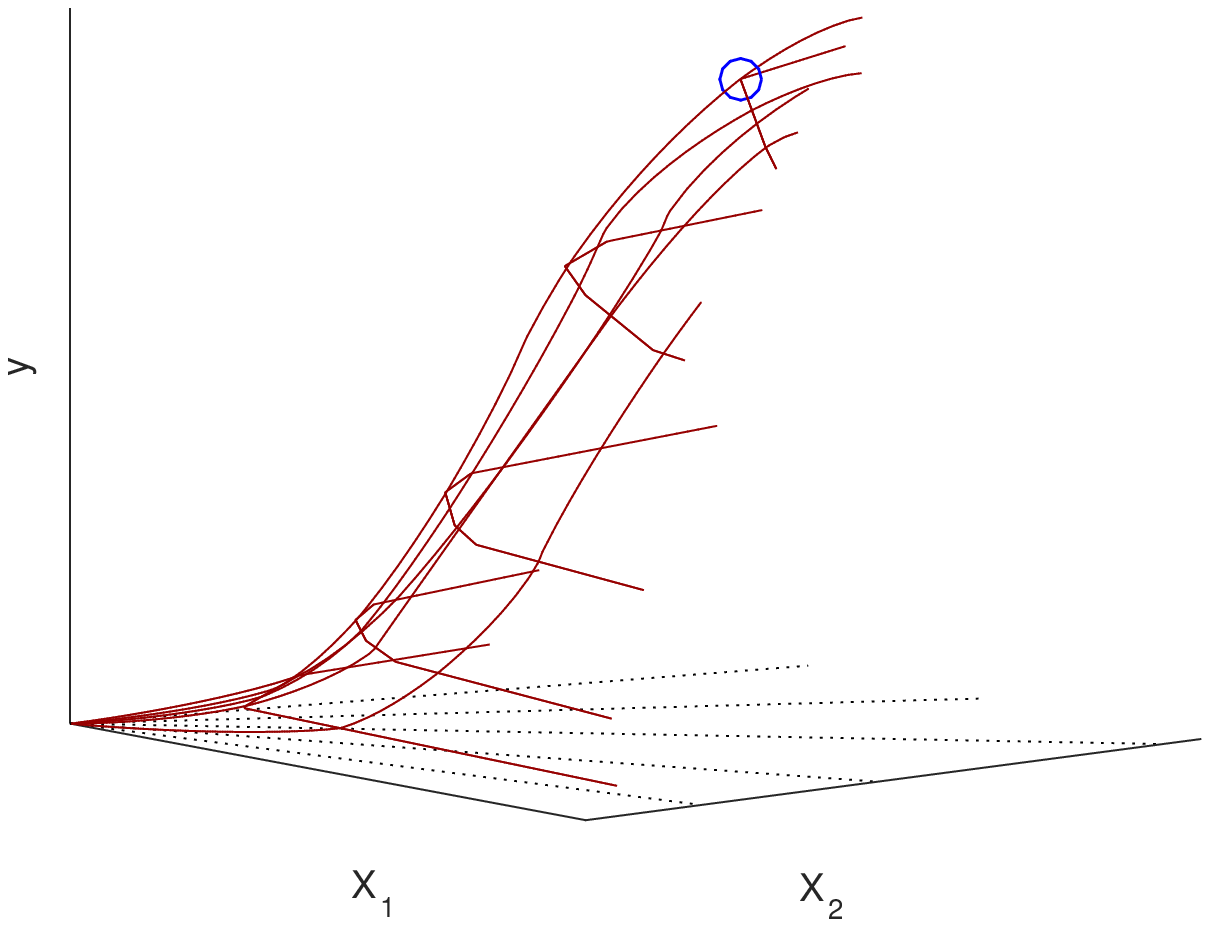}\label{fig:modification_b}}
	\caption{Modification of S-shape estimates}
	\label{fig:modification}
\end{figure}

We now have computed both S-shape and isoquant estimates. If the S-shape estimates do not violate the input convexity assumption, then the functional estimates of the S-shape functions and the input isoquants should match at each isoquant $y$-level. However if the S-shape estimates violate the input convexity assumption, then the S-shape estimates will not match the isoquant estimates at some isoquant $y$-level as shown in Figure \ref{fig:modification}\subref{fig:modification_a} with a blue circle. Here we propose to solve a quadratic programming problem which aims to minimize the gap between S-shape and isoquant estimates.

In this problem, we try to modify the S-shape estimates while fixing an inflection point at the same position as the original S-shape estimates. The objective function computes the weighted average of two deviations: 1) a gap between original S-shape estimates and revised S-shape estimates, and 2) a gap between revised S-shape estimates and the isoquant estimates at each isoquant $y$-level. Intuitively, we want to obtain the revised S-shape estimates which is close to the original S-shape estimates while satisfying input convexity. Figure \ref{fig:modification}\subref{fig:modification_b} shows the example that a violation is resolved through this step.

Here, we describe the mathematical formulation. We start from redefining the evaluation points on a ray, $\bm{\theta}^{(r)}$ as
\begin{equation}
    \begin{aligned}
        &   r_{g}^{(r)}\in\{r_1^{r},\ldots,r_G^{r}\}
        & & \forall g=1,\ldots,G\\
        &   r_{g^{(i)}}^{(r)}\in\{r^{(1)(r)},\ldots,r^{(I)(r)}\}
        & & \forall i=1,\ldots,I\\
        &   r_{g'}^{(r)}\in\{r_1^{r},\ldots,r_G^{r}\}\cup\{r^{(1)(r)},\ldots,r^{(I)(r)}\}
        & & \forall g'=1,\ldots,G'\\
    \end{aligned}
\end{equation}
where $G'=G+I.$ $r_{g^{(i)}}^{(r)}$ is the intersecting points obtained in section \ref{subsubsec:input_info} and they are added to the set of evaluation points, $r_{g'}^{(r)}$. We aim to minimize the gap between S-shape and isoquant estimates by solving the following quadratic programming problem:

\label{subsubsec:min_gap}
\begin{equation}
\label{eq:gap}
    \begin{aligned}
        & \min_{\tilde{a}_g^{(r)}}
        & & w^{S}\cdot\frac{1}{R\cdot G}\sum_{r=1}^{R}\sum_{g=1}^{G}\left(\tilde{a}_{g}^{(r)}-a_{g}^{(r)}\right)^2 + w^{I}\cdot\frac{1}{R\cdot I}\sum_{r=1}^{R}\sum_{i=1}^{I}\left(\tilde{a}_{g^{(i)}}^{(r)}-y^{(i)}\right)^2\\
        & \mbox{subject to}
        & & \frac{\tilde{a}_{g+2}^{(r)}-\tilde{a}_{g+1}^{(r)}}{r_{g+2}^{(r)}-r_{g+1}^{(r)}} \geq \frac{\tilde{a}_{g+1}^{(r)}-\tilde{a}_{g}^{(r)}}{r_{g+1}^{(r)}-r_{g}^{(r)}}
        ~~~~~ \forall r \mbox{ and } \forall g=1,\ldots,g_*^{(r)}-2\\
        &
        & & \frac{\tilde{a}_{g+2}^{(r)}-\tilde{a}_{g+1}^{(r)}}{r_{g+2}^{(r)}-r_{g+1}^{(r)}} \leq \frac{\tilde{a}_{g+1}^{(r)}-\tilde{a}_{g}^{(r)}}{r_{g+1}^{(r)}-r_{g}^{(r)}}
        ~~~~~ \forall r \mbox{ and } \forall g=g_*^{(r)}-2,\ldots,G\\
        &
        & & \tilde{a}_{g+1}^{(r)}\geq\tilde{a}_{g}^{(r)}
        ~~~~~~~~~~~~~~~~~~~~~~~~~ \forall r \mbox{ and } \forall g=1,\ldots,G\\
    \end{aligned} 
\end{equation}
where $\tilde{a}_g^{(r)}$ denotes a revised functional estimate at a grid point $g$ on a ray $r$. $w^{S}$ and $w^{I}$ are the weights for the S-shape estimator\footnote{We set $w^{S}=0.1$ and $w^{I}=0.9$ for our simulation and application to make sure the gap between isoquants and S-shape estimates become small for every single iteration.} and the isoquant estimator respectively satisfying $w^{S},w^{I}\in\lbrack0,1\rbrack$ and $w^{S}+w^{I}=1$. The objective function computes the weighted average of two deviations: 1) a gap between original S-shape estimates and revised S-shape estimates, 2) a gap between revised S-shape estimates evaluated at the input vectors located on the estimated isoquant and isoquant level $y^{(i)}$. Intuitively, when we put more weight on the original S-shape estimate, $w^{S}$ is large, the revised S-shape is close to the original S-shape, and input convexity may be violated. In contrast, when we put more weight on the isoquant estimates, $w^{I}$ is large, the revised S-shape can be far from the original S-shape, but the resulting estimate is more likely to satisfy input convexity without any violations. Based on our numerical experiments, we recommend to set a larger value of $w^{I}$ to avoid violations of the input convexity.

Constraints in (\ref{eq:gap}) correspond to constraints in (\ref{eq:SCKLS}). First two constraints impose the convexity and concavity for the RUP law, and the last constraint imposes the estimated function is monotonically increasing.

\subsubsection{Computing functional estimates on observations}
\label{subsubsec:estimates_on_obs}

The last step of an iteration is obtaining the functional estimates $\hat{g}(\bm{x})$ at any given value of input vector $\bm{x}$, and compute MSE against observations $\{\bm{X}_j,y_j\}_{j=1}^{n}$. Let $\left(r_{\bm{x}}, \bm{\phi_{x}}\right)$ denotes a given input vector $\bm{x}$ in spherical coordinates system. 

The simplest way is finding the closest ray to a given input vector $\bm{x}$, and use the S--shape estimates on this particular ray. The procedure requires: 1) Compute the weighted average of the two closest isoquants to $\bm{x}$, and 2) Compute the functional estimates of the closest ray to $\bm{x}$. The first step is explained in appendix \ref{subsubsec:project}. In this step, we obtain projected input data $\{r_{\bm{x}}^{(r)},\bm{\theta}^{(r)}\}_{r=1}^{R}$ which is defined in equation (\ref{eq:proj_point}). 

Then we can select the ray which is the closest to $\bm{\phi_{x}}$. Specifically,
\begin{equation}
    \label{eq:min_ray}
    r^* = \argmin_r \{ D(\bm{\phi_x},\bm{\theta}^{(r)}) \}_{r=1}^{R}
\end{equation}
where $D(\cdot)$ denotes a Euclidean distance function between two angles defined by
\begin{equation}
    \label{eq:dist_func}
    D(\bm{\phi_x},\bm{\theta}^{(r)})=\norm{\bm{\phi_x}-\bm{\theta}^{(r)}}.
\end{equation}
Then we can compute the functional estimates $\tilde{a}_{\bm{x}}^{(r^*)}$ at the closest ray $r^*$ by linear interpolating revised S-shape estimates $\tilde{a}_g^{(r)}$ obtained in (\ref{eq:gap}).

However, this simple solution will make the discontinuity in the functional estimates because it only uses the functional estimates from one particular ray. Here we propose the another way to compute the functional estimates by smoothing the functional estimates on the rays close to an input vector $\bm{x}$ by modifying the 2nd step of the procedure above. Instead of using the one particular ray, we compute the weighted average of S-shape estimates on rays close to a given input $\bm{x}$. 

We can compute the functional estimates $\tilde{a}_{\bm{x}}^{(r)}$ on each ray $r=1\ldots,R$. Subsequently, we can compute the inverse distance weighted average of functional estimates by
\begin{equation}
    \label{eq:inv_ave_est}
    \hat{g}_0(\bm{x})=
    \begin{cases}
        \tilde{a}_{\bm{x}}^{(r)}
        & \exists \text{ }r\mbox{ such that }D(\bm{\phi_x},\bm{\theta}^{(r)})=0\\
        \frac{\sum_{r=1}^{R}p^{(r)}\tilde{a}_{\bm{x}}^{(r)}}{\sum_{r=1}^{R}p^{(r)}} 
        & otherwise
    \end{cases}
\end{equation}
where $p^{(r)}$ is the inverse distance weight defined by
\begin{equation}
    \label{eq:inv_weight}
    p^{(r)}=
    \begin{cases}
        \frac{1}{D(\bm{\phi_x},\bm{\theta}^{(r)})}  & \mbox{if }D(\bm{\phi_x},\bm{\theta}^{(r)})\mbox{ is smaller than the $d$-th minimum of }\{D(\bm{\phi_x},\bm{\theta}^{(r)})\}_{r=1}^R\\
        0 & otherwise
    \end{cases}
\end{equation}
for $\forall r=1,\ldots,R$. Intuitively, we select the rays within the distance to the $d$-th closest ray, and compute the inverse distance weighted average of the S--shape estimates on these rays.

Finally, we can compute the MSE against observations $\{\bm{X}_j,y_j\}_{j=1}^{n}$ as
\begin{equation}
\label{eq:MSE}
    MSE = \frac{1}{n}\sum_{j=1}^{n}\left(y_j-\hat{g}_0(\bm{X}_j)\right)^2.
\end{equation}

\subsection{Updating parameters}
\label{subsec:update}
Finally, we update the parameters for the estimation before moving forward to the next iteration. We first update the parameters defining the number of both isoquants and rays to be estimated. When the gap between isoquants and S-shape estimates is large for a certain number of consecutive iterations, we delete the corresponding isoquant or ray. Specifically for any ray $r$, if $\left(\frac{y^{(i)}-\hat{g}(\bm{X}^{(i)(r)})}{y^{(i)}}\right) > \delta$ for some isoquant $i$ for $T$ consecutive iterations, then delete ray $r$ where $\delta$ is a tolerance value of percentage errors and $T$ is a number of consecutive iterations allowing errors over the tolerance.\footnote{We allow large errors for $T=10$ iterations for our simulation studies.},\footnote{We use $\delta=0.01$ or $\delta=0.05$ in our implementation depending on the noise size of data set.} And similarly defined for isoquant $i$.

We also update the bandwidth between rays, $\bm{\omega}$, used in the SCKLS-based S-shape estimation. We update the value of $\bm{\omega}$ increasing it by $\Delta\bm{\omega}$ in each iteration. As an iteration goes forward, the bandwidth $\bm{\omega}$ becomes larger. We continue iterations until $\bm{\omega}$ becomes large enough that the functional estimates are stable between iterations and then we select the results of the iteration with the lowest $MSE$ among the solutions with
\[
\left(\frac{y^{(i)}-\hat{g}(\bm{X}^{(i)(r)})}{y^{(i)}}\right) \leq \delta ~~~~~ \forall r=1,\ldots,R \mbox{ and } \forall i=1,\ldots,I\\. 
\]

Since the algorithm start from a small value of $\bm{\omega}$, the S-shape function only uses observations close to the ray for the estimation. As the iterative algorithm proceeds, the S-shape estimator includes observations which are more distant from the ray on which the evaluation point under consideration lies. Thus, the estimated functions on each ray becomes more similar as the bandwidth increases. If there still exists a gap between S-shape and input isoquant estimates even with large $\bm{\omega}$, we delete the corresponding isoquant or ray following the rule described above. Thus, the gap between the S-shape estimates and the isoquant estimates can be made arbitrarily small by deleting isoquants. This characteristic of the algorithm will be used to prove the convergence of our iterative algorithm because a production function estimate with only one isoquant estimate is a homothetic production function and our estimation procedure has no gap for estimating functions that satisfying the RUP law and are homothetic in inputs.

\newpage
\section{Comparison of different input isoquant estimation methods} \label{App:compCNLS}
In section \ref{subsec:isoq}, we introduce three different methods to estimate convex input isoquants: Convex Nonparametric Least Squares (CNLS), Directional Convex Nonparametric Least Squares (DCNLS) and Averaging Convex Nonparametric Least Squares (ADCNLS). In this section, we compare the performance of these estimators through Monte Carlo simulations.

We consider the following convex isoquant with 2-input. 
\begin{equation}
    \label{eq:isoq_DGP}
    X_{2} = \mathscr{H}(X_{1})=\frac{a}{X_{1}}
\end{equation}
where $a$ defines the shape of convex isoquant, and we use $a=10$ in this experiment. Two-input satisfying equation \ref{eq:isoq_DGP} is generated by
\begin{equation}
    \label{eq:isoq_true_val}
    \begin{aligned}
    & X_{1j}^* = \sqrt{\frac{a}{\tan{(\eta_j)}}} & \forall j = 1,\ldots,n\\
    & X_{2j}^* = \sqrt{a\cdot\tan{(\eta_j)}} & \forall j = 1,\ldots,n
    \end{aligned}
\end{equation}
where angles $\eta_j$ are randomly generated by $\eta_j\sim unif(0.05,\frac{\pi}{2}-0.05)$. Then we generate samples by adding noise in the direction orthogonal to the true function.
\begin{equation}
    \label{eq:isoq_obs}
    \begin{aligned}
    & X_{1j} = X_{1j}^* + \epsilon_j\cdot\cos\left(\arctan\left({{X_{1j}^*}^2}/{a}\right)\right) & \forall j = 1,\ldots,n \\
    & X_{2j} = X_{2j}^* + \epsilon_j\cdot\sin\left(\arctan\left({{X_{1j}^*}^2}/{a}\right)\right) & \forall j = 1,\ldots,n
    \end{aligned}
\end{equation}
where additive noise $\epsilon_j$ is generated by $\epsilon_j\sim N(0,\sigma_v)$.

\begin{figure}[p]
    \includegraphics[width=\textwidth]{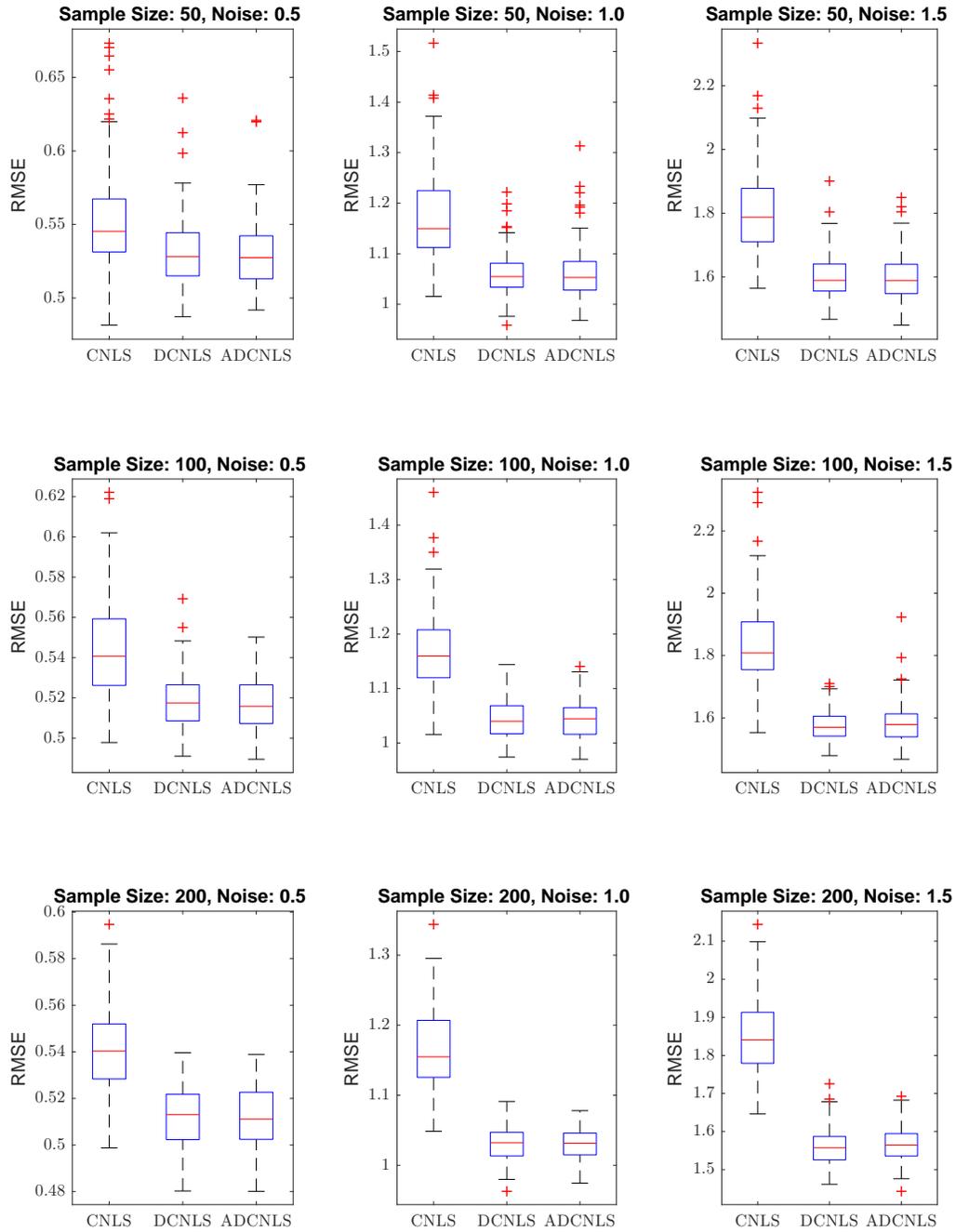}
    \centering
    \caption{Estimation results on the testing set for the isoquant estimation}
    \label{fig:expA1test}
\end{figure}

We consider 9 different scenarios with the different training sample size $n\in(50,100,200)$ and the standard deviation of the noise $\sigma_v\in(0.5,1.0,1.5)$. We use $M=10$ different error directions for estimating ADCNLS where error directions are chosen by equally spaced percentiles of the input ratio $\{X_{2j}/X_{1j}\}_{j=1}^{n}$. We generate 100 training-testing set pairs for each scenario, and draw box plots of RMSE against the true function for each estimator on testing set in Figure \ref{fig:expA1test}. Note that RMSE is computed in the direction orthogonal to the true function. The size of the testing set is 1000, and it is randomly drawn from the same distribution as the training set.

The DCNLS and ADCNLS estimators perform better than the CNLS estimator because these estimation  methods assume errors are contained in both input dimensions. Although these two estimators still have misspecification of error directions, it helps to reduce the bias caused by the misspecification of error directions in the CNLS estimator.

\newpage

\section{Technical proofs}
\label{app:proof}

\subsection{Proof of Theorems in Section \ref{sec:model}}
\subsubsection{Proof of Lemma \ref{lmm:RUP}}
\begin{proof}
    \label{lmm:RUP.proof}
    For simplicity, we focus on the case of  $d=1$. Note that following arguments can be extended for the multiple input case with $d>1$ by studying the function $g_0$ along any direction.
    
    Now, the elasticity of scale is defined as
    \[
    \epsilon(x)=g_0'(x)\frac{x}{g_0(x)}.
    \]
    Next we compute the derivative of the elasticity of scale,
    \begin{equation}
    \label{eq:deriv_elas}
    \epsilon'(x)=\frac{1}{g_0(x)}\left(xg_0''(x)+g_0'(x)\left(1-\epsilon(x)\right)\right).
    \end{equation}
    
    By  Definition \ref{def:rup}, we have following conditions on the elasticity of scale:
    \[
        \epsilon'(x)<0 \text{ for } \forall x
    \]
    \[
        \epsilon(x_A)>1 \text{ and } \epsilon(x_B)<1 \text{ for some } x_A<x_B.
    \]
    
    By using these conditions on Equation (\ref{eq:deriv_elas}) and assumption that $g_0$ is monotonically increasing, we have,
    \[
        g_0''(x)<0 \text{ for } \forall x>x_B.
    \]
    
    Here, by the assumption that there exists a single point of inflection point $x^*$ such that $g_0''(x^*)=0$, we have
    \[
    g_0''(x)>0 \text{ for }\forall x<x^*
    \]
    \[
    g_0''(x)=0 \text{ for } x=x^*
    \]
    \[
    g_0''(x)<0 \text{ for }\forall x>x^*
    \]
    which implies the function $g_0(\cdot)$ is a S-shaped function define in Definition \ref{def:s-shape}.
\end{proof}



\subsubsection{Proof of Theorem \ref{thm:1}}
\begin{proof}
    
    First, note that the S-shape function is defined for any expansion path and a ray from the origin is a subset of the set of expansion paths. So a single inflection point exist on each 2-D sectional of the production function by definition of an S-shape function. The result to be shown, the set of inflection points lie on the same input isoquant with aggregate input level $x_A^*$, can be stated mathematically as
    \[
	 x_{A}^*=\argmaxE_{x_A \in \alpha \bm{x}} \left( \dfrac{dF_0(x_A)}{dx_A}\,\middle|\,x_A=\mathscr{H}_{0}(\bm{x}) \right)
	\]
	 for a ray vector $\alpha \bm{x}$, where $\bm{x}=(x_1,\ldots,x_d)$ and the origin define a ray in input space and $x_{A}^*$ is the inflection point on that ray. 
	By the definition of homothetic we have $g_0(\bm{x}) = F_0(\mathscr{H}_{0,k}(\bm{x}))$. We substitute $x_A=\mathscr{H}_{0}(\bm{x})$ and take the derivative of $g_0$ with respect to $x_A$, which is just $\dfrac{dF_0(x_A)}{dx_A}$. Notice this is independent of the ray from the origin selected. Thus, we have the first part of the claim.

    For the second part, we know that if the S-shape function definition holds for a ray from the origin then it holds for any ray from the origin and the inflection point will be located on the same isoquant. Now we just need to show for an arbitrary (non-radial) expansion path the that RUP law holds.
    
    By the definition of an expansion path, we see that as we move from input vector $\bm{X}_{m-1}$ to $\bm{X}_{m}$ we move between two input isoquants which are in the same sequential order as they would be for an expansion path along a ray from the origin, thus the passum coefficient is decreasing, given us the desired result.
\end{proof}

\subsection{Proof of Theorems in Section \ref{sec:property}}
\subsubsection{Proof of Theorem \ref{thm:nonhomo_consistency1}}
\begin{proof}
	First, it follows from \citet{fan2016} that the pilot estimator satisfies
	\[
	\sup_{\bm{x} \in \bm{S}} |\tilde{g}_0(\bm{x})- g_0(\bm{x})| = O_p(n^{-2/(d+4)}\log n).
	\]
	Without loss of generality, in the following, we could assume that the event 
	\begin{align}
	\label{eq:proof_pilot}
	\sup_{\bm{x} \in \bm{S}} |\tilde{g}_0(\bm{x})- g_0(\bm{x})| \le {C}_1 n^{-2/(d+4)}\log n
	\end{align} 
	holds for some large enough positive $C_1$.
	
	For any given $y \in (y_\circ, y^\circ)$, we could always find an $i \in \{1,\ldots,I\}$ (that depends on $n$) such that $y^{(i)} \le y < y^{(i+1)}$. Moreover, write $\Delta_n \equiv y^{(i+1)}-y^{(i)} = O(I^{-1})$, which is of order greater than $O(n^{-2/(d+4)}\log n)$.
	
	Let $\mathcal{I}_i \subset \{1,\ldots,n\}$ be the index set with $\{\bm{X}_j,\tilde{y}_j\}_{j \in \mathcal{I}_i}$ projected to the isoquant level $y^{(i)}$. Then $\mathcal{I}_i$ contains all the indices $j$ such that $\tilde{g}_0(\bm{X}_j) \in \Big[y^{(i)}- \Delta_n/2, y^{(i)} + \Delta_n/2\Big)$. In view of (\ref{eq:proof_pilot}), we could conclude that for sufficiently large $n$,
	\begin{enumerate}
		\item $\mathcal{I}_i$ contains all the indices $j$ such that ${g}_0(\bm{X}_j) \in \Big[y^{(i)}- \Delta_n/4, y^{(i)} + \Delta_n/4\Big]$.
		\item  All indices $j$ contained in $\mathcal{I}_i$ satisfy ${g}_0(\bm{X}_j) \in \Big[y^{(i)}- \Delta_n, y^{(i)} + \Delta_n\Big]$.
	\end{enumerate}
	Furthermore, recall that $\bm{C}'$ is a compact set that belongs to the interior of $\bm{C}_y$. For every $j \in \mathcal{I}_i$, and every $\bm{X}_{j,-d} \in \bm{C}'$, it follows from Assumption~\ref{ass:1} and Definition~\ref{def:inputconvex} that
	\begin{align*}
	\Big|\bm{X}_{j,d} - \mathscr{H}_{0,d}(\bm{X}_{j,-d}; y)\Big|  &= \Big|\mathscr{H}_{0,d}(\bm{X}_{j,-d}; g_0(\bm{X}_{j})) - \mathscr{H}_{0,d}(\bm{X}_{j,-d}; y)\Big| \\
	& \le C_2 \Big|g_0(\bm{X}_{j}) - y^{(i)}\Big| = C_2 \Big(\Big|g_0(\bm{X}_{j}) -\tilde{g}_0(\bm{X}_{j}) \Big| + \Big|\tilde{g}_0(\bm{X}_{j}) - y\Big|\Big) \\
	&\le C_2 \Big(\Big|g_0(\bm{X}_{j}) -\tilde{g}_0(\bm{X}_{j}) \Big| + \Big|y^{(i+1)} - y^{(i)}\Big|\Big) \le C_3 \Delta_n.
	\end{align*}
	for some $C_2$ and $C_3$ (that only depend on $g_0$).
	
	Let $\mathcal{G}_{d-1}$ be the class of functions $h: \mathbb{R}^{d-1} \rightarrow \mathbb{R}$ that are convex and decreasing.  When applying CNLS on $\{\bm{X}_{j,-d},\bm{X}_{j,d}\}_{j \in \mathcal{I}_i}$, we have that
	\begin{align*}
	\inf_{h \in \mathcal{G}_{d-1}} \sum_{j \in  \mathcal{I}_i} \Big(\bm{X}_{j,d}- h(\bm{X}_{j,-d})\Big)^2 \le  \sum_{j \in  \mathcal{I}_i} \Big(\bm{X}_{j,d}- \mathscr{H}_{0,d}(\bm{X}_{j,-d}; y)\Big)^2 \le C_3^2 \Delta_n^2 |\mathcal{I}_i|.
	\end{align*}
	Note that $|\mathcal{I}_i|$ is bounded above by the number of observations satisfy ${g}_0(\bm{X}_j) \in \big[y^{(i)}- \Delta_n, y^{(i)} + \Delta_n\big]$, which we denote by $|\tilde{\mathcal{I}}_i|$. Let $m$ be the marginal density of ${g}_0(\bm{X}_1)$. Then, as $n^{-1/2} = o(\Delta_n)$ it follows from Donsker's theorem (see, for example, \cite{van1996weak}) that 
	$|\tilde{\mathcal{I}}_i|/(2\Delta_n) \rightarrow m(y)$ almost surely. Here we also used the fact that $m$ is continuous, so $m(y^{(i)})\rightarrow m(y)$ as $n\rightarrow\infty$. Note that the above result actually holds simultaneous for any $y \in [y_\circ+\eta, y^\circ-\eta]$ for any pre-specified small $\eta >0$. 
	This implies that $|\mathcal{I}_i|$ is at most $O(\Delta_n)$, so as $n \rightarrow \infty$,
	\begin{align}
		\label{eq:proof_1}
	\inf_{h \in \mathcal{G}_{d-1}} \sum_{j \in  \mathcal{I}_i} \Big(\bm{X}_{j,d}- h(\bm{X}_{j,-d})\Big)^2 \le O(\Delta_n^3).
	\end{align}
	
	Now suppose that $\sup_{\bm{x}_{-d} \in \bm{C}'} |\hat{\mathscr{H}}_{0,d}(\bm{x}_{-d};y)-{\mathscr{H}}_{0,d}(\bm{x}_{-d};y)| > \epsilon$ for some  $\epsilon$ that is smaller than the Hausdorff distance between $\bm{C}'$ and $\bm{C_y}$, and that the supremum occurs at $x^*_{-d}$. Then, by the monotonicity constraint and the fact that ${\mathscr{H}}_{0,d}$ is Lipschitz continuous, we could find some $C_4$ (that only depends on $g_0$ but not $x^*_{-d}$), such that
	\[
	\sup_{\bm{x}_{-d} \in B(x^*_{-d}, C_4 \epsilon)} |\hat{\mathscr{H}}_{0,d}(\bm{x}_{-d};y^{(i)})-{\mathscr{H}}_{0,d}(\bm{x}_{-d};y)| > \epsilon/2,
	\]
	where $B(x,r)$ is the closed ball centered at $x$ of radius $r$. For a detailed construction of this fact, see  \cite{chensamworth2016} or \cite{yagi2018shape}. This means that for sufficiently large $n$,
	\begin{align*}
	 &\sum_{j \in  \mathcal{I}_i} \Big(\bm{X}_{j,d}- \hat{\mathscr{H}}_{0,d}(\bm{X}_{j,-d};y^{(i)})\Big)^2 \\
	 &\ge	 \sum_{\big\{j \in \mathcal{I}_i \;|\;  \bm{X}_{j,-d} \in B(x^*, C_4\epsilon)\big\}} \Big(\bm{X}_{j,d}-\mathscr{H}_{0,d}(\bm{X}_{j,-d};y)+\mathscr{H}_{0,d}(\bm{X}_{j,-d};y)- \hat{\mathscr{H}}_{0,d}(\bm{X}_{j,-d};y^{(i)})\Big)^2\\
	 &\ge	 \sum_{\big\{j \in \mathcal{I}_i \;|\;  \bm{X}_{j,-d} \in B(x^*, C_4\epsilon)\big\}} \big(C_3\Delta_n+\epsilon/2\big)^2\\
	 &\ge  (\epsilon/4)^2 \Big|\Big\{j: {g}_0(\bm{X}_j) \in \big[y^{(i)}- \Delta_n/4, y^{(i)} + \Delta_n/4\big] \mbox{ and }  \bm{X}_{j,-d} \in B(x^*, C_4\epsilon)\Big\}\Big|.
	\end{align*}
	Now note that 
 	\begin{align*}
	 &\Big\{\bm{x}: {g}_0(\bm{x}) \in \big[y^{(i)}- \Delta_n/4, y^{(i)} + \Delta_n/4\big],  \bm{x}_{-d} \in B(x^*, C_4\epsilon)\Big\}\\
	 & =  \Big\{\bm{x}: {g}_0(\bm{x}) \ge y^{(i)}- \Delta_n/4 \Big\} \cap \Big\{\bm{x}: {g}_0(\bm{x}) \le y^{(i)}+ \Delta_n/4 \Big\}  \cap  \Big\{\bm{x}:  \bm{x}_{-d} \in B(x^*, C_4\epsilon)\Big\},
	\end{align*}
	where all of these three individual sets are Vapnik--Chervonenkis, regardless of the value of $x^*$, $\epsilon$, $y^{(i)}$ and $\Delta_n$ (details could be found in, for instance, Chapter 2.6 of \cite{van1996weak}). Therefore, the indicator function 
	$
	\mathbf{1}_{\Big\{\bm{x}\Big|{g}_0(\bm{x}) \in \big[y^{(i)}- \Delta_n/4, y^{(i)} + \Delta_n/4\big],  \bm{x}_{-d} \in B(x^*, C_4\epsilon)\Big\}}\
	$
	is VC as well. It then follows from Donsker's theorem that 
	\[
	\Big|\Big\{j: {g}_0(\bm{X}_j) \in \big[y^{(i)}- \Delta_n/4, y^{(i)} + \Delta_n/4\big] \mbox{ and }  \bm{X}_{j,-d} \in B(x^*, C_4\epsilon)\Big\}\Big| = O(\Delta_n).
	\]
	Thus, as $n \rightarrow \infty$, 
	\begin{align}
	\label{eq:proof_2}
	\sum_{j \in  \mathcal{I}_i} \Big(\bm{X}_{j,d}- \hat{\mathscr{H}}_{0,d}(\bm{X}_{j,-d};y)\Big)^2 \ge O(\Delta_n).
	\end{align}
	
	Consequently, comparing \eqref{eq:proof_1} with  \eqref{eq:proof_2} leads a contradiction. Since $\epsilon$ is arbitrary, we conclude that 
	\[
		\sup_{\bm{x}_{-d} \in \bm{C}'} |\hat{\mathscr{H}}_{0,d}(\bm{x}_{-d};y^{(i)})-{\mathscr{H}}_{0,d}(\bm{x}_{-d};y)| \stackrel{p}{\rightarrow} 0.
	\]
	
	Using  the same argument on $\{\bm{X}_{j,-d},\bm{X}_{j,d}\}_{j \in \mathcal{I}_{i+1)}}$, we could obtain that
	\[
	\sup_{\bm{x}_{-d} \in \bm{C}'} |\hat{\mathscr{H}}_{0,d}(\bm{x}_{-d};y^{(i+1)})-{\mathscr{H}}_{0,d}(\bm{x}_{-d};y)| \stackrel{p}{\rightarrow} 0.
	\] 
	Finally, since $\hat{\mathscr{H}}_{0,d}(\bm{x}_{-d};y)$ is a weighted average of $\hat{\mathscr{H}}_{0,d}(\bm{x}_{-d};y^{(i)})$ and $\hat{\mathscr{H}}_{0,d}(\bm{x}_{-d};y^{(i+1)})$ (with corresponding weights $w$ and $1-w$, for some $w \in [0,1]$), we have that
    \[
    \sup_{\bm{x}_{-d} \in \bm{C}'} |\hat{\mathscr{H}}_{0,d}(\bm{x}_{-d};y)-{\mathscr{H}}_{0,d}(\bm{x}_{-d};y)| \stackrel{p}{\rightarrow} 0.
    \]
\end{proof}

\subsubsection{Proof of Theorem \ref{thm:nonhomo_consistency2}}
\begin{proof}

	First, we investigate the behavior of $\hat{g}_0(\alpha\bm{v})$ for $\alpha \in [\delta c_{\bm{v}} ,c_{\bm{v}}-\delta]$ with $\bm{\phi}_{\bm{v}} = \bm{\theta}^{(r)}$ for any particular $r \in \{1,\ldots, R\}$. 
	
	Without loss of generality, we assume that  $\|\bm{v}\|_2=1$. Since $\bm{v}=(v_1,\ldots,v_d)^T$ and $\min_i v_i > 0$, in our asymptotic regime, it suffices for us to consider $\bm{\theta}^{(r)} \in [\eta,\pi/2-\eta]^{d-1}$ with some pre-specified $\eta > 0$. Here to faciliate our theoretical analysis, we focus on the distance-based projection method.
	Recall that SCKLS solves the following optimization problem.
	\begin{equation}
	\begin{aligned}
	& \min_{\bm{a}^{(r)},\bm{b}^{(r)},g_{*}^{(r)}}
	& & \sum_{g=1}^{G}\sum_{j=1}^{n}\left(\tilde{y}_j-\left(a_g^{(r)}+b_g^{(r)}\left(r_j^{(r)}-r_g^{(r)}\right)\right)\right)^2 
	\scalebox{1}{$
		K\left(\frac{D\left(\bm{\phi}_{\bm{X}_j},\bm{\theta}^{(r)}\right)}{\bm{\omega}}\right) k\left(\frac{r_j^{(r)}-r_g^{(r)}}{h}\right)
		$}\\
	& \mbox{subject to}
	& &
	a_g^{(r)}-a_l^{(r)}\leq b_g^{(r)}\left(r_g^{(r)}-r_l^{(r)}\right) 
	~~~~~
	\forall g,l=1,\ldots,g_*^{(r)}-1\\
	&
	& &
	a_g^{(r)}-a_l^{(r)}\geq b_g^{(r)}\left(r_g^{(r)}-r_l^{(r)}\right) 
	~~~~~
	\forall g,l=g_*^{(r)},\ldots,G\\
	&
	& &
	b_g^{(r)}\geq 0
	~~~~~
	\forall g,l=1,\ldots,G\\
	\end{aligned}
	\end{equation}
	where the angle of $\bm{v}$ for any $\bm{v} \in \mathbb{R}^d$ with $\|\bm{v}\|_2=1$ is  $\bm{\phi}_{\bm{v}}$ (and its inverse function as $\bm{\phi}^{-1}_{\cdot}$). Note that $K(\cdot)$ is a bounded kernel and $\bm{\omega}\rightarrow 0$, we only need to consider the pairs of observations $(\bm{X}_j,y_j)$ with $D(\bm{\phi}_{\bm{X}_j}, \bm{\theta}^{(r)}) \le C h^{d-1}$ for some $C > 0$. This means that $\sup_{j \in \{1,\ldots,n\}}\Big|r_j-\|\bm{X}_j\|\Big| \stackrel{p}{\rightarrow} 0$ and  $\sup_{j \in \{1,\ldots,n\}} \|r_j \bm{v} - \bm{X}_j\|  \stackrel{p}{\rightarrow} 0$. This together with the facts that  $\sup_{\bm{x} \in \bm{S}} |\tilde{g}_0(\bm{x}) - g_0(\bm{x})| \stackrel{p}{\rightarrow} 0$ and $g_0$ is a Lipschitz continuous function yield
	\[
	\sup_{j \in \{1,\ldots,n\}} |\tilde{y}_j - g_0(r_j \bm{v})| \stackrel{p}{\rightarrow} 0. 
	\]
	
	Write
	\[ S(\bm{a},\bm{b}) =\frac{1}{G}\sum_{g=1}^{G}\frac{1}{nh^d}\sum_{j=1}^{n}\left(\tilde{y}_j-\left(a_g+b_g\left(r_j^{(r)}-r_g^{(r)}\right)\right)\right)^2 
	\scalebox{1}{$
		K\left(\frac{D\left(\bm{\phi}_{\bm{X}_j},\bm{\theta}^{(r)}\right)}{h}\right) k\left(\frac{r_j^{(r)}-r_g^{(r)}}{h}\right).
		$}
	\]
	Set $\bm{a}_0=\Big(g_0(r_1\bm{v}),\ldots, g_0(r_g\bm{v})\Big)'$ and $\bm{b}_0=\Big(\frac{dg_0(\bm{v}x)}{dx}(r_1),\ldots, \frac{dg_0(\bm{v}x)}{dx}(r_1) \Big)'$. Note that for any given $g$, there are at most $O(nh^d)$ observations with (i) non-zero (i.e. positive) and bounded value of $K\left(\frac{D\left(\bm{\phi}_{\bm{X}_j},\bm{\theta}^{(r)}\right)}{h}\right) k\left(\frac{r_j^{(r)}-r_g^{(r)}}{h}\right)$, and (ii) $\Big|\tilde{y}_j-a_{0,g}-b_{0,g}(r_j^{(r)}-r_g^{(r)})\Big|\stackrel{p}{\rightarrow} 0$ uniformly. The last part follows  from $	\sup_{j \in \{1,\ldots,n\}} |\tilde{y}_j - g_0(r_j\bm{v})| \stackrel{p}{\rightarrow} 0$ and $|r_j^{(r)}-r_g^{(r)}| \rightarrow 0$ for those with positive value of  $k\left(\frac{r_j^{(r)}-r_g^{(r)}}{h}\right)$. This means that $S(\bm{a}_0,\bm{b}_0)  \stackrel{p}{\rightarrow} 0$. 
	
	Let $(\tilde{\bm{a}}, \tilde{\bm{b}})$ be the minimizer of $S(\cdot,\cdot)$ without any constraints, and $(\hat{\bm{a}}, \hat{\bm{b}})$ be an minimizer of $S(\cdot,\cdot)$ with the constraints. Since
	\[
	0 \le S(\tilde{\bm{a}}, \tilde{\bm{b}}) \le S(\hat{\bm{a}}, \hat{\bm{b}}) \le S(\bm{a}_0,\bm{b}_0),
	\]
	we have that $|S(\hat{\bm{a}}, \hat{\bm{b}}) - S(\tilde{\bm{a}}, \tilde{\bm{b}})| \stackrel{p}{\rightarrow} 0$. Now notice that $S(\cdot,\cdot)$ is a quadratic function with respect to its arguments that minimizes at $(\tilde{\bm{a}}, \tilde{\bm{b}})$, therefore,
	\[
	S(\hat{\bm{a}}, \hat{\bm{b}}) - S(\tilde{\bm{a}}, \tilde{\bm{b}}) =\frac{1}{G}\sum_{g=1}^G (\hat{a}_g-\tilde{a}_g, h(\hat{b}_g-\tilde{b}_g)) \boldsymbol{\Sigma}_g (\hat{a}_g-\tilde{a}_g, h(\hat{b}_g-\tilde{b}_g))',
	\]
	where 
	\[
	\boldsymbol{\Sigma}_g = \frac{1}{n h^d}\sum_{j=1}^n \Big(1, \frac{r_j^{(r)}-r_g^{(r)})}{h}\Big)\Big(1, \frac{r_j^{(r)}-r_g^{(r)}}{h}\Big)'  	\scalebox{1}{$
		K\left(\frac{D\left(\bm{\phi}_{\bm{X}_j},\bm{\theta}^{(r)}\right)}{h}\right) k\left(\frac{r_j^{(r)}-r_g^{(r)}}{h}\right).
		$}
	\]
	It can be shown following the argument of Lemma~5 of \citet{fan2016} that
	\[
	\min_{g \in \big\{\lceil(\delta c_{\bm{v}}^{-1}/2) G \rceil,\ldots, \lfloor(1-\delta c_{\bm{v}}^{-1}/2) G \rfloor \big\}} \lambda_2(\boldsymbol{\Sigma}_g) > C'
	\]
	in probablity for some $C' > 0$, where $\lambda_2(\cdot)$ returns the smallest eigenvalue of an $2 \times 2$ matrix. As such,
	\[
 \frac{1}{G}\sum_{g=1}^G (\hat{a}_g-\tilde{a}_g, h(\hat{b}_g-\tilde{b}_g)) \boldsymbol{\Sigma}_g (\hat{a}_g-\tilde{a}_g, h(\hat{b}_g-\tilde{b}_g))' \ge \frac{1}{GC'}\sum_{g=\lceil(\delta c_{\bm{v}}^{-1} /2) G \rceil}^{\lfloor(1-\delta c_{\bm{v}}^{-1}  /2) G \rfloor} (\hat{a}_g-\tilde{a}_g)^2.
	\]
	Consequently, $\frac{1}{G}\sum_{g=\lceil(\delta c_{\bm{v}}^{-1} /2) G \rceil}^{\lfloor(1-\delta c_{\bm{v}}^{-1}  /2) G \rfloor} (\hat{a}_g-\tilde{a}_g)^2 \stackrel{p}{\rightarrow} 0$.
	
	Now applying the same argument to $S({\bm{a}}_0, \bm{b}_0) - S(\tilde{\bm{a}}, \tilde{\bm{b}})$, we have that $\frac{1}{G}\sum_{g=\lceil(\delta c_{\bm{v}}^{-1} /2) G \rceil}^{\lfloor(1-\delta c_{\bm{v}}^{-1} /2) G \rfloor} (\tilde{a}_g-{a}_{0,g})^2 \stackrel{p}{\rightarrow} 0$. It follows from the triangular inequality (NB. since the above quantities can be viewed as the squares of the differences in a $L_2$ norm) that
	\begin{align}
	\label{eq:proof_ray_eq2}
	\frac{1}{G}\sum_{g=\lceil(\delta c_{\bm{v}}^{-1} /2) G \rceil}^{\lfloor(1-\delta c_{\bm{v}}^{-1} /2) G \rfloor} (\hat{a}_g-{a}_{0,g})^2 \stackrel{p}{\rightarrow} 0.
	\end{align}
	
	For any $\epsilon > 0$, if $\sup_{\alpha \in [\delta,c_{\bm{v}}-\delta]}|\hat{g}_0(\alpha \bm{v}) - g_0(\alpha \bm{v})| > \epsilon$, then since both $\hat{g}_0(\bm{v}\cdot)$ and $g_0(\bm{v}\cdot)$ are increasing, with $g_0(\bm{v}\cdot)$ also being Lipschitz (denoting its constant by $M$), we are always able to find an interval $\mathcal{I}$ over  $[\delta/2,c_{\bm{v}}-\delta/2]$ with length of at least $\frac{\epsilon}{2M}$ such that  $\inf_{\alpha \in \mathcal{I}}|\hat{g}_0(\alpha \bm{v}) - g_0(\alpha \bm{v})| > \epsilon/2$. For a detailed construction, see also \citet{yagi2018shape}. Since we take equal-spacing evaluation points with $G\rightarrow \infty$, we have that as $n \rightarrow \infty$,
	\[
	\frac{1}{G}\sum_{g=\lceil(\delta c_{\bm{v}}^{-1} /2) G \rceil}^{\lfloor(1-\delta c_{\bm{v}}^{-1} /2) G \rfloor} (\hat{a}_g-{a}_{0,g})^2 \ge \frac{\sum_{g=1}^G \mathbf{1}_{\{r_g^{(r)} \in \mathcal{I}\}}}{G} \Big(\frac{\epsilon}{2}\Big)^2  \rightarrow \frac{\epsilon}{2M c_{\bm{\theta}^{(r)}}}\Big(\frac{\epsilon}{2}\Big)^2  > 0,
	\]
	contradicting the fact of \eqref{eq:proof_ray_eq2}. As here $\epsilon$ is arbitrary, we can conclude that 
	\begin{align}
	\label{eq:proof_ray_eq3}
	\sup_{\alpha \in [\delta,c_{\bm{v}}-\delta]}|\hat{g}_0(\alpha \bm{v}) - g_0(\alpha \bm{v})| \stackrel{p}{\rightarrow} 0.
	\end{align}
	Moreover, $\hat{g}_0(\bm{v}\cdot)$ is S-shaped by construction.
	
	Finally, a closer inspection of the above proof suggests that  \eqref{eq:proof_ray_eq3} holds uniformly for all $\bm{v}$ such that 
	\[
	\bm{\phi}_{\bm{v}} \in \Big\{\bm{\theta}^{(r)}\Big|r=1,\ldots,R,\quad \bm{\theta}^{(r)} \in [\eta,\pi/2-\eta]^{d-1}\Big\}.
	\]
	As such, for any given $\bm{v}$ with $\min_i v_i > 0$ and $\|\bm{v}\|_2=1$, we could pick $\eta$ in such a way that  $\bm{\phi}_{\bm{v}} \in [2\eta,\pi/2-2\eta]^{d-1}$. As $n \rightarrow \infty$ (so $R\rightarrow\infty$ as well), we could always find $r^* = \argmin_{r \in \{1,\ldots,R\}} D(\bm{\phi}_{\bm{v}}, \bm{\theta}^{(r)})$ satisfying  $D(\bm{\phi}_{\bm{v}}, \bm{\theta}^{(r^*)}) \rightarrow 0$, and thus $\bm{\theta}^{(r^*)} \in [\eta,\pi/2-\eta]^{d-1}$. Write $\bm{v}^*={\bm{\phi}^{-1}_{\bm{\theta}^{{(r)}^*}})}$. We have that $\|{\bm{v}}-{\bm{v}^*} \|\rightarrow 0$ and 
	$|c_{\bm{v}}-c_{\bm{v}^*}| \rightarrow 0$. Therefore,
	\begin{align*}
	\sup_{\alpha \in [\delta,c_{\bm{v}}-\delta]}|\hat{g}_0(\alpha \bm{v}) - g_0(\alpha \bm{v})| &\le  
		\sup_{\alpha \in [\delta,c_{\bm{v}}-\delta]}|\hat{g}_0(\alpha \bm{v}^*) - g_0(\alpha \bm{v}^*)|  
	+ 	\sup_{\alpha \in [\delta,c_{\bm{v}}-\delta]}| g_0(\alpha \bm{v}^*) - g_0(\alpha \bm{v})| \\
	&=: \mathrm{(M1)} + \mathrm{(M2)}.
	\end{align*}
	$\mathrm{(M1)}\stackrel{p}{\rightarrow} 0$ due to the facts that $|c_{\bm{v}}-c_{\bm{v}^*}| \rightarrow 0$, $\bm{\theta}^{(r^*)} \in [\eta,\pi/2-\eta]^{d-1}$ and \eqref{eq:proof_ray_eq3}. In addition,  $\mathrm{(M2)}{\rightarrow} 0$ because $g_0$ is continuous and that $\|{\bm{v}}-{\bm{v}^*} \|\rightarrow 0$. Consequently,
	\[
		\sup_{\alpha \in [\delta,c_{\bm{v}}-\delta]}|\hat{g}_0(\alpha \bm{v}) - g_0(\alpha \bm{v})|\stackrel{p}{\rightarrow} 0.
	\]

\end{proof}

\subsubsection{Proof of Theorem~\ref{thm:homo_consistency1}}
\begin{proof}

At each level $y^{(i)}$ for $i=\lceil \delta I\rceil, \ldots, \lfloor (1-\delta) I\rfloor$, we denote $\lambda_y$ the scalar such that $\lambda(1,\ldots,1)^T$ is on the true isoquant of the level $y$. 

First, we show that $\hat{\lambda}_{y^{(i)}} \stackrel{p}{\rightarrow} \lambda_{y^{(i)}}$. By definition, 
\begin{align*}
\lambda_{y^{(i)}} &= \mathscr{H}_{0,d}(\lambda_{y^{(i)}}(1,\ldots,1)^T; y^{(i)})\\
\hat{\lambda}_{y^{(i)}} &= \hat{\mathscr{H}}_{0,d}(\hat{\lambda}_{y^{(i)}}(1,\ldots,1)^T; y^{(i)}).
\end{align*} 
Here the existence of $\lambda_{y^{(i)}}$ is guaranteed by the fact $\mathrm{argmax}_{\bm{x} \in \bm{S}} g_0(\bm{x}) = (c,\ldots,c)^T$, which follows from the monotonicity of the isoquants, and the fact that $\bm{S}$ (i.e. the domain of the product function) is $[0,c]^d$.	 
It is clear that $\lambda(1,\ldots,1)^T \in \mathbb{R}^{d-1}$ lies in the interior of $\bm{C}_{y^{(i)}}$, so we have that there exists some small $\epsilon > 0$ such that both $(\lambda-\epsilon)(1,\ldots,1)^T$ and $(\lambda+\epsilon)(1,\ldots,1)^T$ (all are $(d-1)$-dimensional vector) lie in the interior of $\bm{C}_{y^{(i)}}$. By Theorem~\ref{thm:nonhomo_consistency2}, 
\begin{align*}
\hat{\mathscr{H}}_{0,d}((\lambda_{y^{(i)}}+\epsilon)(1,\ldots,1)^T; y^{(i)}) &\stackrel{p}{\rightarrow} {\mathscr{H}}_{0,d}((\lambda_{y^{(i)}}+\epsilon)(1,\ldots,1)^T; y^{(i)}) < \lambda_{y^{(i)}} < \lambda_{y^{(i)}} + \epsilon\\ \hat{\mathscr{H}}_{0,d}((\lambda_{y^{(i)}}-\epsilon)(1,\ldots,1)^T; y^{(i)}) &\stackrel{p}{\rightarrow} {\mathscr{H}}_{0,d}((\lambda_{y^{(i)}}-\epsilon)(1,\ldots,1)^T; y^{(i)}) > \lambda_{y^{(i)}} > \lambda_{y^{(i)}} - \epsilon. 
\end{align*}
Due to the monotonicity constraint of $\hat{\mathscr{H}}_{0,d}$, $\hat{\mathscr{H}}_{0,d}(\lambda(1,\ldots,1)^T; y^{(i)})$ is decreasing with respect to $\lambda$. Therefore, we have that $\hat{\lambda}_{y^{(i)}} \in [\lambda_{y^{(i)}} - \epsilon, \lambda_{y^{(i)}} + \epsilon]$. Since $\epsilon$ is arbitrary,  $\hat{\lambda}_{y^{(i)}} \stackrel{p}{\rightarrow} \lambda_{y^{(i)}}$.

Second, a closer inspection of Theorem~\ref{thm:nonhomo_consistency2} shows that it holds uniformly for all $y^{(i)}$ with $i=\lceil \delta I\rceil, \ldots, \lfloor (1-\delta) I\rfloor$. It then follows that
\[
\max_{i=\lceil \delta I\rceil, \ldots, \lfloor (1-\delta) I\rfloor}|\hat{\lambda}_{y^{(i)}} - \lambda_{y^{(i)}}|  \stackrel{p}{\rightarrow} 0.
\]
In fact, since $\lambda_y$ is continuous with respect to $y$, we have that
\begin{align}
\label{Eq:proof_homo_1_1}
\max_{\Big\{j: \bm{X}_j \in \mathcal{I}_i \Big| i \in \{\lceil \delta I\rceil, \ldots, \lfloor (1-\delta) I\rfloor\}\Big\}}|\hat{\lambda}_{y^{(i)}} - \lambda_{g_0(\bm{X}_j)}|  \stackrel{p}{\rightarrow} 0.
\end{align}

Note that that $\mathscr{H}_{0,d}\big(\bm{X}_{j,-d};g_0(\bm{X}_j)\big)=\bm{X}_{j,d}$. Setting 
\[
\mathcal{I} = \Big\{j: \bm{X}_j \in \mathcal{I}_i \Big| i \in \{\lceil \delta I\rceil, \ldots, \lfloor (1-\delta) I\rfloor\}\Big\}
\]
and for every observation $\bm{X}_j \in \mathcal{I}_i$ with $i\in \{\lceil \delta I\rceil, \ldots, \lfloor (1-\delta) I\rfloor\}$, rewriting $\tilde{\bm{X}}_j = \hat{\lambda}_{y^{(i)}}^{-1} \bm{X}_{j}$. Then, because of the homotheticity,
\[
\mathscr{H}_{0,d}\Big(\tilde{\bm{X}}_{j,-d}; F_0(\hat{\lambda}_{y^{(i)}}^{-1} \mathscr{H}_0(\bm{X}_j)) \Big) = \mathscr{H}_{0,d}\Big(\hat{\lambda}_{y^{(i)}}^{-1} \bm{X}_{j,-d};F_0(\hat{\lambda}_{y^{(i)}}^{-1} \mathscr{H}_0(\bm{X}_j))\Big)= \hat{\lambda}_{y^{(i)}}^{-1} \bm{X}_{j,d} = \tilde{\bm{X}}_{j,d}.
\]
Note that here we also defined $\mathscr{H}_{0,d}(\cdot; y)$ at any $y>0$ on the domain of $(0,\infty)^d$ in a meaningful manner due to the homothetic condition. It then follows from \eqref{Eq:proof_homo_1_1} and the homotheticity (with the identifiability condition of $\mathscr{H}_0((1,\ldots,1)^T)=1$) that 
\[
F_0(\hat{\lambda}_{y^{(i)}}^{-1} \mathscr{H}_0(\bm{X}_j) = F_0(\hat{\lambda}_{y^{(i)}}^{-1} \mathscr{H}_0(\lambda_{g_0(\bm{X}_j)}(1,\ldots,1)^T) = F_0(\hat{\lambda}_{y^{(i)}}^{-1} \lambda_{g_0(\bm{X}_j)}) \rightarrow F_0(1) 
\]
in probability (uniformly over $j$). Consequently, we have that
\[
\max_{j \in \mathcal{I}} \Big|\mathscr{H}_{0,d}\Big(\tilde{\bm{X}}_{j,-d}; F_0(1) \Big) - \tilde{\bm{X}}_{j,d}\Big| \stackrel{p}{\rightarrow} 0.
\]

We are now in the position to show the consistency of our isoquant estimator under homotheticity. When applying CNLS, we have that 
\begin{align}
\label{Eq:proof_homo_1_2}
\inf_{h \in \mathcal{G}_{d-1}} \frac{1}{n} \sum_{j \in \mathcal{I}} \Big(\tilde{\bm{X}}_{j,d}- h(\bm{X}_{j,-d})\Big)^2 
\le   \frac{1}{n} \sum_{j \in \mathcal{I}}  \Big(\tilde{\bm{X}}_{j,d}- \mathscr{H}_{0,d}(\tilde{\bm{X}}_{j,-d}; y)\Big)^2 \stackrel{p}{\rightarrow} 0.
\end{align}
The rest of the proof is similar to that of Theorem~\ref{thm:nonhomo_consistency1}. To give more details, suppose that 
$$
\sup_{\bm{x}_{-d} \in \bm{C}'} |\hat{\mathscr{H}}_{0,d}(\bm{x}_{-d};F_0(1))-{\mathscr{H}}_{0,d}(\bm{x}_{-d};F_0(1))| > \epsilon
$$
for some  $\epsilon$ that is smaller than the Hausdorff distance between $\bm{C}'$ and $\bm{C}_y$, and that the supremum occurs at $\bm{x}^*_{-d}$. Then, by the monotonicity constraint and the fact that ${\mathscr{H}}_{0,d}$ is Lipschitz continuous, we could find some $C \in (0,1)$ (that only depends on $g_0$ but not $\bm{x}^*_{-d}$), such that
\[
	\sup_{\bm{x}_{-d} \in B(x^*_{-d}, C \epsilon)} \Big|\hat{\mathscr{H}}_{0,d}(\bm{x}_{-d};F_0(1))-{\mathscr{H}}_{0,d}(\bm{x}_{-d};F_0(1))\Big| > \epsilon/2,
\]
where $B(\bm{x},r)$ is the closed ball centered at $\bm{x}$ of radius $r$.  This means that for sufficiently large $n$,
	\begin{align}
	\notag& \frac{1}{n} \sum_{j \in \mathcal{I}}  \Big(\tilde{\bm{X}}_{j,d}- \hat{\mathscr{H}}_{0,d}(\bm{X}_{j,-d};F_0(1))\Big)^2 \\
	\notag&\ge	 \frac{1}{n}\sum_{\big\{j \in \mathcal{I} \;|\;  \tilde{\bm{X}}_{j,-d} \in B(\bm{x}^*{-d}, C\epsilon)\big\}} \Big(\tilde{\bm{X}}_{j,d}-\mathscr{H}_{0,d}(\tilde{\bm{X}}_{j,-d};F_0(1))+\mathscr{H}_{0,d}(\tilde{\bm{X}}_{j,-d};F_0(1))- \hat{\mathscr{H}}_{0,d}(\tilde{\bm{X}}_{j,-d};F_0(1))\Big)^2\\
	\notag&\ge	 \frac{1}{n} \sum_{\big\{j \in \mathcal{I} \;|\;  \tilde{\bm{X}}_{j,-d} \in B(\bm{x}^*_{-d}, C\epsilon)\big\}} \big(\epsilon/4\big)^2\\
	\label{Eq:proof_homo_1_3} &=  (\epsilon/4)^2  \frac{1}{n}  \Big|\big\{j \in \mathcal{I} \;|\;  \tilde{\bm{X}}_{j,-d} \in B(\bm{x}^*_{-d}, C\epsilon)\big\}\Big|.
	\end{align}
	For any given $\bm{x}^*_{-d} \in \bm{C}'$, write $\bm{x}^* = \begin{bmatrix}\bm{x}^*_{-d} \\ \mathscr{H}_{0,d}(\bm{x}^*_{-d};F_0(1))\end{bmatrix}$. Note that for $\bm{z} = \lambda \begin{bmatrix}\bm{x}_{-d} \\ \mathscr{H}_{0,d}(\bm{x}_{-d};F_0(1))\end{bmatrix}$ with any $\lambda >0$,
	$
	\lambda_{g_0(\bm{z})}^{-1}\bm{z} = \begin{bmatrix}\bm{x}_{-d} \\ \mathscr{H}_{0,d}(\bm{x}_{-d};F_0(1))\end{bmatrix}.
	$
	This, combined with \eqref{Eq:proof_homo_1_1} implies that
	\begin{align*}
	&\mathcal{I}^{\bm{x}^*_{-d}}:= \Bigg\{j \in \{1,\ldots,n \} \;\Bigg|\; \bm{X}_j \in \Big\{\lambda \bm{x}\;\Big|\; \lambda \in \Big[(\delta+\epsilon)c_{\bm{x}^*},(1- \delta - \epsilon) c_{\bm{x}^*}\Big], \bm{x} \in B\Big(\bm{x}^*_{-d},C\epsilon\Big) \Big\}\Bigg\} \\
	&\subset \big\{j \in \mathcal{I} \;|\;  \tilde{\bm{X}}_{j,-d} \in B(\bm{x}^*_{-d}, C\epsilon)\big\}.
	\end{align*}
	Since the class of sets $\Big\{\lambda \bm{x}\;\Big|\; \lambda \in \Big[(\delta+\epsilon)c_{\bm{x}^*},(1- \delta - \epsilon) c_{\bm{x}^*}\Big], \bm{x} \in B\Big(\bm{x}^*_{-d},C\epsilon\Big) \Big\}\Bigg\}$  over all $x^*_{-d} \in \bm{C}'$ is Glivenko-Cantelli (as its elements are necesarily bounded an convex), we have that in probability,
	\begin{align}
	\label{Eq:proof_homo_1_4}
	\inf_{x^*_{-d} \in \bm{C}'} \lim_{n \rightarrow \infty}\frac{1}{n}  \Big|\big\{j \in \mathcal{I} \;|\;  \tilde{\bm{X}}_{j,-d} \in B(\bm{x}^*_{-d}, C\epsilon)\big\}\Big| > 0.
	\end{align}
	Plugging  \eqref{Eq:proof_homo_1_4} into \eqref{Eq:proof_homo_1_3} and comparing it with \eqref{Eq:proof_homo_1_2} leads to a contradiction, and thus 
	$
	\sup_{\bm{x}_{-d} \in \bm{C}'} |\hat{\mathscr{H}}_{0,d}(\bm{x}_{-d};F_0(1))-{\mathscr{H}}_{0,d}(\bm{x}_{-d};F_0(1))| \le \epsilon.
	$
	Finally, as $\epsilon > 0$ could be picked arbitrarily, the proof for the consistency of the estimated isoquant is complete.
\end{proof}

\subsubsection{Proof of Theorem \ref{thm:homo_consistency2}}
\begin{proof}	
For the case of linear isoquants, recall that we aim to find
\begin{align}
\label{eq:proof4_eq}
(\hat{\boldsymbol{\beta}}_0, \hat{F}_0)  \in  \argmin_{ \|\boldsymbol{\beta}\|_1 = 1, \boldsymbol{\beta} \ge \mathbf{0}, F \in \mathcal{F}} \sum_{j=1}^n \big(Y_j - F(\boldsymbol{\beta}^T \bm{X}_j)\big)^2
\end{align}
where $\mathcal{F}$ as the class of increasing and S-shaped functions from $[0,\infty) \rightarrow \mathbb{R}$. This is exactly the single index model, with the link function following the S-shape and increasing constraints, and the index following the non-negativity constraint. Let $\mathcal{F}'$ be the class of increasing functions. Obviously $\mathcal{F} \subset \mathcal{F}'$. If we replace $\mathcal{F}$ by $\mathcal{F}'$ in \eqref{eq:proof4_eq}, then the problem becomes the monotone single index regression, as investigated as a special case in \citet{chensamworth2016}. 

With the additional S-shape constraint and non-negativity index constraint, we are actually considering a smaller class of candidate functions, so all the arguments in the proof of Theorem~2 of \citet{chensamworth2016} would go through with minor modifications. Therefore, we have that 
\[
\sup_{\bm{x}\in \bm{C}} |\hat{F}_0(\hat{\boldsymbol{\beta}}_0^T\bm{x}) - g_0(\bm{x})| \stackrel{p}{\rightarrow} 0,
\] 
for any compact $\bm{C}$ that belongs to the interior of $\bm{S}$.  It then follows from the identifiability of the single index model that $\hat{\boldsymbol{\beta}}_0 \stackrel{p}{\rightarrow} \boldsymbol{\beta}_0$. 

For the case of power isoquants, write $G_0(\cdot) = F_0(\exp(\cdot))$ and $\bm{z}=\log(\bm{x})$, and thus
\[
g_0(\bm{x})=F_0(\bm{x}^{\boldsymbol{\beta}_0})= F_0(\exp({\boldsymbol{\beta}_0}^T\log(\bm{x})))=G_0({\boldsymbol{\beta}}_0^T \bm{z}).
\]
Therefore, our estimator can be rewritten as 
\begin{align*}
(\hat{\boldsymbol{\beta}}_0, \hat{G}_0)  \in  \argmin_{ \|\boldsymbol{\beta}\|_1 = 1, \boldsymbol{\beta} \ge \mathbf{0}, G \in \mathcal{G}} \sum_{j=1}^n \big(Y_j - G(\boldsymbol{\beta}^T \bm{z}_j)\big)^2
\end{align*}
where $\bm{z}_j = \log(\bm{X}_j)$, and where $\mathcal{G}$ is a sub-class of increasing functions. This could again be viewed as the single index model, which means that we could again follow the proof of Theorem~2 of \citet{chensamworth2016} to have that
\[
\sup_{\bm{x}\in \bm{C}} |\hat{F}_0(\hat{\boldsymbol{\beta}}_0^T\bm{x}) - g_0(\bm{x})| \stackrel{p}{\rightarrow} 0,
\] 
for any compact $\bm{C}$ that belongs to the interior of $\bm{S}$. Consequently, $\hat{\boldsymbol{\beta}}_0 \stackrel{p}{\rightarrow} \boldsymbol{\beta}_0$. 
\end{proof}

\newpage
\section{Comparison Between S-shape Definition and the RUP Law}
\label{app:ce_rup}
In this section, we provide an example in which a production function that satisfies the RUP law, Definition \ref{def:rup}, contains multiple inflection points. 

Consider the following univariate example.
\begin{example}
    \[
    g(x)=x^{(1.8)}\exp{\left(-x\right)}\exp{\left(\frac{-x\sin{(100x)}}{10000}\right)}
    \]
\end{example}

\noindent Then we can compute the elasticity of scale and its derivative.
\[
\epsilon(x)=1.8-x\Big\{\frac{\cos{(100x)}}{100}+1\Big\},\\
\]

\begin{figure}[p]
    \includegraphics[width=0.6\textwidth]{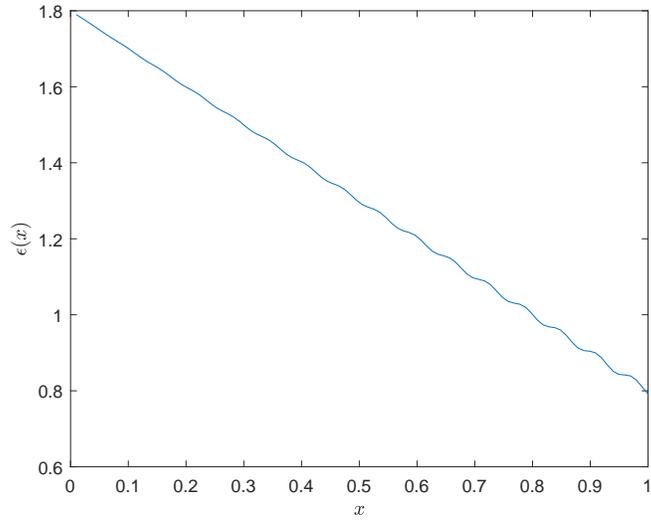}
    \centering
    \caption{The elasticity of scale}
    \label{fig:ex1e}
\end{figure}

\begin{figure}[p]
	\centering
	\subfloat[Production function]{\includegraphics[width=0.33\textwidth]{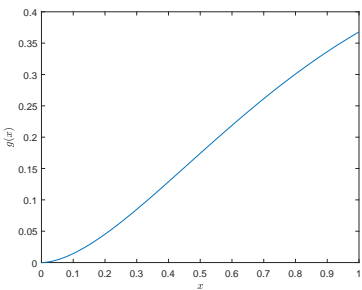}\label{fig:ex1g0}}
	\hfill
	\subfloat[First derivative]{\includegraphics[width=0.33\textwidth]{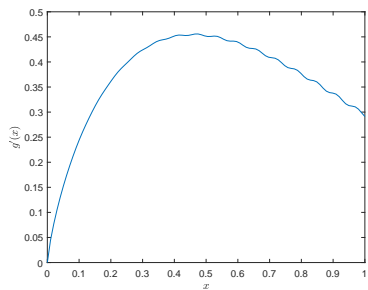}\label{fig:ex1g1}}
	\hfill
	\subfloat[Second derivative]{\includegraphics[width=0.33\textwidth]{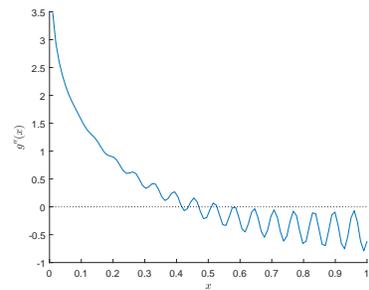}\label{fig:ex1g2}}
	\caption{Production function and its derivatives}
	\label{fig:ex1g}
\end{figure}

\noindent Figure \ref{fig:ex1e} shows the elasticity of scale, $\epsilon(x)$, is monotonically decreasing on $x\in[0,1]$ from 1.8 to 0.8, which satisfies Definition \ref{def:rup}. Figure \ref{fig:ex1g} shows that the production function and its first and second derivative respectively. In Figure \ref{fig:ex1g} \subref{fig:ex1g0}, the production function looks S-shape; however, Figure \ref{fig:ex1g} \subref{fig:ex1g2} shows that the production function has a multiple inflection points as there are multiple intersections between its second derivative $g''(x)$ and constant function at $x=0$. So this is a counterexample of S-shape with the RUP law. Thus, to avoid having multiple inflection points, we added the condition on the second derivative of the function $g_0(\cdot)$ as shown in Definition \ref{def:s-shape}.

\newpage

\section{Bootstrapping to quantify uncertainty of the estimator}
\label{app:uncertainty}
We provide the bootstrapping procedure described in \citet{yagi2018shape} to measure the uncertainty of the estimator by computing the confidence interval. We can also use this procedure to validate whether the shape constraints are fulfilled by the true function $g_0$ or not as shown in \citet{yagi2018shape}.

The bootstrapping procedure has three steps:
	\begin{enumerate}
		\item Estimate the error at each $\bm{X}_j$ by $\tilde{\epsilon}_j = y_j-\tilde{g}_n(\bm{X}_j)$ for $j = 1,\ldots,n$, where $\tilde{g}$ is the unconstrained nonparametric estimator such as local linear.
		
		\item  The wild bootstrap method is used to construct a confidence interval. Let $B$ be the number of Monte Carlo iterations. For every $k = 1,\ldots,B$, let $\bm{u}_k = (u_{1k},\ldots,u_{nk})'$ be a random vector with components sampled independently from the Rademacher distribution, i.e. $P(u_{jk} = 1) = P(u_{jk} = -1) = 0.5$. Furthermore, let $y_{jk} = y_j + u_{jk}\ \tilde{\epsilon}_j$.  Then, the wild bootstrap sample is
    \[
    \{\bm{X}_j, y_{jk}\}_{j=1}^n.
    \]
	\item Obtain the functional estimates with with the bootstrap sample $\hat{g}_{0k}(\bm{x})$ for every $k = 1,\ldots,B$. Then we order the bootstrap estimates and obtain the lower and upper bound by taking the corresponding percentile of the bootstrap estimates. For instance, when we compute the 95\% confidence interval on $\bm{x}$, we set 2.5 and 97.5 percentile of the bootstrap samples $\{\hat{g}_{0k}(\bm{x})\}_{k=1}^{B}$ as the lower and upper bound respectively.
\end{enumerate}	


\newpage

\section{Productivity dispersion among different models}
\label{app:prod_disp}

There are many different models and methods to compute productivity. Here we compare these models by compute the productivity dispersion observed across firms within the industry. We will use three methods to calculate aggregate inputs. The first two methods are described in \cite{syverson2004product} and referred to as growth accounting methods, but we will briefly summarize them here. Aggregated input is estimated by
\begin{equation}
\label{eq:prod_Syverson}
    g_0(L_{jt},K_{jt}) = L_{jt}^{\alpha_L}K_{jt}^{\alpha_K}
\end{equation}
where 
$\alpha_L$ and $\alpha_K$ are factor elasticities used as weights to aggregate the various inputs. These factor elasticities can be approximated either by industry level cost shares or by individual firm cost shares. Since we have individual firm cost shares in our data set, we calculate both.\footnote{Because of the various units of measures used for different inputs, the scale of TFP is not easily interpretable. Thus, we normalize each firms TFP by the median TFP for the industry, following \cite{syverson2004product}.} A third option is to fit a Cobb--Douglas regression,
\[
\ln y_{jt} = \ln g(\bm{X}) + \epsilon = \beta_0 + \beta_K \ln K_{jt} + \beta_L \ln L_{jt} + \epsilon
\]
\[
y_{jt} = g_0(L_{jt},K_{jt})\exp({\epsilon_{jt}}) = \beta_0 \ln L_{jt}^{\beta_L} K_{jt}^{\beta_K} \exp({\epsilon_{jt}})
\]
We calculate the estimates of the Cobb--Douglas production function and substitute them for $\hat{g}(\cdot)$ in Equation \ref{eq:TFP} to calculate TFP. 

Table \ref{tab:ProdVar} summarizes the results of the three methods. Using the industry and firm cost shares results in a 90-10 percentile ratio of 3.97 and 3.56, respectively. This is considerable larger than the the value of 2.68 and 1.91 \cite{syverson2004product} reports as an average across a variety of four digit Standard Industry Classification (SIC) industries in the U.S. economy. 
We find firms in the 90th percentile of the productivity distribution makes almost four times as much output with the same measured inputs as the 10th percentile firm. Using a Cobb--Douglas production function and optimizing the selection of the factor elasticities to best fit the data results in an approximately 35\% drop in productivity ratio compared to growth accounting method using industry level cost shares. 

\begin{table}[htbp]
  \centering
  \caption{The ratio of the 90th to 10th percentile productivity level for four different methods}
    \begin{tabular}{C{2in}C{1in}}
    \toprule
          & \multicolumn{1}{c}{90-10 percentile range}  \\
    \midrule
    Industry Cost Shares & 3.971 \\ 
 Firm Cost Shares & 3.559 \\
Cobb--Douglas & 2.963 \\    
    \bottomrule
    \end{tabular}%
  \label{tab:ProdVar}%
\end{table}%

\newpage

\section{Comprehensive results of productivity decomposition}
\label{app:prod_decomp}
\begin{figure}[t]
    \includegraphics[width=0.99\textwidth]{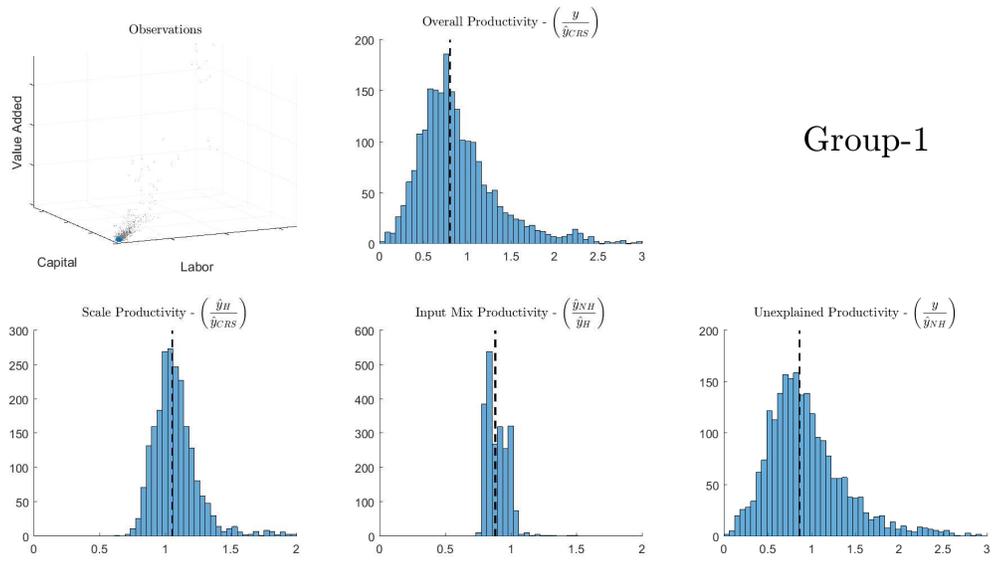}
    \centering
    \captionsetup{justification=centering}
    \caption{Productivity decomposition (Group--1)}
    \label{fig:group_1}
\end{figure}

\begin{figure}[t]
    \includegraphics[width=0.99\textwidth]{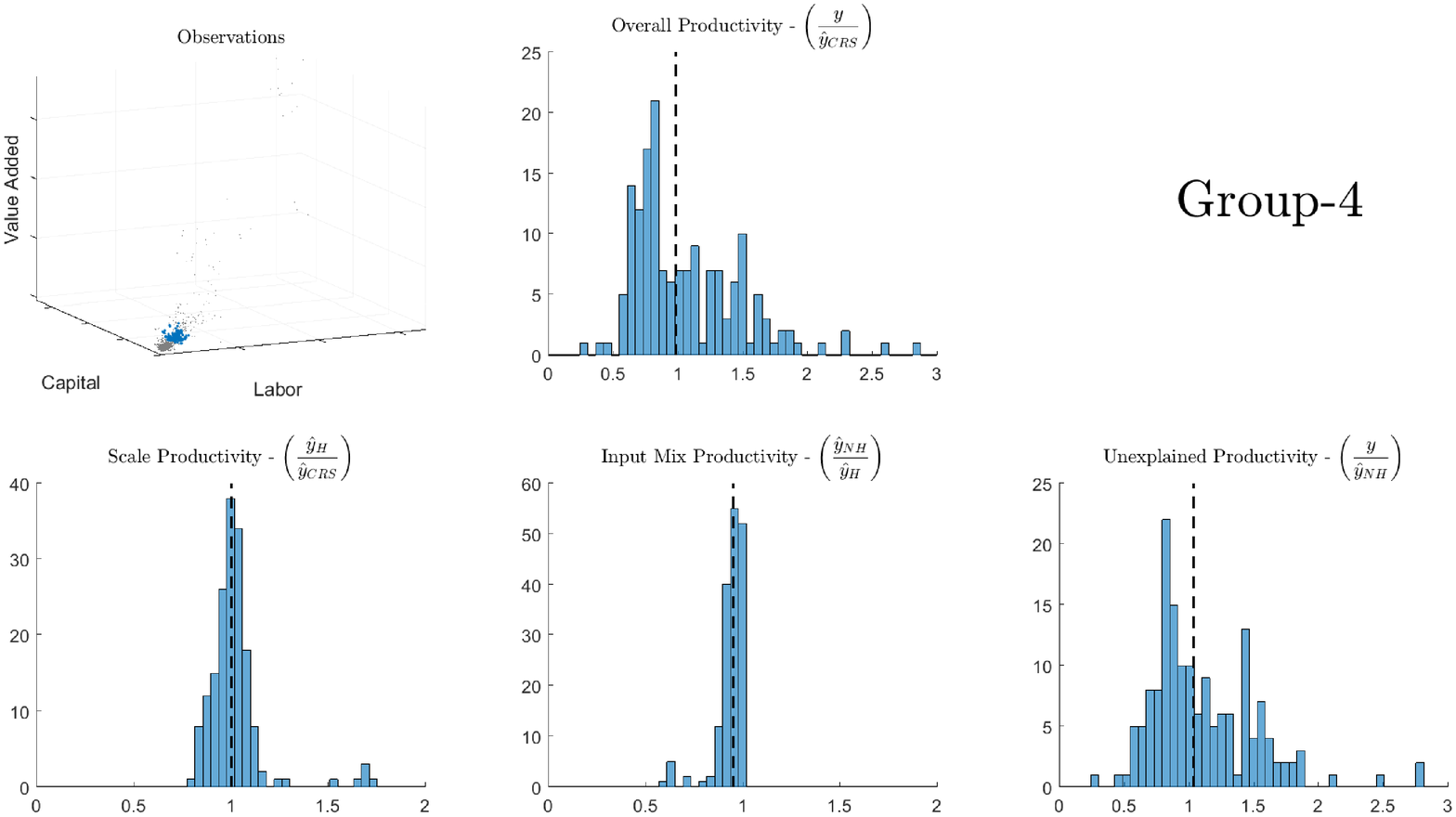}
    \centering
    \captionsetup{justification=centering}
    \caption{Productivity decomposition (Group--4)}
    \label{fig:group_4}
\end{figure}

\begin{figure}[t]
    \includegraphics[width=0.99\textwidth]{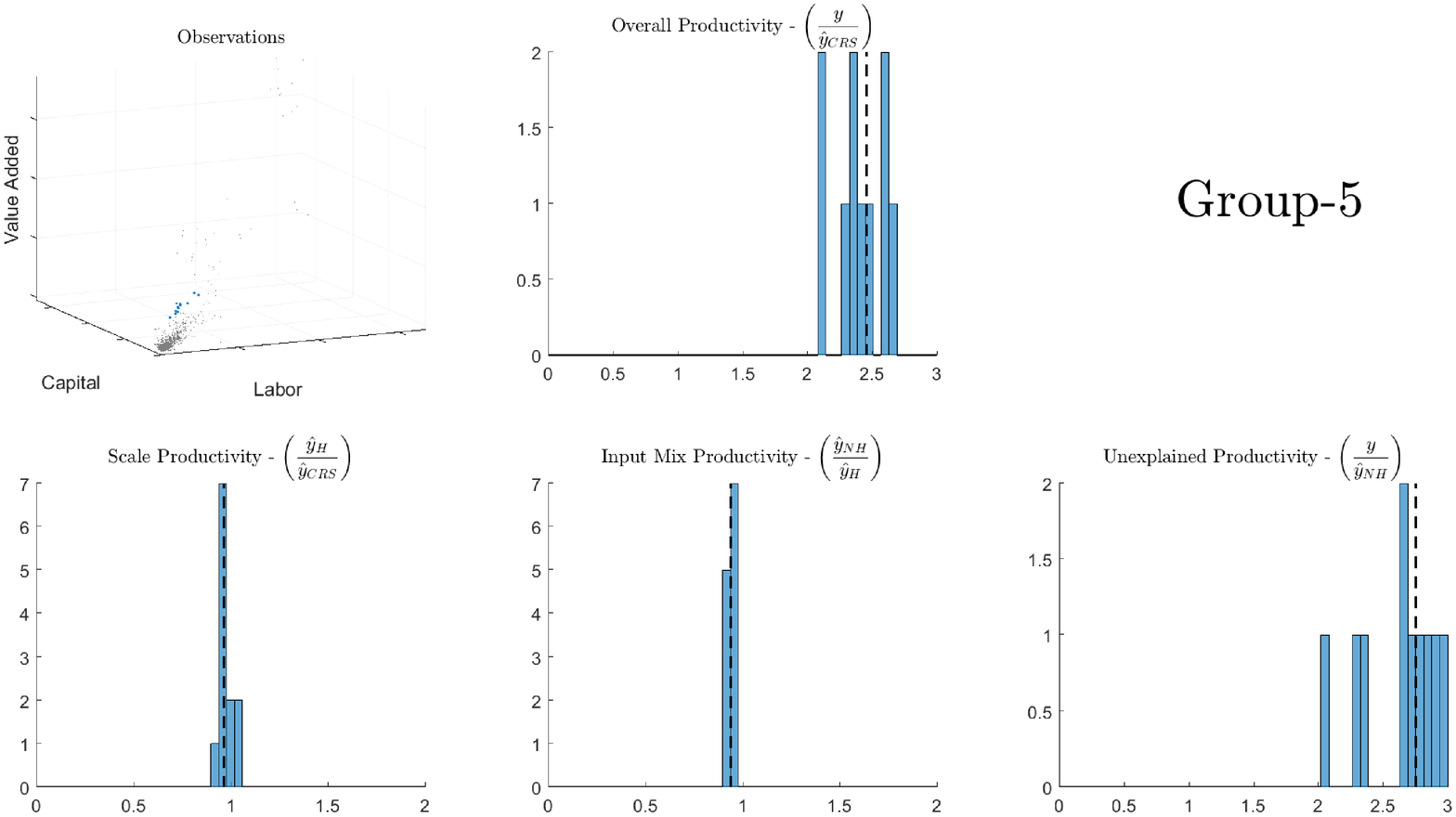}
    \centering
    \captionsetup{justification=centering}
    \caption{Productivity decomposition (Group--5)}
    \label{fig:group_5}
\end{figure}

\begin{figure}[t]
    \includegraphics[width=0.99\textwidth]{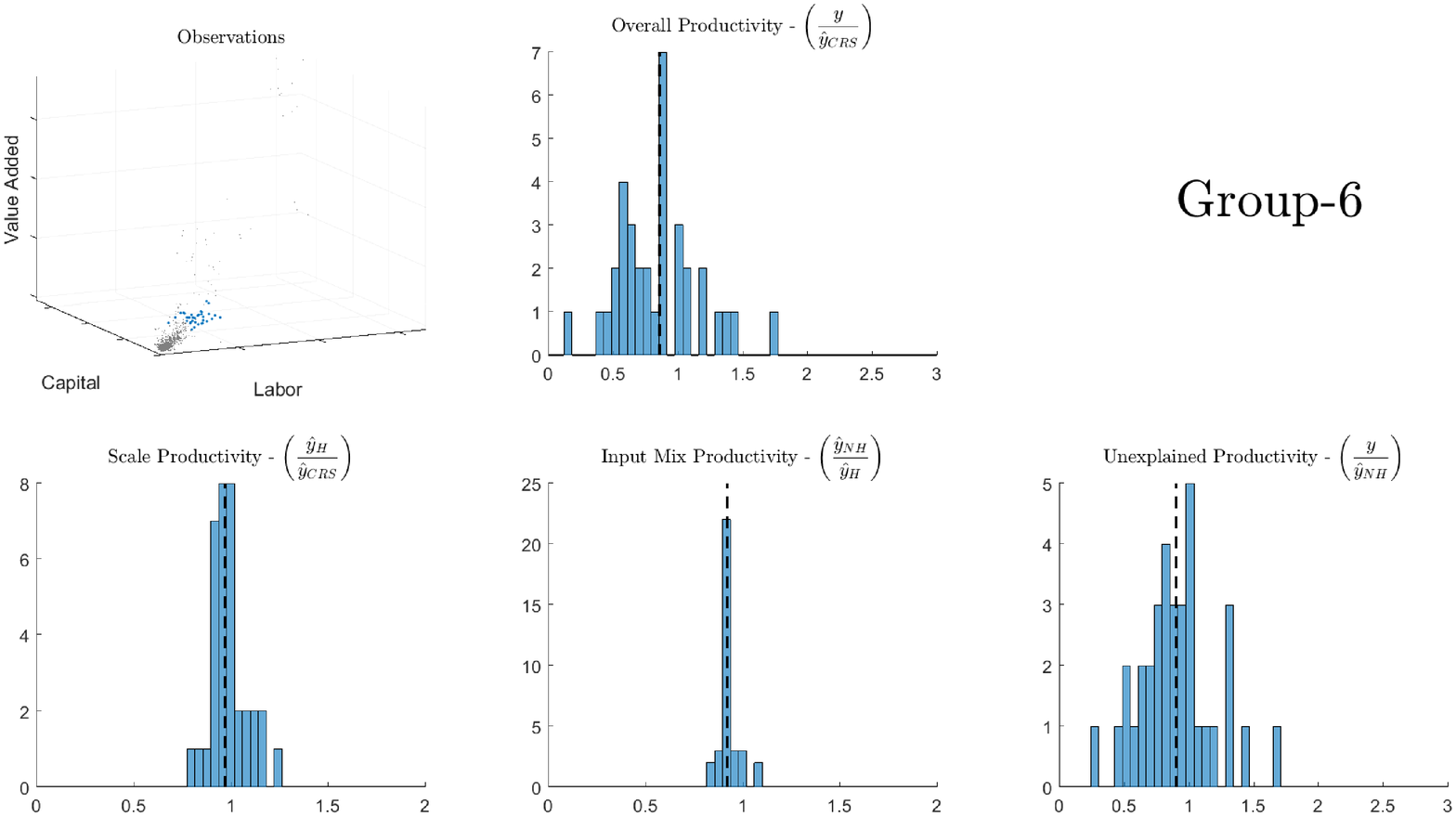}
    \centering
    \captionsetup{justification=centering}
    \caption{Productivity decomposition (Group--6)}
    \label{fig:group_6}
\end{figure}

\begin{figure}[t]
    \includegraphics[width=0.99\textwidth]{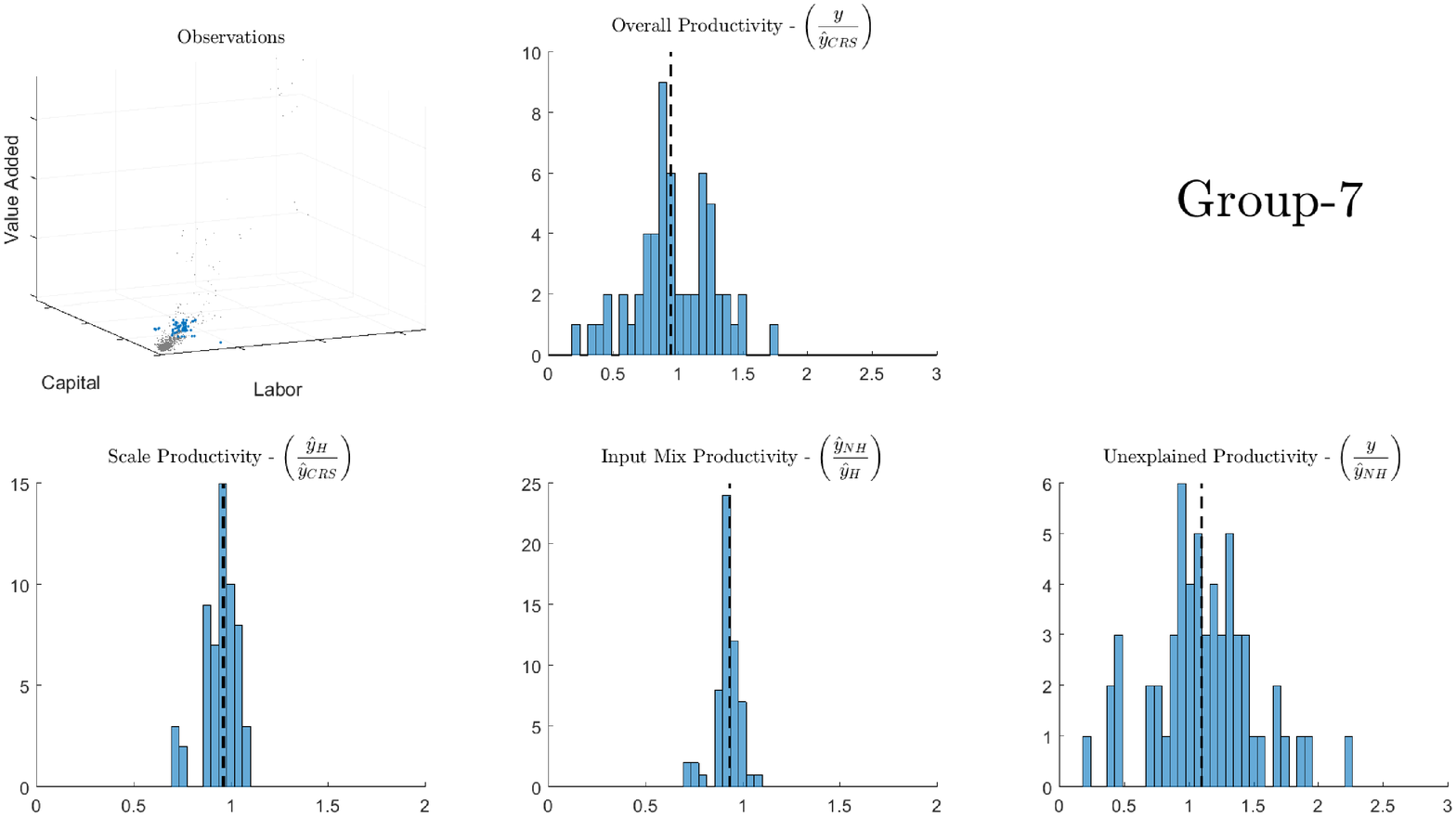}
    \centering
    \captionsetup{justification=centering}
    \caption{Productivity decomposition (Group--7)}
    \label{fig:group_7}
\end{figure}

\begin{figure}[t]
    \includegraphics[width=0.99\textwidth]{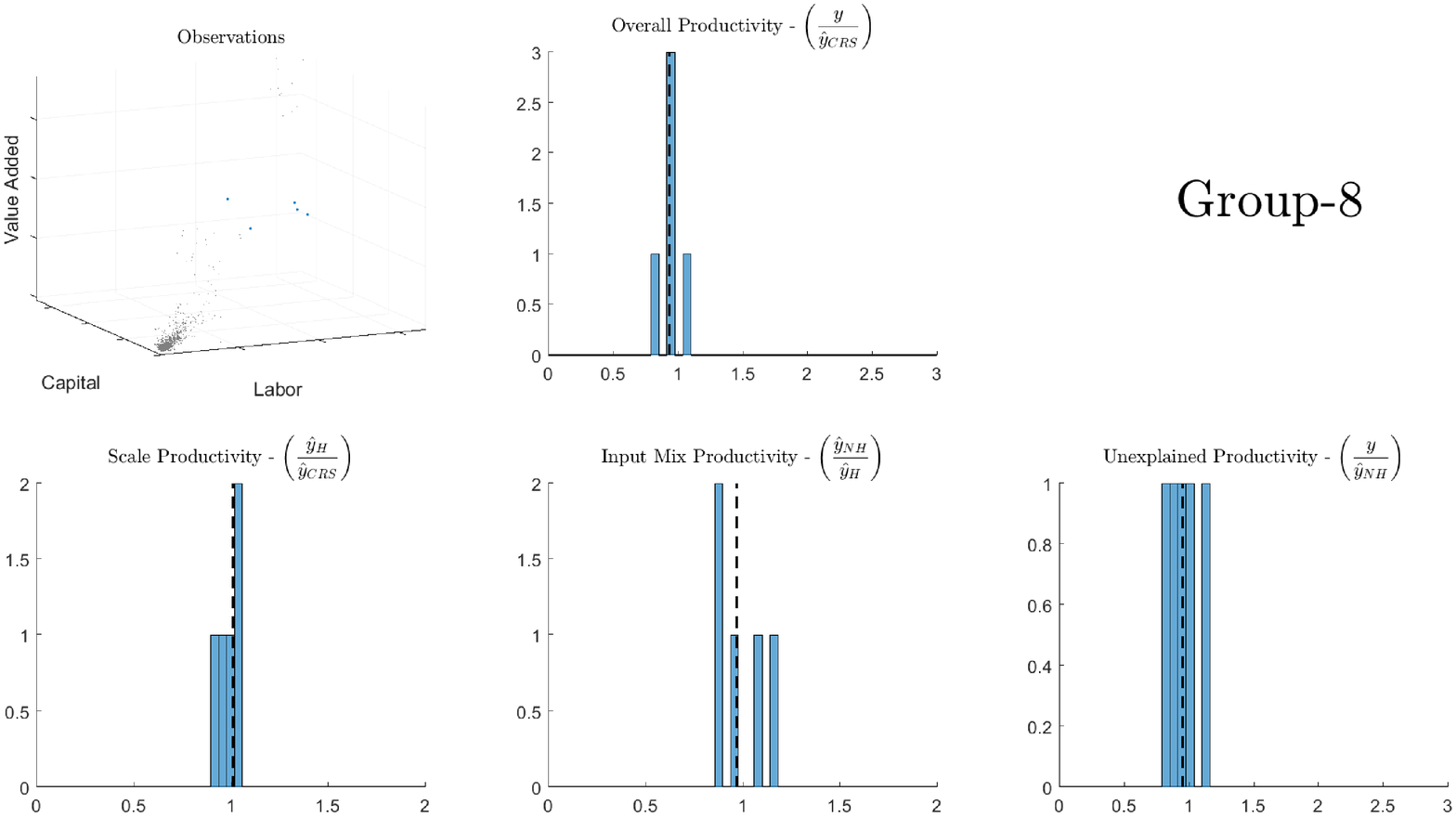}
    \centering
    \captionsetup{justification=centering}
    \caption{Productivity decomposition (Group--8)}
    \label{fig:group_8}
\end{figure}

\begin{figure}[t]
    \includegraphics[width=0.99\textwidth]{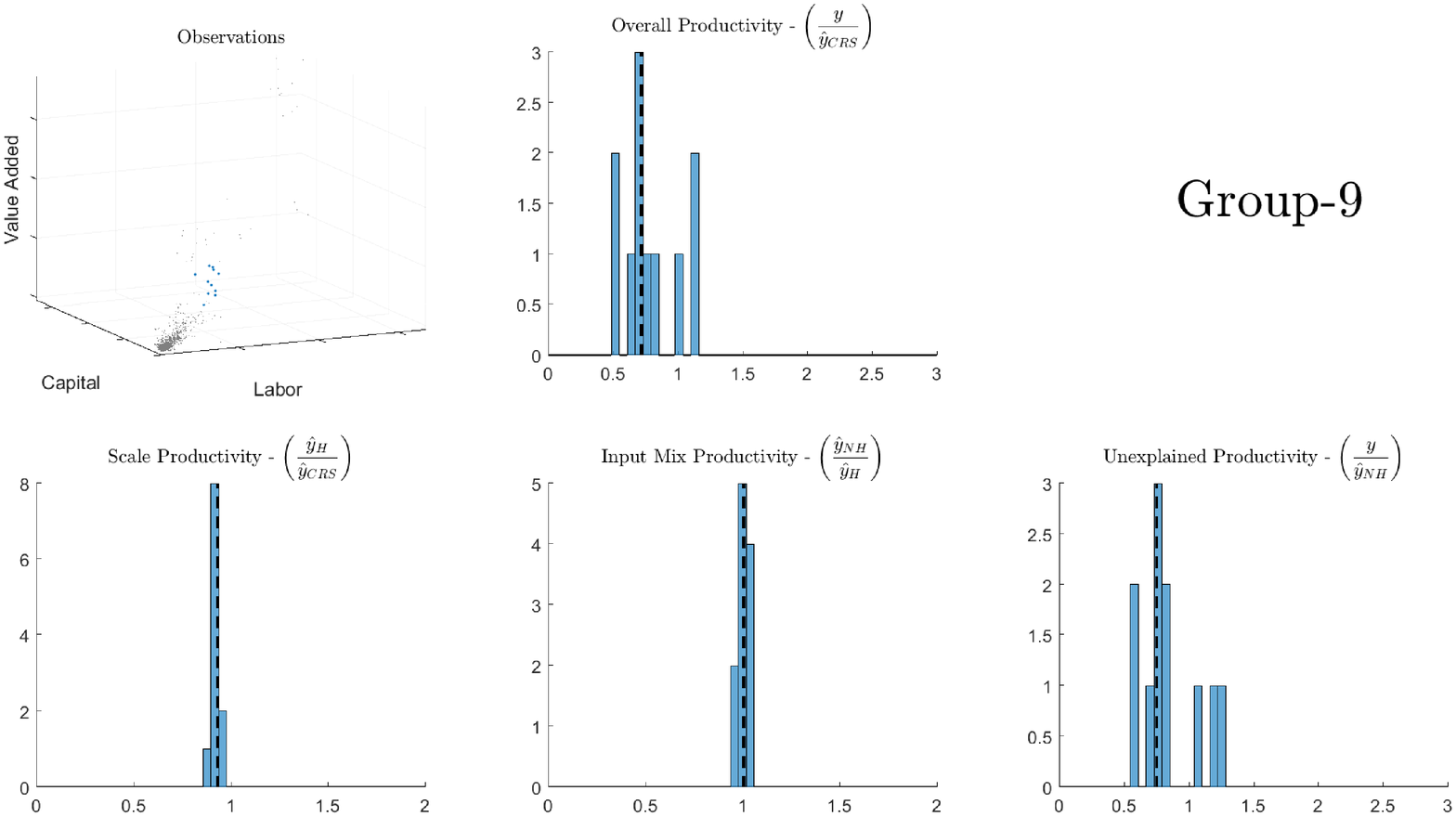}
    \centering
    \captionsetup{justification=centering}
    \caption{Productivity decomposition (Group--9)}
    \label{fig:group_9}
\end{figure}

\begin{figure}[t]
    \includegraphics[width=0.99\textwidth]{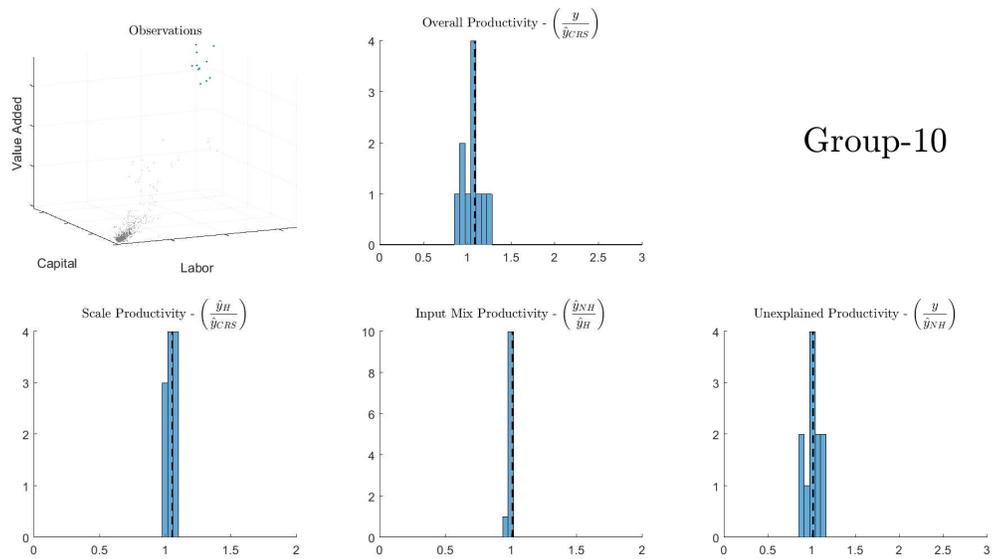}
    \centering
    \captionsetup{justification=centering}
    \caption{Productivity decomposition (Group--10)}
    \label{fig:group_10}
\end{figure}

\begin{figure}[t]
    \includegraphics[width=0.99\textwidth]{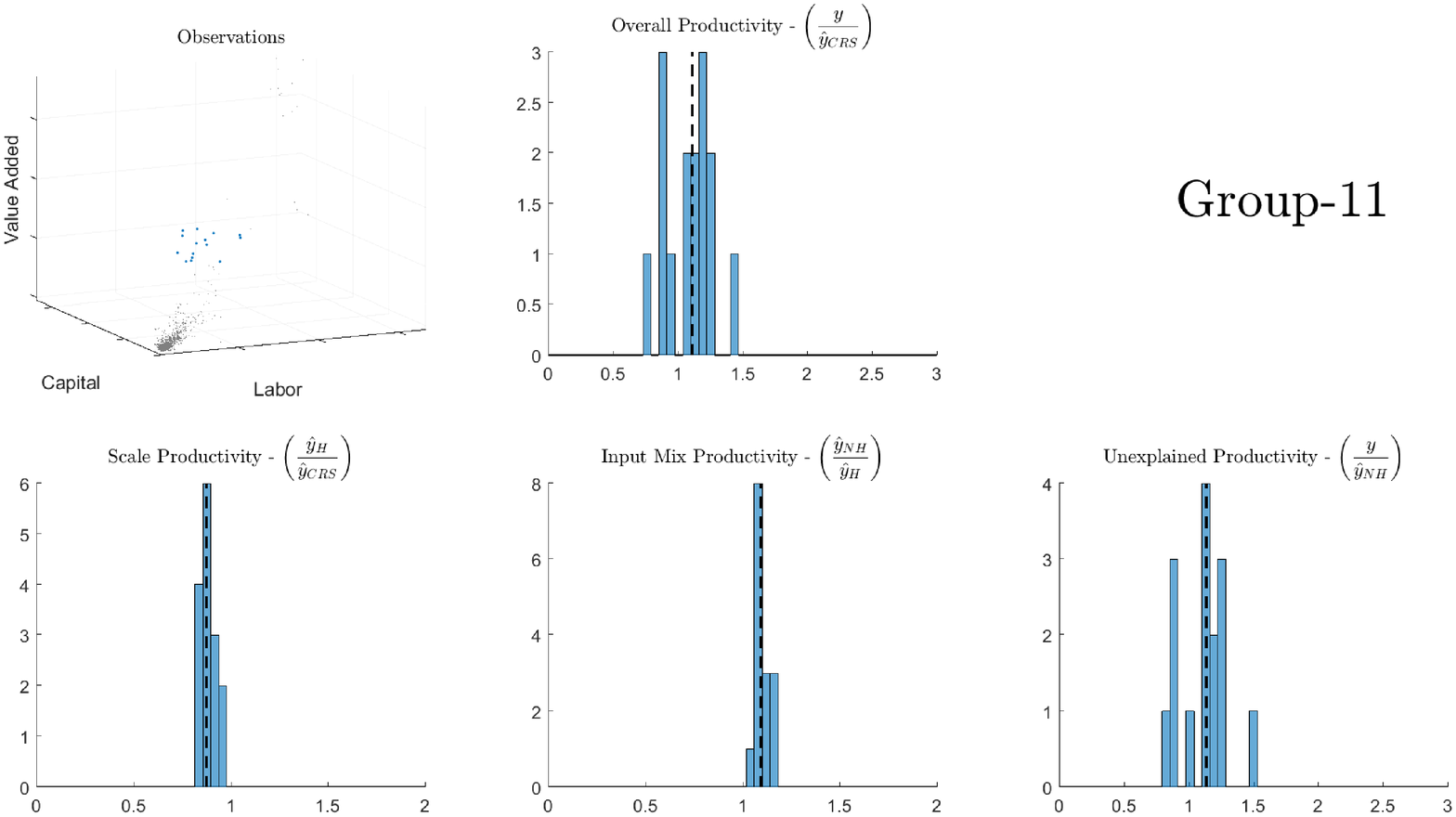}
    \centering
    \captionsetup{justification=centering}
    \caption{Productivity decomposition (Group--11)}
    \label{fig:group_11}
\end{figure}

\begin{figure}[t]
    \includegraphics[width=0.99\textwidth]{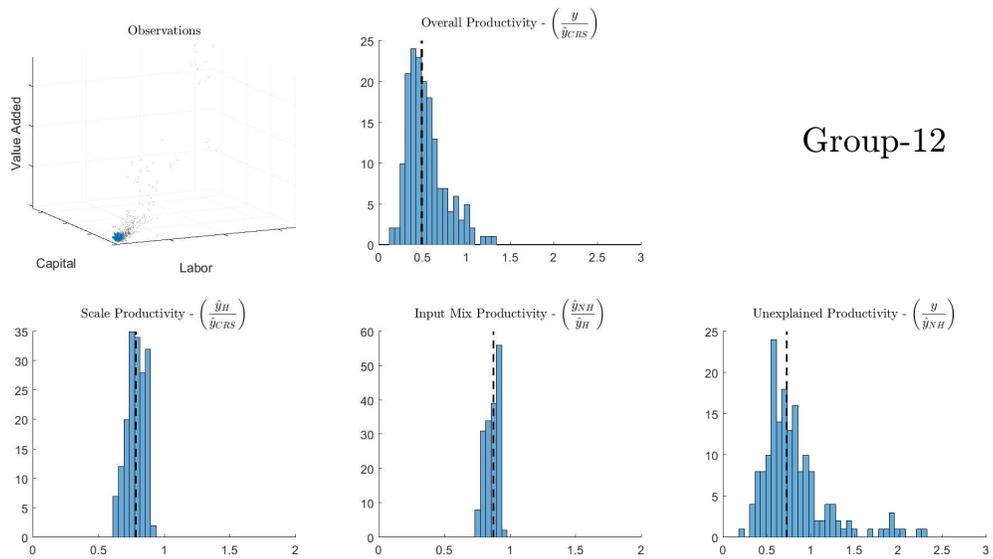}
    \centering
    \captionsetup{justification=centering}
    \caption{Productivity decomposition (Group--12)}
    \label{fig:group_12}
\end{figure}

We show the comprehensive productivity decomposition results for all 12 groups we defined by using K--means clustering in Section \ref{subsec:appsetup}. The groups are arranged in the ascending order of capital intensity. Figures \ref{fig:group_1} thorugh \ref{fig:group_4} are the group with labor intensive firms with relatively low value added amount. Labor intensive firms operate at small scales, both scale productivity and input mix productivity are close to one for these groups. Figure \ref{fig:group_5} is composed by the firms with medium size and large value added. These firms likely have better management strategies than other firms, and thus have a much higher productivity level than firms in other groups. Figure \ref{fig:group_6} and \ref{fig:group_7} show the productivity decomposition of the medium size firms with relatively high capital intensity. Since these firms are capital intensive, they are able to increase productivity by either increasing their scale size or if the firm cannot expand production, then adjusting their input mix to become more labor intensive will improve productivity. We can see their performance in the measures of scale productivity and input mix productivity that are slightly lower than one. Figure \ref{fig:group_8} through \ref{fig:group_11} show the groups of capital intensive firms operated with a large scale. Since these firms are capital intensive, they benefit from operating at a large scale. Finally, \ref{fig:group_12} is the most capital intensive group, but the firms in this group are operating at relatively low scales of production. Thus we can observe that both decomposed productivity measures are significantly lower than one, which indicates that the firms in this group should increase their scale size or adjust their input ratio to increase the productivity.

\begin{figure}[t]
    \includegraphics[width=0.99\textwidth]{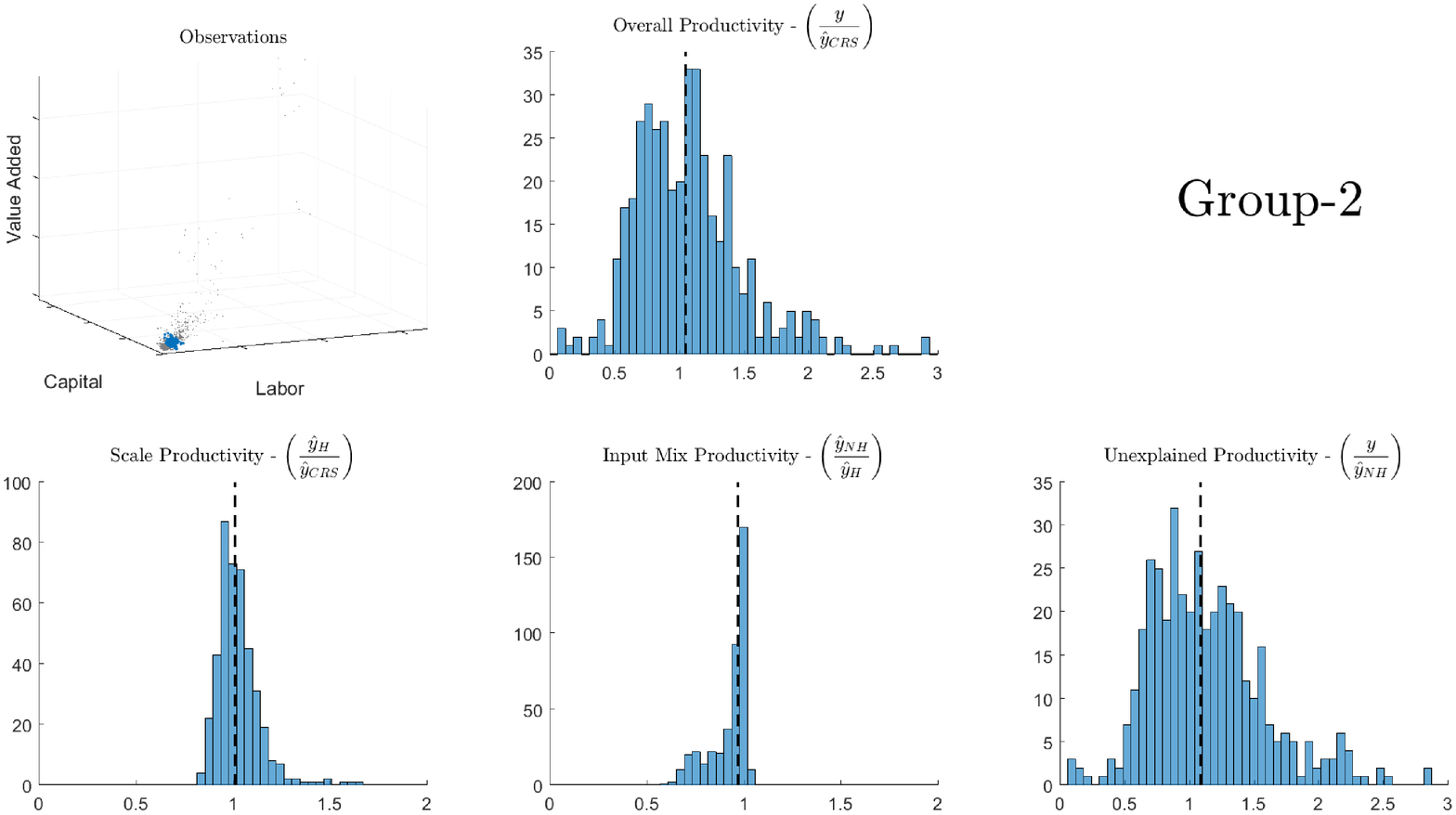}
    \centering
    \captionsetup{justification=centering}
    \caption{Productivity decomposition (Group--2)}
    \label{fig:group_2}
\end{figure}
\begin{figure}[t]
    \includegraphics[width=0.99\textwidth]{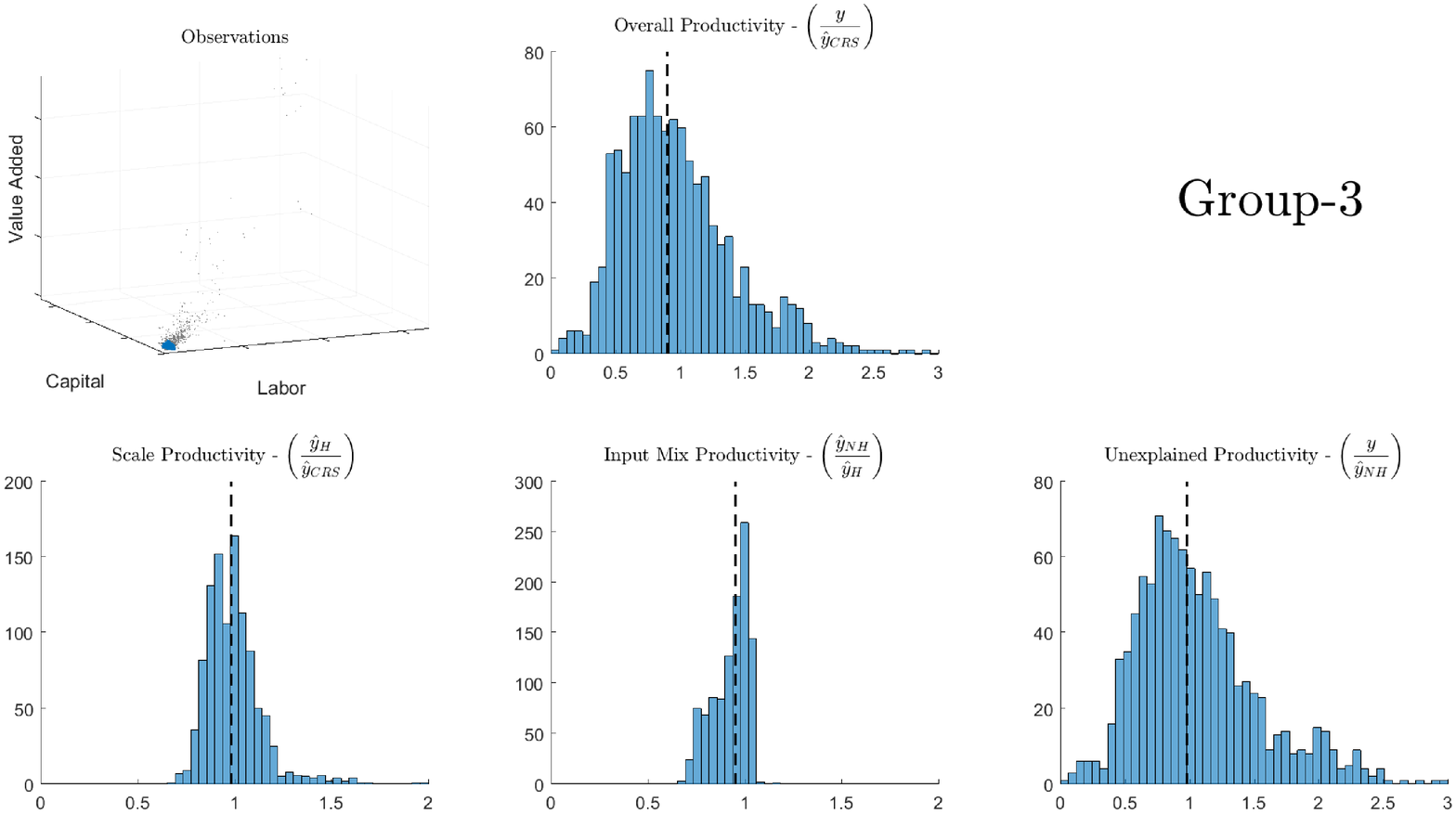}
    \centering
    \captionsetup{justification=centering}
    \caption{Productivity decomposition (Group--3)}
    \label{fig:group_3}
\end{figure}

\clearpage
\bibliographystyle{chicago}	
\bibliography{reference}

\end{document}